\documentclass[journal]{IEEEtran}

\usepackage[pdftex]{graphicx}
\graphicspath{{../pdf/}{../jpeg/}{images/}}
\DeclareGraphicsExtensions{.pdf,.jpeg,.png}

\usepackage{xcolor}

\usepackage{adjustbox}

\usepackage{amsmath,amssymb,amsfonts}
\usepackage{bm}
\usepackage{nccmath}

\usepackage{xfp}
\usepackage{algorithmic}

\usepackage[caption=false,font=footnotesize]{subfig}

\usepackage{url}

\usepackage{xspace} 
\newcommand{\initiator}{initiator\xspace}
\newcommand{\responder}{responder\xspace}

\usepackage{siunitx} 
\usepackage{array}
\usepackage{tabu}
\usepackage{tabularx}
\usepackage{makecell}
\usepackage{multicol}
\usepackage{multirow} 
\usepackage{booktabs}
\usepackage{threeparttable}

\definecolor{codegray}{gray}{0.9}
\definecolor{mygray}{rgb}{0.4,0.4,0.4}
\definecolor{mygreen}{rgb}{0,0.8,0.6}
\definecolor{myorange}{rgb}{1.0,0.4,0}

\usepackage[normalem]{ulem} 

\usepackage{soul}

\newcommand{\modi}[1]{{\color{black}#1}}
\usepackage{tikz}

\newcommand{\copyrighttext}{%
	\footnotesize \textcopyright 2021 IEEE. Personal use of this material is permitted.
	Permission from IEEE must be obtained for all other uses, in any current or future
	media, including reprinting/republishing this material for advertising or promotional
	purposes, creating new collective works, for resale or redistribution to servers or
	lists, or reuse of any copyrighted component of this work in other works.
	}
\newcommand{\copyrightnotice}{%
	\begin{tikzpicture}[remember picture,overlay]
	\node[anchor=south,yshift=10pt] at (current page.south) {\fbox{\parbox{\dimexpr\textwidth-\fboxsep-\fboxrule\relax}{\copyrighttext}}};
	\end{tikzpicture}%
}

\begin{document}
	\title{High-Accuracy Ranging and Localization with Ultra-Wideband Communications for Energy-Constrained Devices}

	\author{Laura~Flueratoru,~\IEEEmembership{Graduate Student Member,~IEEE,}
		Silvan~Wehrli,~\IEEEmembership{Member,~IEEE,}
		Michele~Magno,~\IEEEmembership{Senior Member,~IEEE,}
		Elena~Simona~Lohan,~\IEEEmembership{Senior Member,~IEEE,}
		and~Dragoș~Niculescu
		\thanks{This work was supported by funding from European Union's Horizon 2020 Research and Innovation programme under the Marie Sk\l{}odowska Curie grant agreement No.\ $813278$ (A-WEAR: A network for dynamic wearable applications with privacy constraints, http://www.a-wear.eu/). This article is based on a previous conference paper~\cite{flueratoru2020energy} presented at the IEEE Global Communications Conference (GLOBECOM), Dec.\ 2020. (\textit{Corresponding author: Laura Flueratoru.})}%
		\thanks{Laura Flueratoru is with the Department of Computer Science, University Politehnica of Bucharest, Romania, and with the Department of Electrical Engineering, Tampere University, Finland (e-mail: laura.flueratoru@upb.ro).}
		\thanks{Silvan Wehrli is with 3db Access AG (e-mail: silvan.wehrli@3db-technologies.com).}%
		\thanks{Michele Magno is with the Project-Based Learning Center, ETH Z\"urich, Switzerland (e-mail: michele.magno@pbl.ee.ethz.ch).}%
		\thanks{Elena Simona Lohan is with the Department of Electrical Engineering, Tampere University, Finland (e-mail: elena-simona.lohan@tuni.fi).}%
		\thanks{Dragoș Niculescu is with the Department of Computer Science, University Politehnica of Bucharest, Romania (e-mail: dragos.niculescu@upb.ro).}%
	}
	
	\markboth{}%
	{Shell \MakeLowercase{\textit{et al.}}: High-Accuracy Ranging and Localization with Ultra-Wideband Communications for Energy-Constrained Devices}
	
	\maketitle
	\copyrightnotice

	\begin{abstract}
Ultra-wideband (UWB) communications have gained popularity in recent years for being able to provide distance measurements and localization with high accuracy, which can enhance the capabilities of devices in the Internet of Things (IoT). Since energy efficiency is of utmost concern in such applications, in this work we evaluate the power and energy consumption, distance measurements, and localization performance of two types of UWB physical interfaces (PHYs), which use either a low- or high-rate pulse repetition (LRP and HRP, respectively). The evaluation is done through measurements acquired in identical conditions, which is crucial in order to have a fair comparison between the devices. We performed measurements in typical line-of-sight (LOS) and non-line-of-sight (NLOS) scenarios. Our results suggest that the LRP interface allows a lower power and energy consumption than the HRP one. Both types of devices achieved ranging and localization errors within the same order of magnitude and their performance depended on the type of NLOS obstruction. We propose theoretical models for the distance errors obtained with LRP devices in these situations, which can be used to simulate realistic building deployments and we illustrate such an example. This paper, therefore, provides a comprehensive overview of the energy demands, ranging characteristics, and localization performance of state-of-the-art UWB devices.
	\end{abstract}
	
	\begin{IEEEkeywords}
		Ultra-Wideband (UWB), Distance Measurement, Ranging, Accuracy, Energy Efficiency.
	\end{IEEEkeywords}
	
	\IEEEpeerreviewmaketitle
	
	\section{Introduction}

\IEEEPARstart{U}{ltra-wideband} (UWB) communications have become increasingly popular in recent years for their high-accuracy ranging and localization capabilities, which makes them promising candidates for providing location services to devices in the Internet of Things (IoT), industrial deployments, or wireless sensor networks in general. More recently, UWB chipsets have been included in smartphones and it is estimated that \SI{50}{\percent} of the smartphones on the market will incorporate UWB chipsets by 2027~\cite{uwb-market-outlook}. Given the fast adoption of UWB technology and its integration and interaction with devices in the IoT, it is crucial to evaluate both its ranging and localization performance \textit{and} its energy efficiency in order to determine its suitability for different types of applications.

UWB devices provide time-of-flight (ToF) measurements with sub-nanosecond accuracy which can be used to estimate the distance between two devices. Distance measurements (or ranges) are the basis of the true-range multilateration algorithm which is used in many localization applications~\cite{shen2007ultra}. Therefore, evaluating ranging errors is often the first step in analyzing
the localization accuracy of UWB localization systems.

The energy consumption of UWB devices depends on their architecture. The IEEE Standard for Low-Rate Wireless Networks 802.15.4~\cite{802-15-4} specifies two UWB physical interfaces (PHYs), that use high- and low-rate pulse repetition (HRP and LRP, respectively). Transmitting pulses at low rates enables a more energy-efficient implementation of LRP PHYs, using non-coherent receivers, than the ones based on coherent receivers, which are typically used in HRP PHYs. Coherent receivers use the phase of the signal in the detection process, while non-coherent receivers can estimate the channel coefficients with lower synchronization constraints based on the envelope of the signal. This makes LRP UWB devices suitable for energy-constrained devices. So far, it has not been clear whether this advantage comes with a cost in the ranging and localization performance.

Although coherent and non-coherent UWB receivers have been compared from a theoretical standpoint in literature, these studies have relied on simulations rather than measurements~\cite{hazra2014survey, durisi2004performance}. Previous work that analyzed the ranging accuracy of UWB devices through measurements~\cite{tian_human_2019, silva_ranging_2020, schenck_information_2018} has focused mostly on the Decawave DW1000 IC~\cite{dw1000_user_manual}, which implements the HRP PHY. Few works have analyzed commercially available LRP UWB devices and they targeted mostly their ranging accuracy without a detailed analysis of their power and energy consumption~\cite{ruiz2017comparing, shahi2012deterioration}. To the best of our knowledge, only one paper included a comparison of HRP and LRP devices (developed by Decawave and Ubisense, respectively)~\cite{ruiz2017comparing} but only on their ranging and localization performance, without regards to the energy efficiency of the devices.
Therefore, the current literature on comparisons of the two types of PHYs, on the one hand, and on commercially available UWB LRP devices, on the other hand, is very scarce. In particular, LRP devices deserve more attention since they can be implemented with energy-efficient receivers and can therefore potentially enable ranging and localization applications on ultra-low-power devices.

In our previous work~\cite{flueratoru2020energy}, we compared for the first time the power and energy consumption and the ranging performance of LRP and HRP devices using two commercially-available UWB devices: the Decawave DW1000 IC (HRP)\footnote{Decawave has recently been acquired by the semiconductor company Qorvo~\cite{decawave_qorvo}, hence is in the process of changing its name to Qorvo. Since this change is rather recent, we still refer to the company and devices as ``Decawave,'' this being the name under which they are still widely known.} and the 3db Access 3DB6830C IC~\cite{3db-access} (LRP)\footnote{We will refer to the 3db 3DB6830C (Release 2016) and the Decawave DW1000 (Release 2014) as the 3db and Decawave ICs, respectively.}. For the ranging performance, we used a database of distance measurements acquired with 3db devices and compared their statistics with results obtained with Decawave devices from the literature. This paper goes one step further and compares the ranging and localization performance of LRP and HRP devices using real measurements acquired in \textit{identical} settings. This last detail is crucial for a fair comparison of the devices since different environments can have a different impact on distance measurement errors. In addition, the current work offers a more in-depth analysis of the typical ranging errors of LRP devices in several scenarios.

Since indoor localization is often subject to multipath and shadowing phenomena, we analyzed the statistics of ranging errors in line-of-sight (LOS) and three non-line-of-sight (NLOS) scenarios, where the obstruction between the transmitter and the receiver was caused by a person, a gypsum wall (also called drywall panel), or a concrete wall. We derived statistical models for the error distributions obtained from measurements, which can be used to simulate realistic ranging and localization scenarios that would otherwise take days or weeks to implement and evaluate. We argue, in particular, that there are still unsolved problems about deploying a UWB-based localization system inside a building. Finding adequate LOS and NLOS error models, such as the ones proposed in our work, and using them to simulate the expected localization errors can help in this regard. For instance, many existing works~\cite{ruiz2017comparing, ledergerber_calibrating_2018, delamare2019static, monica2014experimental, segura2010experimental, kempke_surepoint_2016} consider only setups where the anchors are placed inside the same room because they yield the highest localization accuracy. However, this constraint is often hard to enforce in real deployments. For one, in highly compartmentalized spaces (for instance, office buildings) this would lead to a high anchor density, which in turn increases the deployment costs, the complexity (in terms of synchronization constraints, multiple access, anchor selection and placement strategy, etc.), and the total energy consumption of the localization system. Second, we show that this constraint might not be even needed, for instance when rooms are divided by shallow walls which cause only small localization errors. 

This paper, therefore, provides a comprehensive outlook on the typical power consumption and ranging performance of two state-of-the-art UWB devices, as well as their expected localization accuracy based on both real measurements and simulations. Our proposed error models can be used in future works to simulate custom building deployments and our measurements are publicly available\footnote{\url{https://doi.org/10.5281/zenodo.4686379}} to facilitate future research.

To summarize, the main contributions of this paper are the following:
\begin{itemize}
	\item We analyze the average power consumption of 3db Access (LRP) and Decawave (HRP) devices in the receive, transmit, and idle modes and compute their energy consumption per distance measurement.
	\item We evaluate and compare the accuracy and precision of distance measurements of 3db Access and Decawave devices based on measurements recorded in identical settings in LOS and NLOS scenarios caused by drywall, a concrete wall, and the human body.
	\item We analyze the ranging performance of 3db Access devices on different channels (at $6.5$, $7$, and \SI{7.5}{\giga\hertz}) and propose channel diversity strategies that can improve the ranging accuracy.
	\item We implement localization systems based on the two types of devices and evaluate their performance experimentally in both LOS and NLOS settings.
	\item We provide statistical models for the LOS/NLOS ranging errors of 3db devices and evaluate their performance in a simulated building deployment when anchors are either in the same room or in adjacent rooms separated by a gypsum or a concrete wall.
\end{itemize}

The rest of the paper is organized as follows. In Section~\ref{sec:system}, we analyze the theoretical differences between LRP and HRP PHYs and introduce the basics of UWB ranging and localization. We present the experimental setup in Section~\ref{sec:setup} and evaluate the power consumption, distance measurement errors, and localization performance of the ICs based on measurements in Section~\ref{sec:evaluation}. In Section~\ref{sec:simulation}, we model the ranging errors obtained experimentally and show how they can be used to simulate a localization application. In Section~\ref{sec:discussion}, we present the state-of-the-art in UWB localization and propose several directions for future work. Finally, we draw the conclusions in Section~\ref{sec:conclusion}.	
\section{Background}
\label{sec:system}

In Section~\ref{ssec:uwb-architecture}, we first introduce UWB devices and the main types of receiver architectures used in commercial devices. UWB devices can perform ToF measurements with sub-nanosecond accuracy and can therefore measure the distance between two devices with centimeter-level accuracy. Distances between two devices have value in themselves (e.g., to find lost objects) but also as a first step in multilateration algorithms for localization. In this paper, we compare both the ranging and localization performance of two types of UWB devices. We introduce ranging and localization concepts with UWB devices in Section~\ref{ssec:ranging-methods} and~\ref{ssec:background-localization}, respectively.

\subsection{UWB Device Architectures}
\label{ssec:uwb-architecture}

The IEEE 802.15.4 standard~\cite{802-15-4} defines two types of physical interfaces with low and high pulse repetition frequency: LRP and HRP, respectively. The Decawave DW1000 UWB chip is compliant with the HRP PHY defined in the IEEE 802.15.4 standard~\cite{802-15-4}. It is perhaps the most widely-used UWB device, so we chose it to represent the HRP PHY class. Decawave has recently released a new UWB chipset, the DW3000~\cite{dw3000_website}. However, the new-generation chipsets are currently available only as engineering samples, which is why we focused on the old release. The 3db IC is compliant with the LRP PHY specified in the IEEE 802.15.4z amendment~\cite{802-15-4z}. The chip is already being used for secure keyless car access but it has not been evaluated in high-accuracy applications yet.

UWB transmissions have to satisfy two constraints imposed by international regulations~\cite{federal2002revision}: a maximum average power spectral density (PSD) of \SI{-41.3}{\decibel m \per\mega\hertz} (averaged over \SI{1}{\milli\second}) and a maximum peak power spectral density of \SI{0}{\decibel m \per 50\mega\hertz}. UWB devices can, therefore, transmit over a fixed period either few pulses at high power levels or many pulses with lower transmit power. The first situation falls under the LRP specification and is employed by 3db devices, while the latter is known as HRP and is used by Decawave. If optimally employed, both of these technologies benefit from an \textit{equal} average transmitted RF energy.

Since the HRP PHY transmits individual pulses with lower energy than the LRP, the received pulse energy is also lower for the same path loss (same distance). Therefore, the HRP PHY needs more sophisticated techniques to extract weaker pulses from the receiver noise, typically performed with correlations over many samples. For this reason, Decawave devices use coherent receivers. Because 3db devices implement the LRP PHY, they can use a \textit{non-coherent} receiver based on energy detection (ED) for signals modulated with binary frequency-shift keying (BFSK).

Coherent receivers use phase information in the detection process. They typically have low sensitivity to inter-symbol and co-user interference and benefit from the multipath diversity of the UWB channel~\cite{weisenhorn2004robust}. At the same time, the receiver architecture demands high computational resources and hardware complexity~\cite{arslan2006ultra}. For optimal reception, the coherent receiver needs to estimate the multipath delays, their complex-valued channel coefficients, and the pulse shape distortion~\cite{arslan2006ultra}. 
A precise estimation of the carrier phase is crucial for recovering the baseband pulse since inaccuracies will result in signal power loss and crosstalk interference in signals modulated using phase-shift keying (PSK)~\cite{proakis2001digital}. For a carrier frequency of \SI{8}{\giga\hertz}, a time shift of half of the pulse period flips the phase of the signal, so coherent UWB systems generally tolerate rotations only within $\pi/4$ of the signal phase (around \SI{30}{\pico\second}). These requirements increase the power consumption of coherent demodulators~\cite{weisenhorn2004robust}.

Non-coherent receivers estimate channel coefficients based on the envelope rather than on the phase and amplitude of the received signal, so they have lower synchronization constraints. The timing requirements of a non-coherent receiver are dependent only on the pulse envelope, which is related to the pulse bandwidth. For instance, if the pulse bandwidth is \SI{500}{\mega\hertz}, the non-coherent receiver needs to operate with a timing resolution of \SI{1}{\nano\second} and it does not need high RF carrier synchronization. Therefore, non-coherent receivers can be more energy-efficient than coherent ones but have a higher bit error probability~\cite{proakis2001digital}. Another disadvantage of the non-coherent architecture is that it cannot be used for precise angle-of-arrival (AoA) measurements with closely-spaced antennas.

\subsection{Ranging Methods}
\label{ssec:ranging-methods}

The distance between two devices can be estimated based on the time of flight (ToF) of the signal. Using the transmission time ($T_1$) of the signal measured by the sender and the arrival time ($T_2$) at the receiver, we can compute the distance as~\cite{shen2007ultra}:
\begin{equation}
d = (T_2 - T_1) \cdot c,
\end{equation}
where $c$ is the speed of light and $T_p \triangleq T_2 - T_1$ is the propagation time of the signal. To accurately estimate the distance, the devices need to be tightly clock synchronized, as a small mismatch of \SI{1}{\nano\second} can introduce a distance error of around \SI{30}{\centi\meter}.	Because synchronizing the sender and the receiver is usually unfeasible in practice, more messages are exchanged in order reduce such errors, such as in the single- or the double-sided two-way ranging (SS-TWR and DS-TWR, respectively).

\begin{figure}[t]
	\centering
	\label{fig:ss-twr}%
	\includegraphics[width=0.26\textwidth]{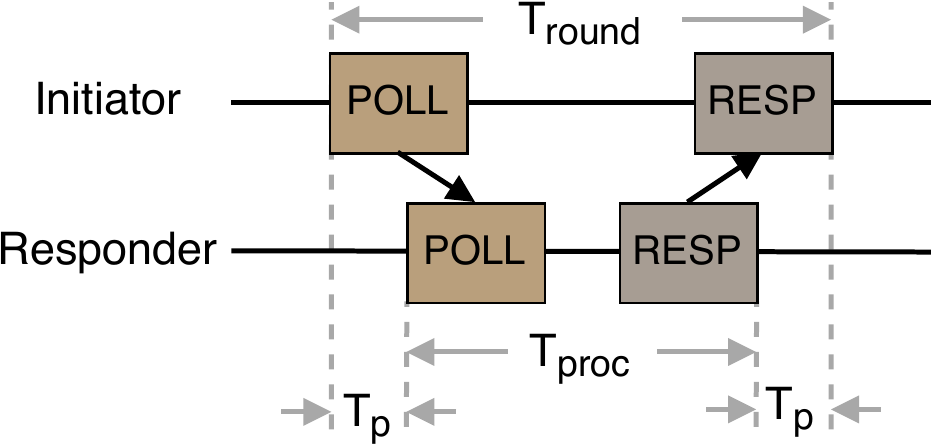}
	\caption{Message exchange in the single-sided two-way ranging.}
	\label{fig:twr}
\end{figure}

The SS-TWR uses two messages per distance estimate, as shown in \figurename~{\ref{fig:ss-twr}}. The propagation time is:
\begin{equation}
T_p = \frac{T_{round} - T_{proc}}{2},
\end{equation}
where $T_{round}$ is the time spent in one message exchange and $T_{proc}$ is the processing time on the \responder side. It can be shown that the error in estimating $T_p$ is~\cite{neirynck2016alternative}:
\begin{equation}
e_{T_p} = e_1 \cdot T_p + \frac{1}{2}T_{proc}(e_1 - e_2),
\end{equation}
where $e_1$ and $e_2$ are the clock drift errors of the \initiator and \responder, respectively. The main source of errors in the SS-TWR are $T_{proc}$, which is in the range of hundreds of microseconds, and the clock drift, which can be up to \SI[separate-uncertainty = true]{\pm20}{ppm} in systems compliant with the IEEE 802.15.4 standard~\cite{802-15-4}.

In the LRP PHY, a location-enhancing information postamble is introduced at the end of each message to estimate the clock drift error~\cite{802-15-4}. Besides, the processing time of LRP messages is shorter than the one of HRP. It is also more convenient to minimize the number of exchanged messages in the TWR since this reduces the time needed to obtain one distance measurement. Therefore, the SS-TWR is usually the method of choice for LRP devices.

The DS-TWR uses an additional message exchange to minimize clock drift errors. Although this is the ranging method typically employed in Decawave devices~\cite{dw1000_user_manual}, the Decawave MDEK1001 kit that we used throughout this paper applies the SS-TWR~\cite{dwm1001_system} to improve the energy efficiency and reduce the air time. Although the HRP PHY does not include a postamble nor does it have a fixed processing time, the tags can estimate their clock drift with respect to the anchors based on periodic beacons transmitted by the anchors~\cite{dwm1001_system}. 

\subsection{True-Range Multilateration}
\label{ssec:background-localization}

The true-range multilateration algorithm estimates the location of a mobile device (also called a tag) using distance measurements between the tag and fixed devices with known locations (also called anchors). The special case for 2D localization using three anchors is known as trilateration.

Let $d_i$ denote the distance between anchor $A_i$ and the tag, which can be written as:
\begin{equation}
d_i = \lVert \bm{x}_{A_i} - \bm{x} \rVert + v_i, \quad i = 1, ..., N
\end{equation}
where $\bm{x}$ is the location of the tag, $\bm{x}_{A_i}$ is the location of anchor $A_i$, and $v_i$ is the measurement noise. The noise terms of all anchors are assumed independent.

In vector form, the measurement equation becomes:
\begin{equation}
\bm{y} = h(\bm{x}) + \bm{v},
\end{equation}
where $\bm{y}$ is the measurement vector (containing all measurements $d_j, j = 1, ..., N$), $\bm{v}$ the error vector, and $h$ the vector-valued measurement function. The equation can be solved by the least-squared solution $\bm{x}^*$ which minimizes $\lVert \bm{y} - h(\bm{x}) \rVert$~\cite{shen2007ultra}.

Multiple algorithms for solving the nonlinear system of equations were compared in~\cite{sirola2010closed}. The regularized Gauss-Newton multilateration algorithm is an iterative algorithm which has a similar accuracy to several algorithms with closed-form solutions and a low computational complexity suitable for real-time applications~\cite{sirola2010closed}.
For this reason, we used it to implement the localization systems evaluated in Section~\ref{ssec:localization-hrp-lrp}.

The algorithm needs an initial starting position $\bm{x}_0$, which should be chosen as close as possible to the real location for a quick convergence. For the first iteration, the starting position can be set to the solution of a closed-form multilateration algorithm or to the latest location of the tag (if available). At each iteration $k$, the algorithm computes the Jacobian matrix:
\begin{equation}
J_k(\bm{x}) = \bigg[ \frac{\bm{x}_{A_1} - \bm{x}}{\lVert \bm{x}_{A_1} - \bm{x} \rVert}, ..., \frac{\bm{x}_{A_N} - \bm{x}}{\lVert \bm{x}_{A_N} - \bm{x} \rVert} \bigg]^T.
\end{equation}
The solution at step $k+1$ is $\bm{x}_{k+1} = \bm{x}_k + \Delta\bm{x}$, where $\Delta\bm{x}_k$ is the least-squares solution to
\begin{equation}
-(\Sigma^{-\frac{1}{2}} J_k + c\bm{I})\Delta\bm{x}_k = \big( \Sigma^{-\frac{1}{2}} (h(\bm{x}_k) - d) + c(\bm{x} - \bm{x}_r) \big),
\end{equation}
where $\bm{x}_r$ is a regularization point taken as the mean of the anchors' coordinates and $c$ is a regularization coefficient equal to the inverse of the standard deviation of a distribution centered at $\bm{x}_r$. The algorithm stops if the location increment is below a tolerance $\delta$ or if the algorithm reaches the maximum number of iterations.

\section{Evaluation Setup}
\label{sec:setup}
In the following, we describe the device setups used in the power, ranging, and localization measurements.

\subsubsection*{3db Access}
The 3db chip is integrated into an Arduino shield on top of an Arduino M0 board. The communication between the chip and the host MCU is performed via SPI. We use the channel centered at \SI{6.52}{\giga\hertz} and the peak data rate of \SI{247}{\kilo b \per\second}. The \SI{10}{\decibel} bandwidth of a pulse is \SI{380}{\mega\hertz} and, because pulse spectra partially overlap in BFSK modulation, the total system bandwidth is approximately \SI{620}{\mega\hertz}. The packet duration is \SI{400}{\micro\second}. The IC was configured to transmit at the maximum level of \SI{-43.86}{\decibel m\per \mega\hertz}, so within UWB regulations~\cite{federal2002revision}. 

\subsubsection*{Decawave}

We use the Decawave MDEK1001 kit which includes the DW1000 UWB chip integrated into the DWM1001 module. The DWM1001 module also contains a Nordic Semiconductor nRF52832 BLE microprocessor mostly used for network communication and an STM LIS2DH12TR 3-axis motion detector. The kit's default PANS software supports only the mode 14 which uses channel 5 (at \SI{6.49}{\giga\hertz}), a data rate of \SI{6.8}{\mega b \per\second}, a PRF of \SI{64}{\mega\hertz}, and a preamble length of 128 symbols, corresponding to a packet length of \SI{287}{\micro\second}. The default configuration is suitable for short-range communication. The devices have a \SI{3}{\decibel} bandwidth of \SI{499.2}{\mega\hertz} (equivalent to a \SI{10}{\decibel} bandwidth of $\approx$\SI{662}{\mega\hertz}). In all the ranging measurements, one of the devices is configured as a tag in the low-power mode, while the other device is an initiating anchor.

In our previous work~\cite{flueratoru2020energy}, we used Decawave devices integrated in the EVK1000 evaluation kit which allowed more configurations and used the DS-TWR. There, we decided to use the long-range mode (Mode 3), in order to attain a similar range as with 3db devices. Here, we favored the MDEK kit because it implements the SS-TWR which requires less message exchanges and is more energy efficient. The SS-TWR is also implemented by 3db devices, making the operation of the two devices similar.
	
\section{Measurement-Based Evaluation}
\label{sec:evaluation}
In this section, we compare how 3db and Decawave devices compare in terms of power and energy consumption, coverage, distance measurement accuracy and precision, and localization performance. Distance measurements are important in themselves, for instance in proximity detection applications, but also because they are at the basis of true-range multilateration.

Section~\ref{ssec:power-consumption} compares the power consumption of the two devices, Section~\ref{ssec:range} their maximum range, and Section~\ref{ssec:distance-meas} the accuracy and precision of their distance measurements. Section~\ref{ssec:ch-diversity} analyzes measurements acquired on multiple channels. In Section~\ref{ssec:localization-hrp-lrp}, we integrate the devices in localization systems and evaluate their performance.

\subsection{Power Consumption}
\label{ssec:power-consumption}

In this section, we compare the power and energy consumption of the two types of devices. Unfortunately, the Decawave MDEK1001 board allows for measuring the current consumption only of the DWM1001 module, which contains, besides the DW1000 UWB chip, a BLE microprocessor, and a motion detector. We present the power consumption measurements of the DWM1001 module for the sake of completeness, but for the comparison between the two \textit{chipsets}, we rely on the current consumption of the Decawave DW1000 UWB chip from the device datasheet~\cite{dw1000_datasheet}. We used the current consumption reported for mode 14 which is referenced to \SI{3.3}{\volt}~\cite{dw1000_datasheet}. 

We measured the current consumption of the 3db chip and the DWM1001 module with a Keysight DC Power Analyzer. We isolated the most important modes, namely the idle, transmit (\texttt{TX}), and receive (\texttt{RX}) and computed their average current consumption. The input voltages of the 3db chip and the DWM1001 module were \SI{1.25}{\volt} and \SI{3.3}{\volt}, respectively.

Table~\ref{tab:power-meas-stats} presents the average power consumption in each mode of the 3db IC, the DW1000 IC, and the DWM1001 module. As mentioned, for the comparison between the UWB chipsets, we rely on the current consumption of the Decawave IC provided in the \textit{datasheet}~\cite{dw1000_datasheet} and on the \textit{measured} current consumption of the 3db IC. Overall, the average power consumption of the 3db IC in the \texttt{TX}, \texttt{RX}, and idle mode is at least 9  times lower than the one of the Decawave IC. Only in the deep sleep mode the 3db chipset has $1.9\times$ higher power consumption than the Decawave chipset. Note that the average power consumption of the DW1000 chip in the \textit{idle} mode is about $1.45$ times higher than the one of 3db devices in the \textit{receive} mode, also the most power-hungry state.
The results suggest that, indeed, the LRP interface can be more power-efficient than the HRP one.

The power consumption profile is a starting point for evaluating the energy consumption of an UWB-based localization system. A key challenge in a localization system is minimizing the energy consumption of the \textit{tag}, which is usually battery-powered. 
To avoid synchronizing the tag and the anchors, the tag can \textit{initiate} the message exchange and stay in the idle or sleep mode between rangings.
Using the SS-TWR implies, in this case, that the \textit{tag} estimates the distance (or the location). 

To illustrate the energy efficiency of a tag in a localization system, let us consider the most favorable scenario in which the tag is the initiator. We disregard the time spent in the idle mode, since it is subject to the desired location update rate and guard times, which can be chosen freely to a certain extent. We therefore compute the energy  consumption only when the device is in the \texttt{TX} or \texttt{RX} mode. The packet duration of the DW1000 chip in Mode 14 is \SI{287}{\micro\second} and the one of the 3db chip is \SI{400}{\micro\second}. Therefore, a Decawave tag will consume \SI{180}{\micro\joule} per SS-TWR during transmission and reception, while a 3db tag will consume \SI{28}{\micro\joule} (including the transition times), so $6.4$ times less energy. When placed in the long-range mode (for instance, mode 3), the packet duration of Decawave devices increases to \SI{3487}{\micro\second} which is about $10\times$ larger than that of 3db devices, causing them to consume at least $100\times$ more energy~\cite{flueratoru2020energy}. The difference between these modes is the \textit{maximum range} at which the devices can communicate, so in the next section we will compare the range of 3db and Decawave devices.

\begin{table}[t!]
	\caption{The average power consumption of 3db and Decawave devices.}
	\centering
	\begin{threeparttable}
		\begin{tabular}{l
				S[table-format=3.1, round-mode=places, round-precision=1, table-column-width=1.1cm] 
				S[table-format=3.1, round-mode=places, round-precision=1, table-column-width=1.1cm] 
				S[table-format=3.1, round-mode=places, round-precision=1, table-column-width=1.1cm] 
				S[table-format=4.1, round-mode=places, round-precision=1, table-column-width=1.1cm] 
			}
			\toprule
			
			\multirow{2}{*}{}
			& \multicolumn{4}{c}{\makecell{Average power consumption [\si{\milli\watt}]}}  \\ [1ex] \cline{2-5} \\ [-1.5ex]
			
			& \multicolumn{1}{c}{TX} 
			& \multicolumn{1}{c}{RX}
			& \multicolumn{1}{c}{Idle}
			& \multicolumn{1}{c}{Deep sleep}  \\
			
			\midrule
			
			3db Access\tnote{$\dagger$}
			& 20.7 & 40.7 & 6.6 & \multicolumn{1}{c}{$6.25 * 10^{-4}$\tnote{*}} \\
			
			DW1000~\tnote{$\ddagger$}
			& 237.6\tnote{*} & 392.7\tnote{*} & 59.4\tnote{*} & \multicolumn{1}{c}{3.3 * $10^{-4}$\tnote{*}} \\
			
			DWM1001\tnote{$\ddagger$}
			& 297.7 & 507.21 & 47.9 & 3.9 \\
			
			\bottomrule
		\end{tabular}
		
		\begin{tablenotes}
			\item[$\dagger$]{Referenced to \SI{1.25}{\volt}.}
			\item[$\ddagger$]{Referenced to \SI{3.3}{\volt}.}
			\item[*]{Based on the device datasheet.}
		\end{tablenotes}
	\end{threeparttable}
	
	\centering
	\label{tab:power-meas-stats}
\end{table}

\subsection{Range}
\label{ssec:range}

\begin{figure}[t!]
	\centering
	\includegraphics[width=0.375\textwidth]{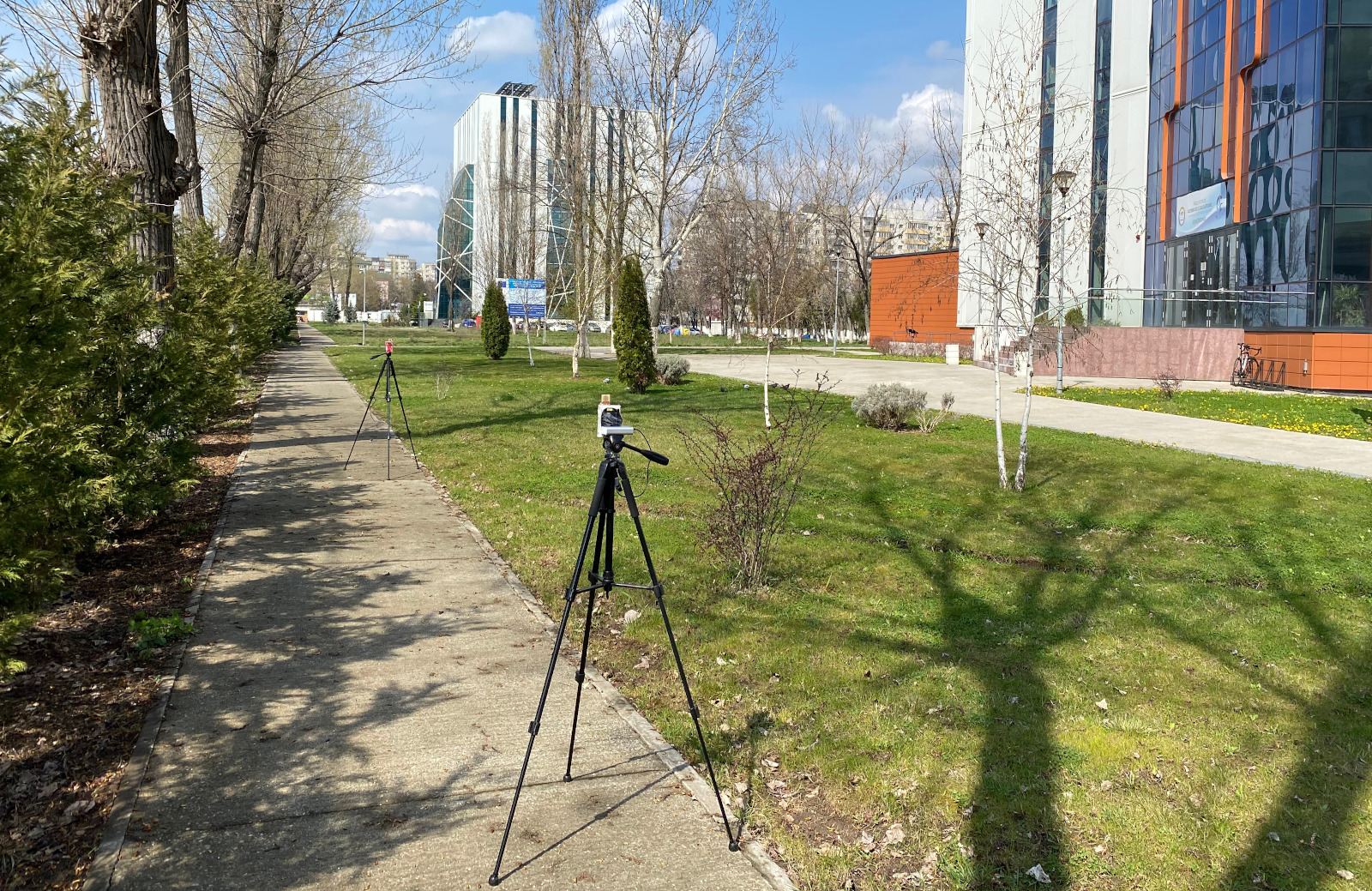}
	
	\caption{Location at which the range of the devices was measured. The devices were placed at distances between \SIrange{5}{220}{\meter} along the pathwalk.}
	\label{fig:setup_range}
\end{figure}

In this section, we want to find the ratio of successful distance measurements between a transmitter (\texttt{TX}) and a receiver (\texttt{RX}) placed at distances between \SIrange{5}{220}{\meter}. Remember that one distance measurement using the SS-TWR involves the successful transmission of two messages, a poll (from \texttt{TX} to \texttt{RX}) and a response (from \texttt{RX} to \texttt{TX}). The devices are said to have a (maximum) range of $d$ meters when the ratio between the number of successful distance measurements and the total number of initiated measurements up to the distance $d$ is higher than a chosen ratio $P = 0.9$. We performed measurements outdoors, on the pathwalk shown in \figurename~\ref{fig:setup_range}, in order to minimize the multipath interference from surrounding objects which is usually higher indoors. At discrete steps, the \texttt{TX} was programmed to send $60$ messages (polls) every \SI{200}{\milli\second}. If the response from the \texttt{RX} does not arrive at the \texttt{TX} either because the \texttt{RX} did not receive the probe or because the response was lost, a timeout occurs and the distance measurement is unsuccessful. We define the packet delivery ratio (PDR) as the number of responses received by the \texttt{TX} over the number of transmitted messages. 

The Decawave PANS software reports only the (successful) responses and produces no output when transmitted packets are not answered. To compute the PDR of Decawave devices, we use the transmission period of \SI{200}{\milli\second} to compute how many messages should have been exchanged between the first and the last successful message at every test point. The PDR is then the number of received messages during that period divided by the number of expected messages. 3db devices report when packets are unanswered and we compute the PDR of 3db devices as described before. 

\begin{figure}[t!]
	\centering
	\includegraphics[width=0.49\textwidth]{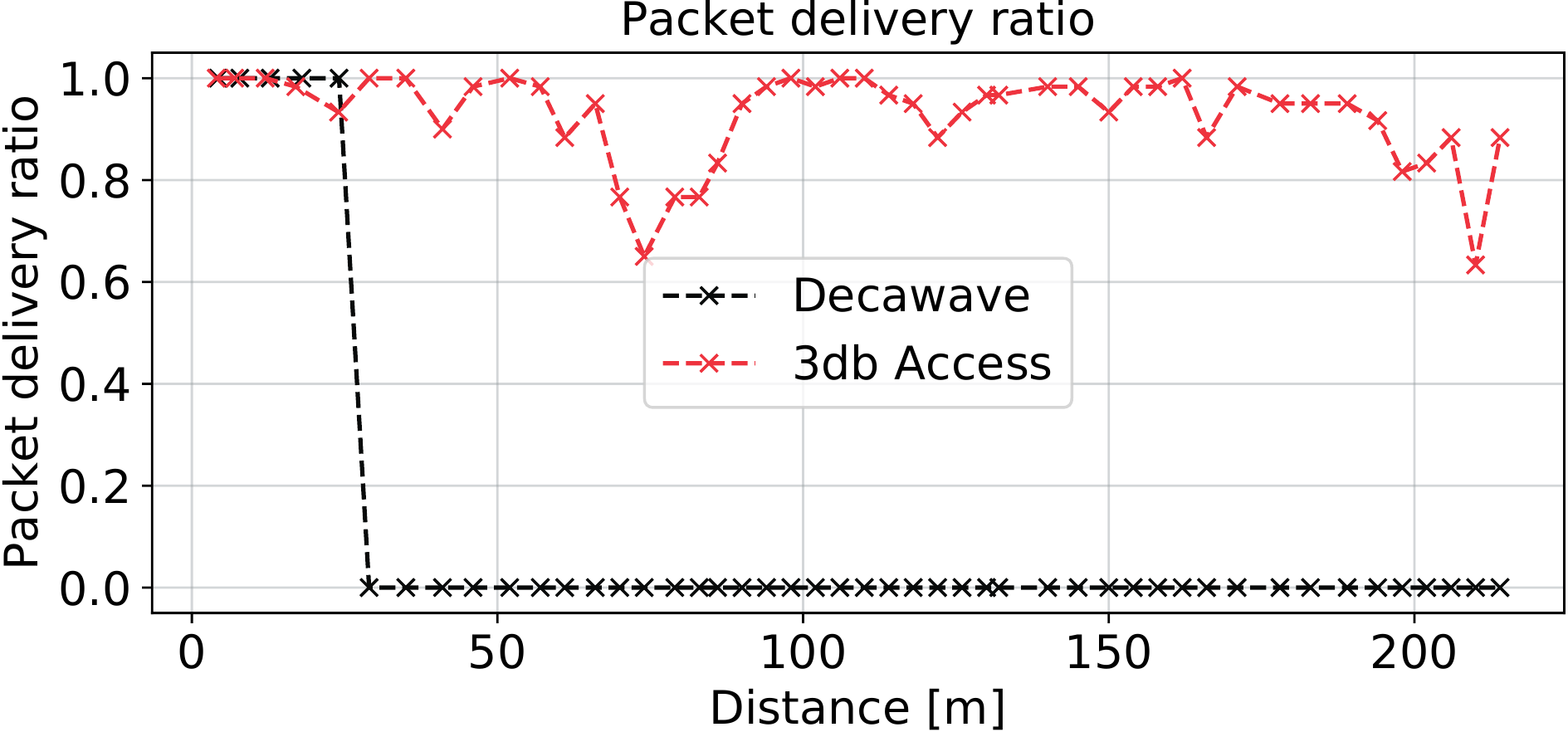}
	
	\caption{The packet delivery ratio of 3db Access and Decawave devices.}
	\label{fig:packet_rx_rate}
\end{figure}

\figurename~\ref{fig:packet_rx_rate} shows the PDR for 3db and Decawave devices. It is important to note that the PDR is highly dependent on the orientation of the devices, since at long distances the irregular radiation pattern of the antennas can cause high packet losses along certain directions. The PDR of Decawave devices dropped to 0 after \SI{25}{\meter}, which is more than half of the expected range of \SI{80}{\meter} of Mode 14 reported in the DW1000 Datasheet~\cite{dw1000_datasheet} (Section 6.3). However, this range was provided for channel 2 at \SI{4}{\giga\hertz}, so the path loss is expected to be higher (and hence the range lower) at the center frequency of \SI{6.5}{\giga\hertz} used in our experiment. 3db devices have a PDR higher than $0.9$ at almost all distances up to \SI{194}{\meter}, except for the higher losses between \SIrange{65}{90}{\meter}. At those distances, we found that the PDR was highly influenced by the relative pose between the devices, most likely due to destructive multipath interference and the antenna radiation pattern. 

In our previous work~\cite{flueratoru2020energy}, we found that 3db devices had a PDR of $0.9$ up to \SI{116}{\meter}, but we did not measure the PDR at longer distances because of the limited space. Similarly, in this experiment, we did not measure the PDR beyond \SI{220}{\meter}. In~\cite{flueratoru2020energy}, we compared the maximum range of 3db devices with the one of Decawave devices operating in the \textit{long-range} mode (Mode 3) and we found that they had a similar PDR over the covered area. Decawave devices in the long-range mode have a packet duration about $10\times$ larger than in the short-range mode and therefore also a higher energy consumption. In practice, this means that a 3db tag could operate over a similar area as a Decawave tag in the long-range mode but with $125\times$ less energy. A Decawave tag in the short-range mode will be more energy-efficient than in the long-range mode but more anchors will be needed to provide coverage over the same area. 

\subsection{Distance Measurements}
\label{ssec:distance-meas}

\newcommand{\LosThreedbMean}{0.02}
\newcommand{\LosThreedbStd}{0.07}
\newcommand{\LosThreedbIqr}{0.09}
\newcommand{\LosDwMean}{0.00}
\newcommand{\LosDwStd}{0.05}
\newcommand{\LosDwIqr}{0.07}
\newcommand{\DrywallThreedbMean}{-0.04}
\newcommand{\DrywallThreedbStd}{0.08}
\newcommand{\DrywallThreedbIqr}{0.12}
\newcommand{\DrywallDwMean}{-0.01}
\newcommand{\DrywallDwStd}{0.09}
\newcommand{\DrywallDwIqr}{0.10}
\newcommand{\ConcreteThreedbMean}{0.46}
\newcommand{\ConcreteThreedbStd}{0.14}
\newcommand{\ConcreteThreedbIqr}{0.19}
\newcommand{\ConcreteDwMean}{0.44}
\newcommand{\ConcreteDwStd}{0.07}
\newcommand{\ConcreteDwIqr}{0.14}
\newcommand{\HumanThreedbMean}{0.55}
\newcommand{\HumanThreedbStd}{0.32}
\newcommand{\HumanThreedbIqr}{0.29}
\newcommand{\HumanDwMean}{0.60}
\newcommand{\HumanDwStd}{0.26}
\newcommand{\HumanDwIqr}{0.46}

\begin{figure}[t!]
	\centering
	\begin{tabular}{cc}
		\adjustbox{valign=b}{\subfloat[]{%
				\label{fig:setup_los}%
				\includegraphics[width=0.35\linewidth]{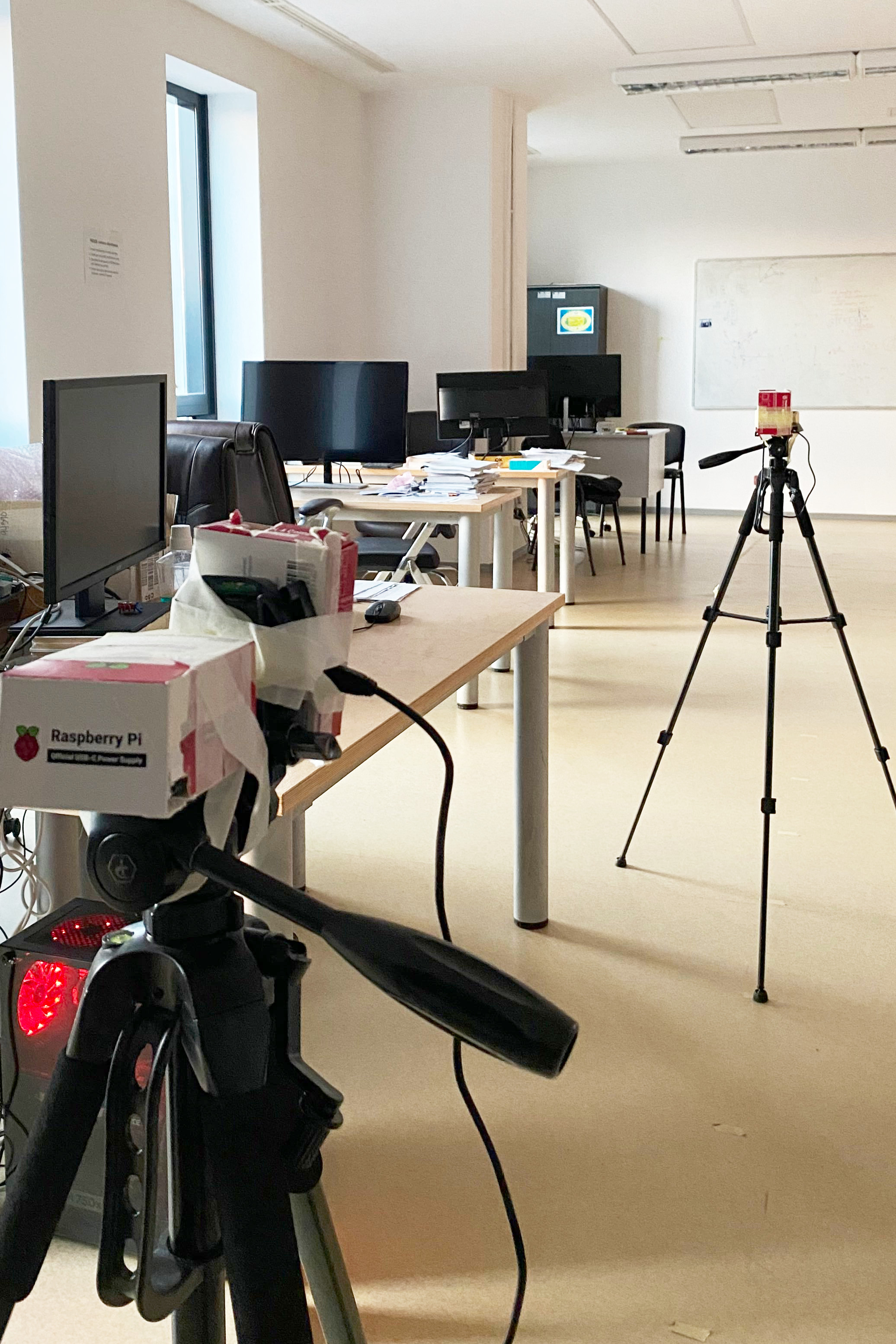}%
		}}
		&      
		\adjustbox{valign=b}{\begin{tabular}{@{}c@{}}
				\subfloat[]{%
					\label{fig:setup_nlos_drywall}%
					\includegraphics[width=0.475\linewidth]{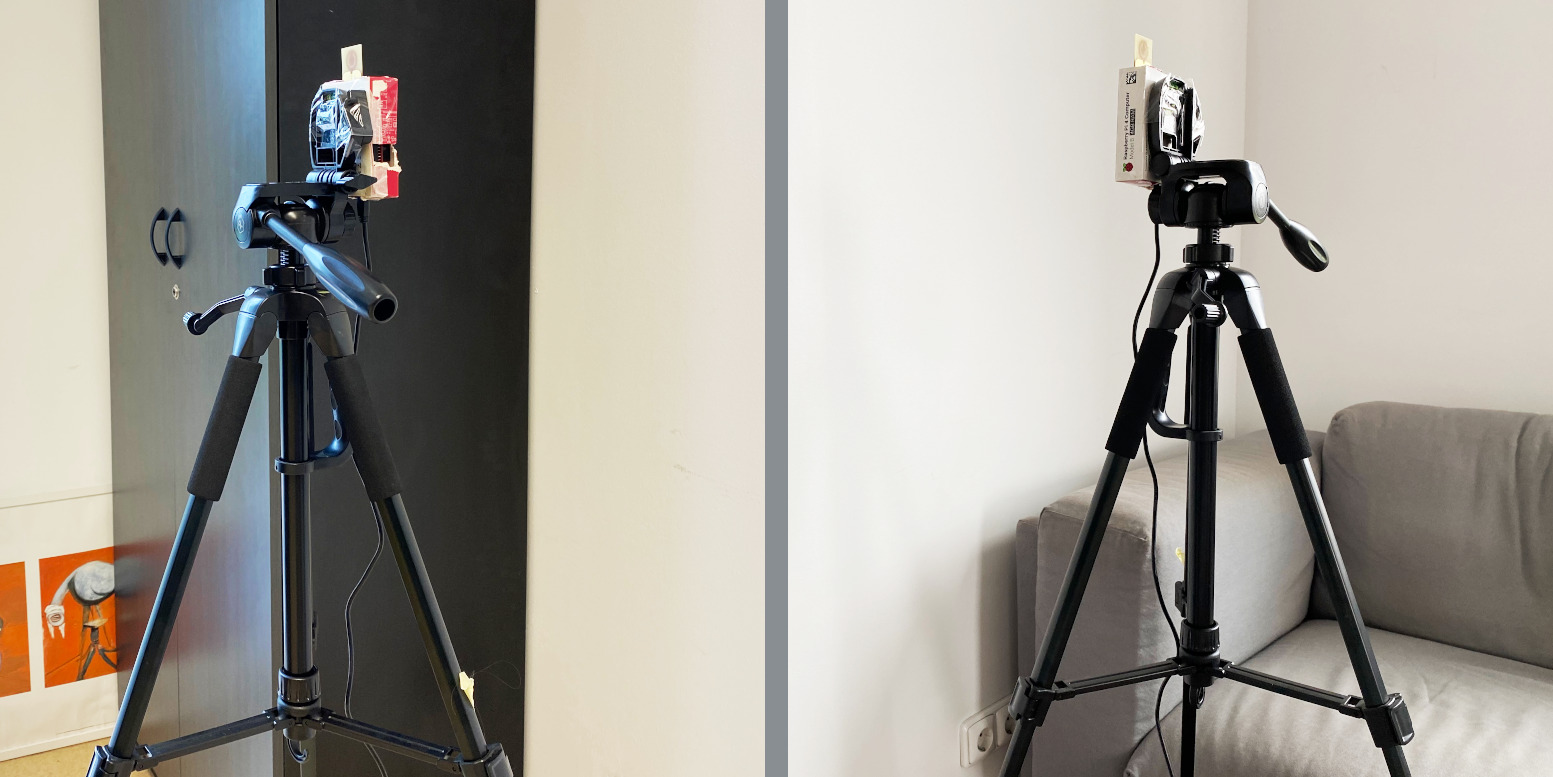}} \\
				\subfloat[]{%
					\label{fig:setup_nlos_wall}%
					\includegraphics[width=0.475\linewidth]{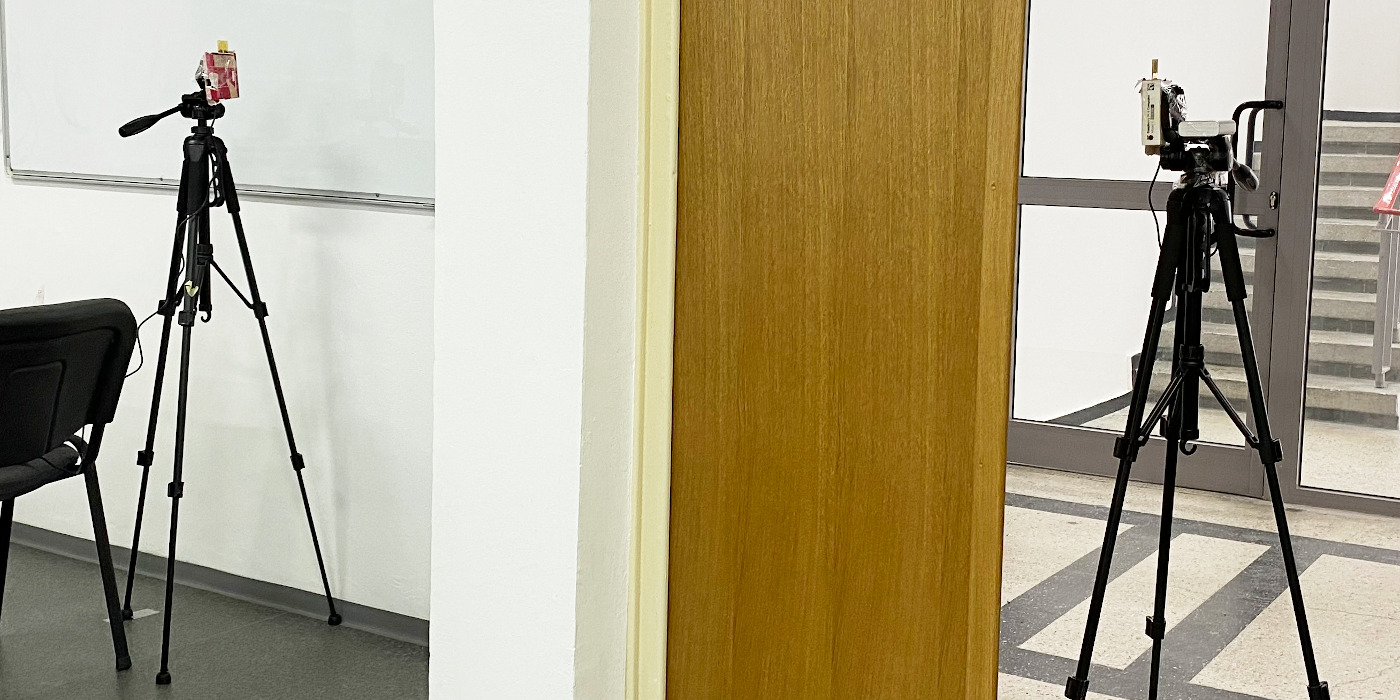}}
		\end{tabular}}
	\end{tabular}
	
	\caption{Setup of ranging measurements in (a) LOS, (b) NLOS with a drywall, and (c) NLOS with a concrete wall. The UWB devices are placed on tripods. The NLOS with human body shadowing setup is identical with the LOS one, except that a person is standing right in front of the transmitter (the device further away).}
	\label{fig:ranging-setup}
\end{figure}

In this section, we compare the distance measurements of 3db and Decawave devices acquired in identical settings. 
We considered four indoor settings: LOS inside a large office and NLOS caused by either a gypsum wall (\SI{12.5}{\centi\meter} thickness), a concrete wall (\SI{29}{\centi\meter} thickness), or a human body. \figurename~\ref{fig:ranging-setup} shows the settings. Decawave and 3db Access devices were placed at exactly the same locations and acquired an equal number of measurements on the \SI{6.5}{\giga\hertz} channel at the same rate (every \SI{0.6}{\second}). In all ranging experiments from this section, the devices were calibrated to account for errors caused by hardware, channel, or distance. The calibration method is described in Appendix~\ref{ssec:distance-calib}. 

For the ranging datasets, at each test point, we recorded measurements for \SIrange{2}{4}{\minute}, which during the calibration phase was deemed enough to obtain a distribution with a mean error within $\pm$\SI{1}{\centi\meter} of the long-term one. The setup for each recording scenario is described in \tablename~\ref{tab:ranging-setup}. 
The 3db devices were configured to acquire measurements on all three channels at 6.5, 7, and \SI{7.5}{\giga\hertz}, cycling through them every \SI{0.2}{\second}. Measurements on all channels will be later used in Section~\ref{ssec:ch-diversity} to investigate whether channel diversity improves the accuracy in certain situations. Because the MDEK1001 devices can use only the \SI{6.5}{\giga\hertz} channel, we compare Decawave and 3db measurements acquired only on this channel.

\begin{table}[t!]
	\caption{Setup of ranging experiments.}
	\centering
	\begin{tabular}{l
			c
			S[table-format=1.1]
			S[table-format=2]
		} 
		\toprule
		Scenario
		& \makecell[c]{Distances\\ $[$\si{\meter}$]$}
		& \multicolumn{1}{c}{\makecell{Sampling\\ period $[$\si{\second}$]$}}
		& \multicolumn{1}{c}{\makecell{Recording\\ time per test\\ point $[$\si{\minute}$]$}} \\ 
		\midrule
		
		LOS & 1, 2, ..., 8 & 0.6 & 2 \\
		
		
		\makecell[l]{NLOS with drywall} & 1, 2, ..., 6 & 0.6 & 4 \\
		
		\makecell[l]{NLOS with concrete wall} & 1, 2, 3 & 0.6 & 2 \\
		
		\makecell[l]{NLOS with human body} & 2, 5, 10 & 0.6 & 2 \\
		\bottomrule
	\end{tabular}
	\centering
	\label{tab:ranging-setup}
\end{table}

\begin{figure*}[t!]
	\centering
	\subfloat{%
		\label{}%
		\includegraphics[width=0.22\linewidth]{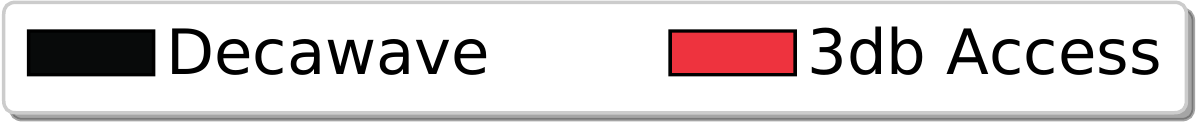}%
	}\\
	\setcounter{subfigure}{0}
	\subfloat[]{%
		\label{fig:los_pdf_dist_error}%
		\includegraphics[width=0.23\linewidth]{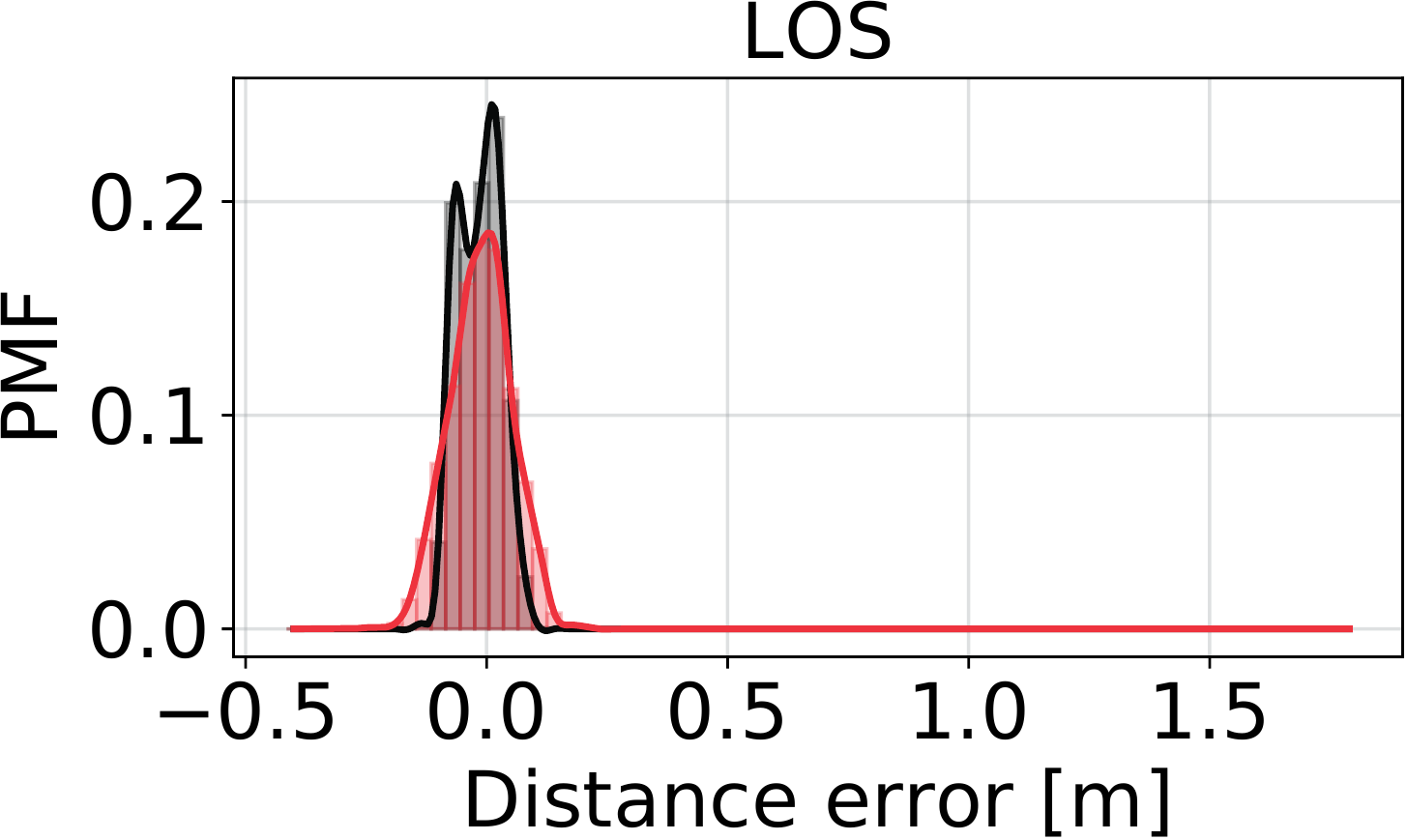}%
	}\quad%
	\subfloat[]{%
		\label{fig:nlos_drywall_pdf_dist_error}%
		\includegraphics[width=0.23\linewidth]{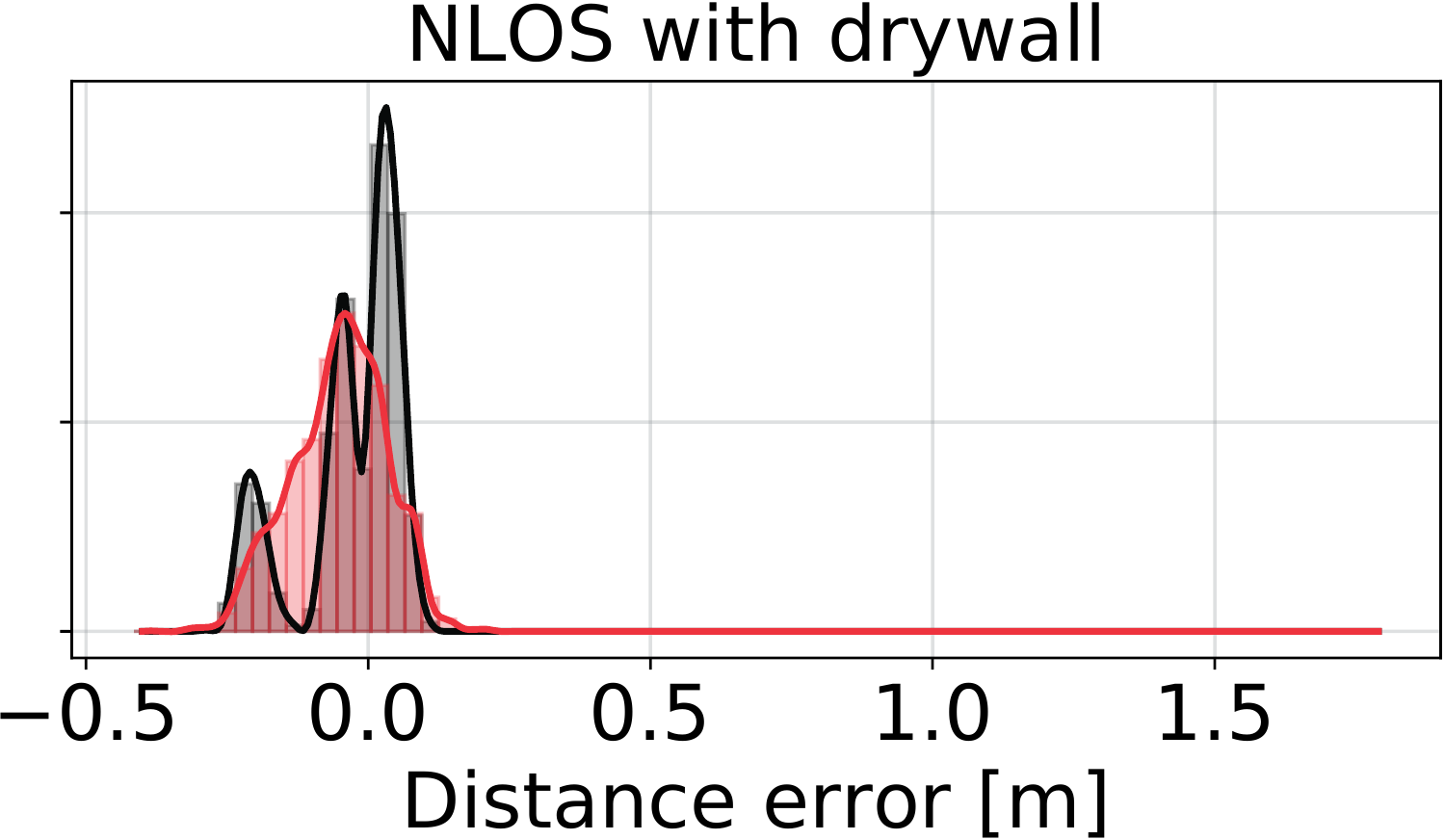}%
	}\quad%
	\subfloat[]{%
		\label{fig:nlos_wall_pdf_dist_error}%
		\includegraphics[width=0.23\linewidth]{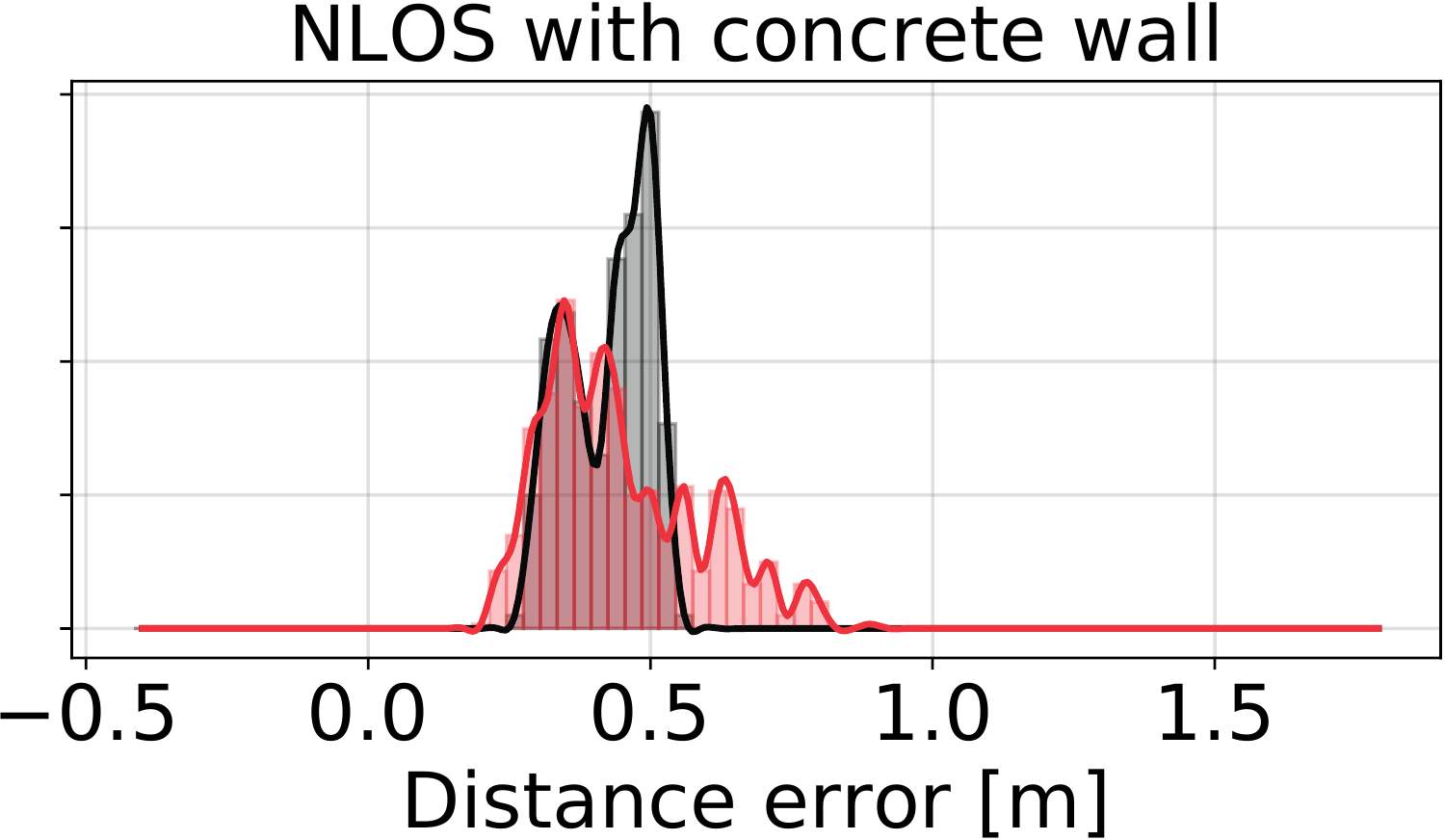}%
	}\quad%
	\subfloat[]{%
		\label{fig:nlos_human_pdf_dist_error}%
		\includegraphics[width=0.23\linewidth]{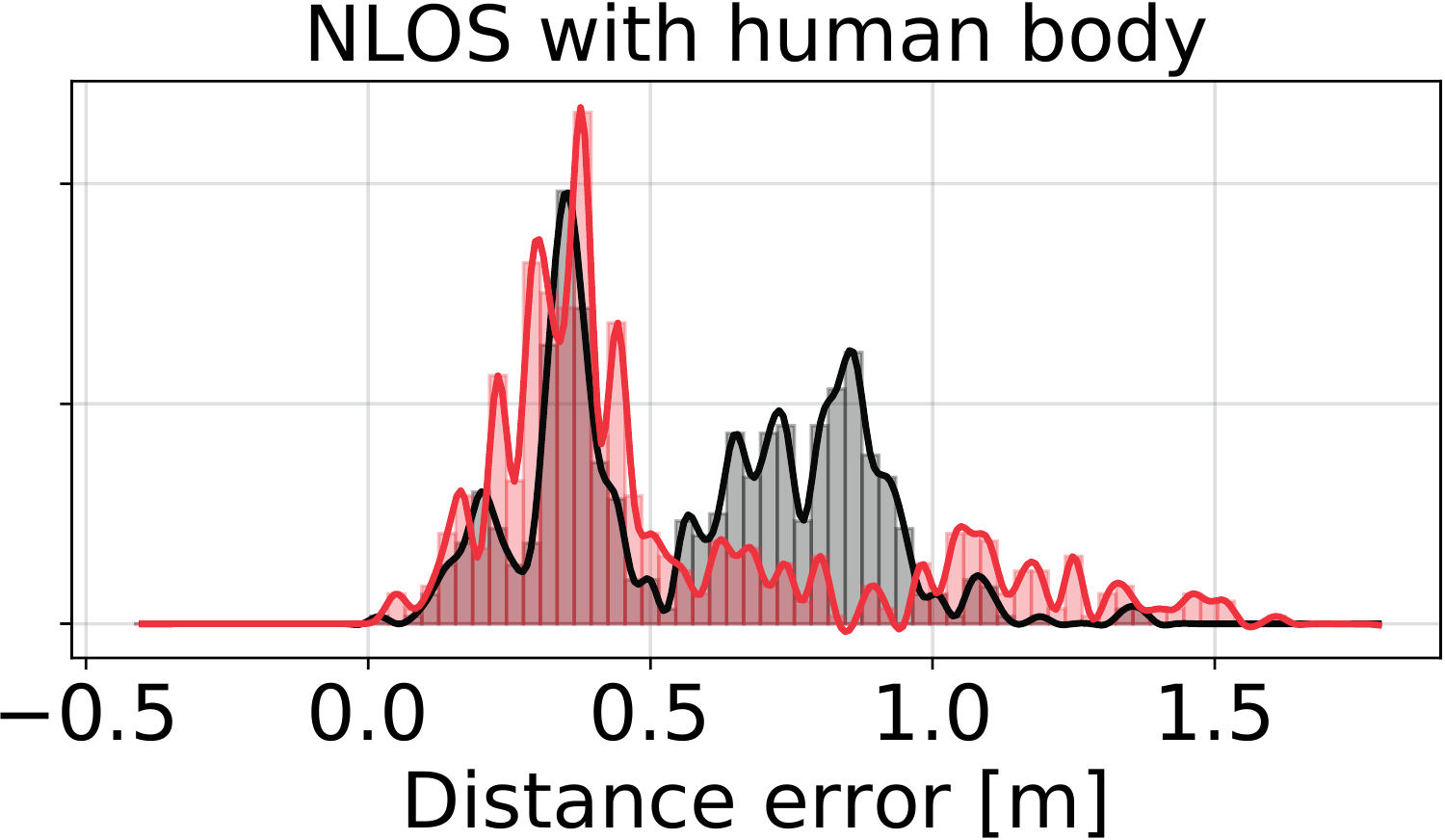}%
	}\\
	\subfloat[]{%
		\label{fig:los_boxplots_error}%
		\includegraphics[width=0.23\linewidth]{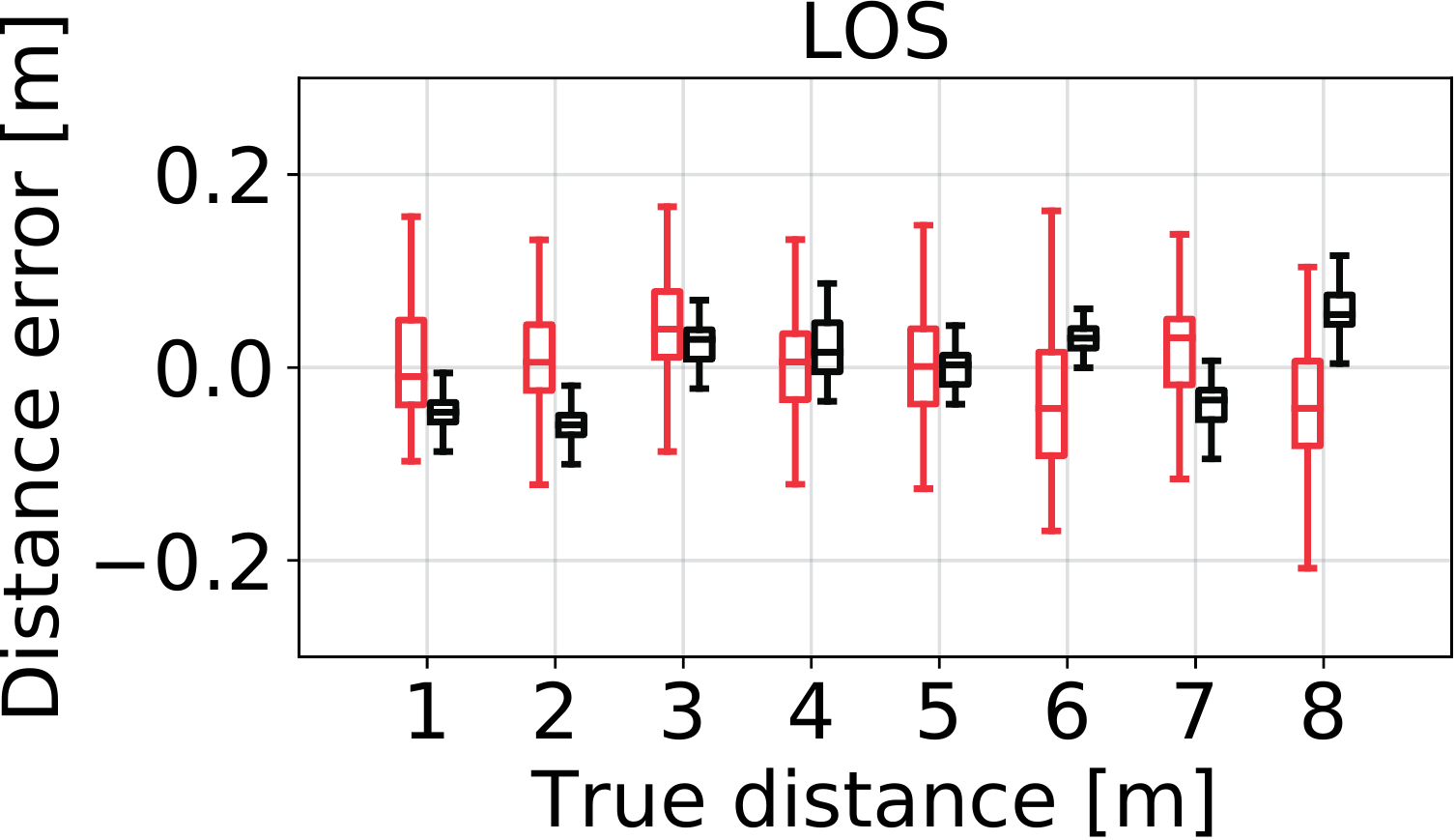}%
	}\quad%
	\subfloat[]{%
		\label{fig:nlos_drywall_boxplots_error}%
		\includegraphics[width=0.23\linewidth]{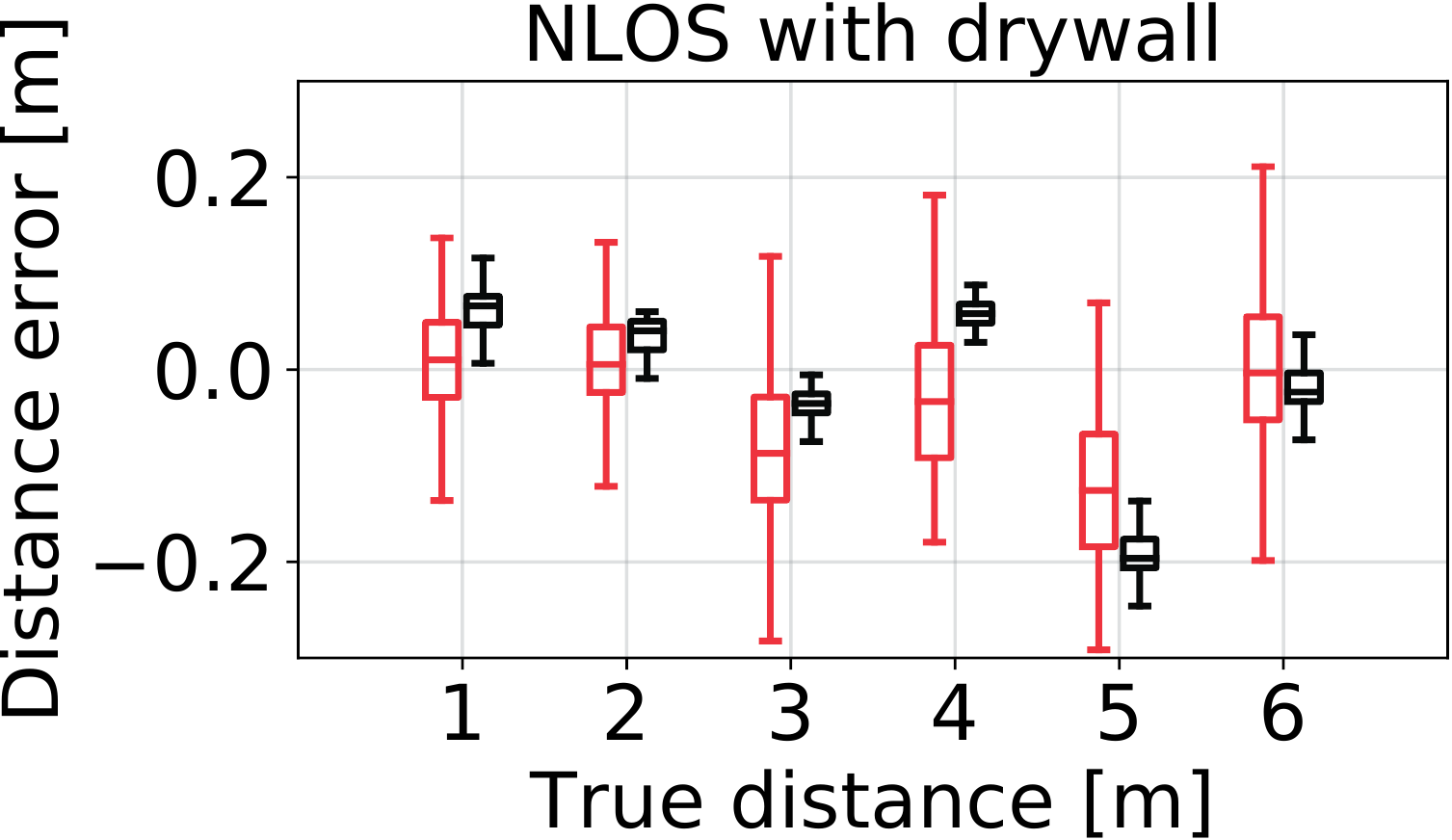}%
	}\quad%
	\subfloat[]{%
		\label{fig:nlos_wall_boxplots_error}%
		\includegraphics[width=0.23\linewidth]{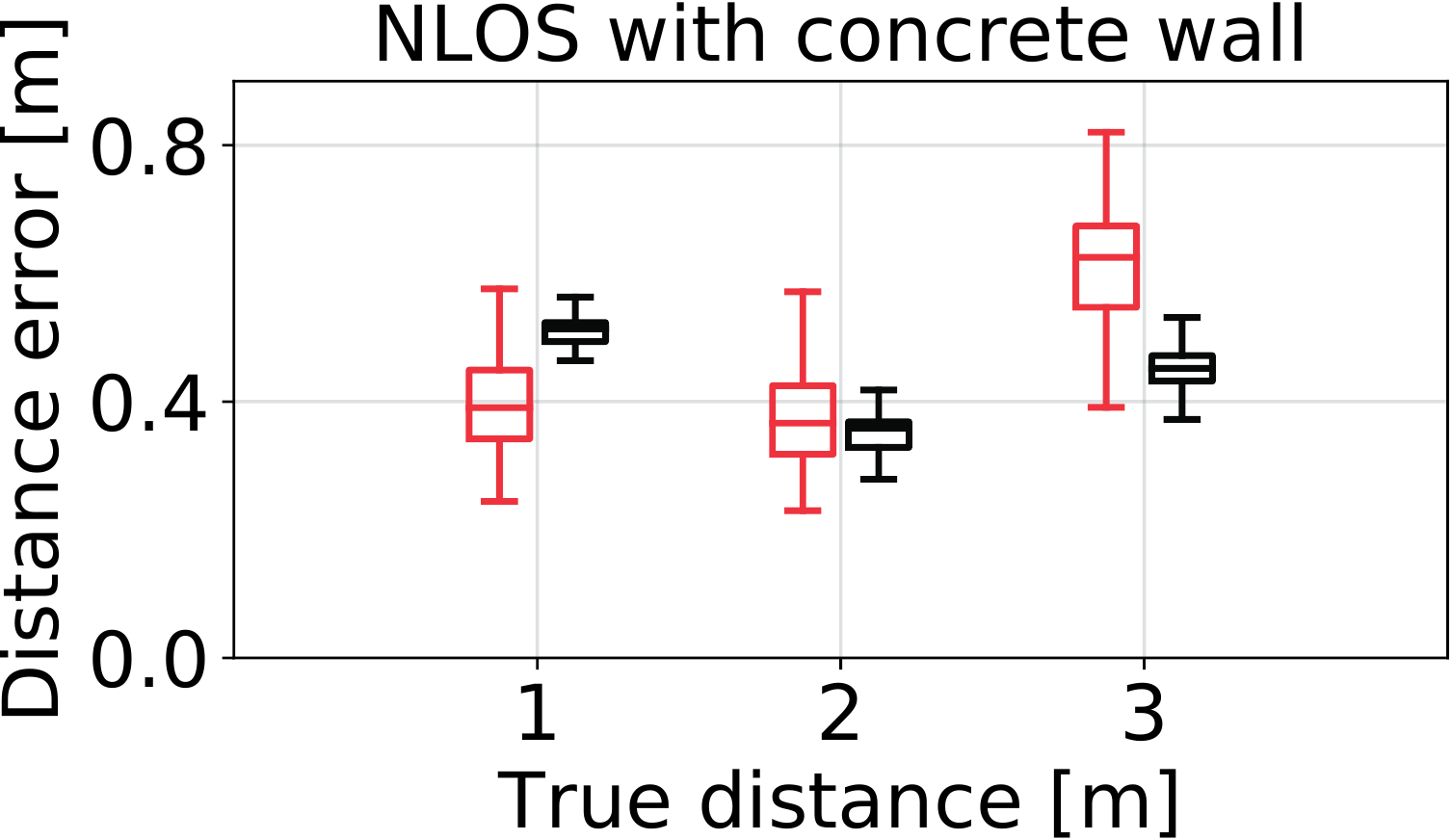}%
	}\quad%
	\subfloat[]{%
		\label{fig:nlos_human_boxplots_error}%
		\includegraphics[width=0.23\linewidth]{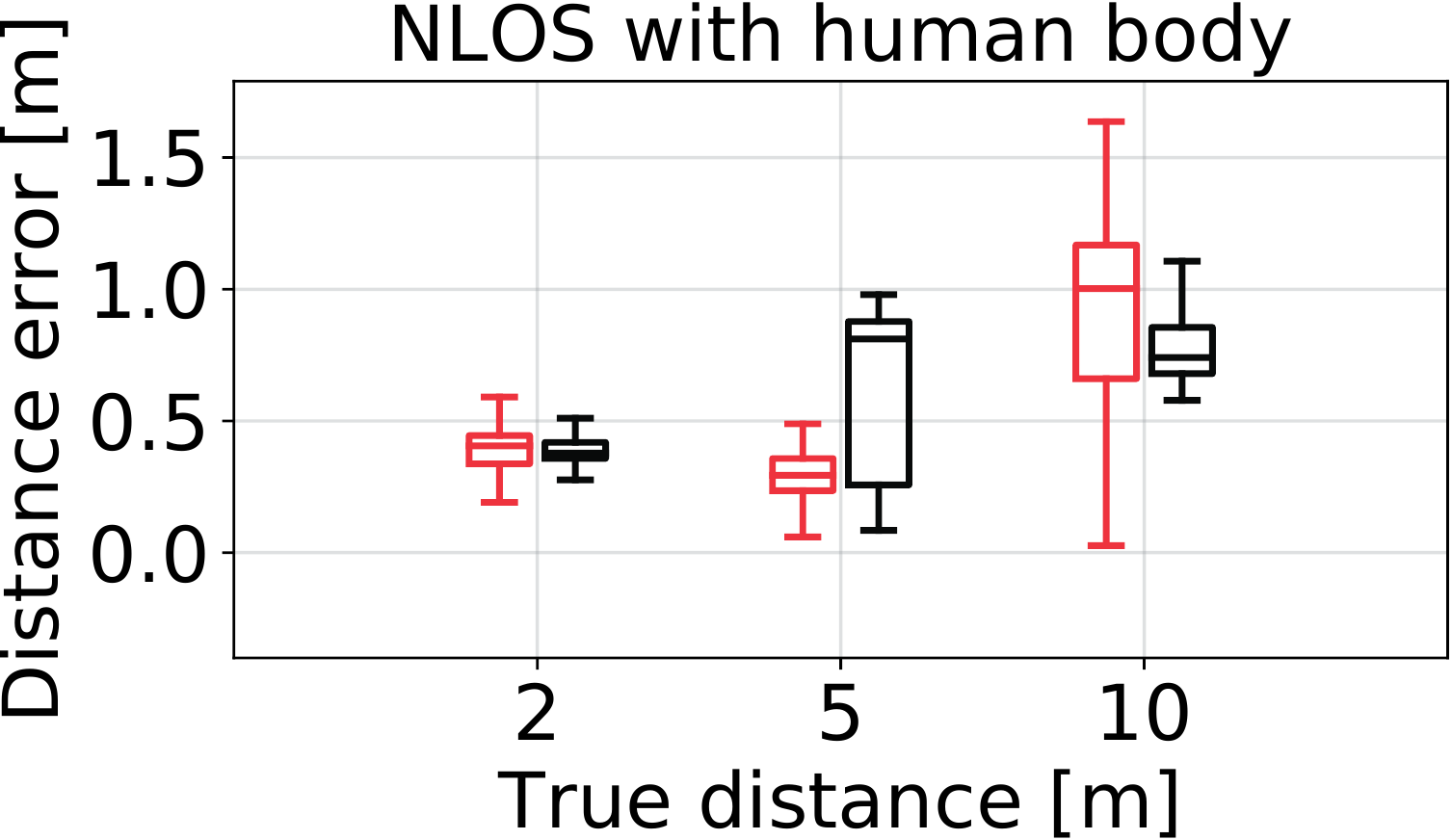}%
	}
	
	\caption{Comparison of the distribution of aggregated ranging errors (\figurename~\ref{fig:los_pdf_dist_error} to \ref{fig:nlos_human_pdf_dist_error}) and of the individual distributions at each distance (\figurename~\ref{fig:los_boxplots_error} to \ref{fig:nlos_human_boxplots_error}) of 3db Access and Decawave devices in LOS, NLOS with drywall, NLOS with concrete wall, and NLOS with human body.}
	\label{fig:pdf_dist_error}
\end{figure*}

\tablename~\ref{tab:distance-error-statistics} presents the mean, standard deviation, and inter-quartile range (IQR) of the distance errors of 3db and Decawave devices, computed as:
\begin{equation}
e_d = \hat{d} - d,
\end{equation}
where $\hat{d}$ is the measured and $d$ is the true distance. \figurename~\ref{fig:los_pdf_dist_error} to \ref{fig:nlos_human_pdf_dist_error} compare the PDF of the aggregated ranging errors at all distances for a particular LOS/NLOS scenario and \figurename~\ref{fig:los_boxplots_error} to \ref{fig:nlos_human_boxplots_error} compare the error distributions of Decawave and 3db at each test point. The boxplots use Tukey's definition.

First, we notice that, at individual test points (\figurename~\ref{fig:los_boxplots_error} to \ref{fig:nlos_human_boxplots_error}), the errors of Decawave devices have a smaller spread than those of 3db devices. After the calibration procedure detailed in Appendix~\ref{ssec:distance-calib}, 3db devices had distance errors of $-0.05\pm$\SI{6.54}{\centi\meter}, while Decawave devices had errors of $0\pm$\SI{3.14}{\centi\meter}. Therefore, on the calibration data set, 3db devices had a bias \SI{5}{\milli\meter} higher and a standard deviation about $2\times$ larger than Decawave devices.
In the LOS scenario, the location and test points were different from the ones in the calibration data set. Hence, we expect errors to be slightly higher than in the calibration setup. Over all test points, 3db Access devices had errors of $\fpeval{\LosThreedbMean * 100} \pm$\SI{\fpeval{\LosThreedbStd * 100}}{\centi\meter} and Decawave devices achieved errors of $\fpeval{\LosDwMean * 100} \pm$\SI{\fpeval{\LosDwStd * 100}}{\centi\meter} in the LOS scenario. The standard deviation of Decawave devices in the LOS scenario is higher than in the calibration data set because, at individual test points, the absolute average error is also higher.

\begin{table}[t!]
	\caption{Statistics of distance measurement errors.}
	\centering
	\begin{tabular}{p{2.2cm} l 
			S[table-format=2.2]
			S[table-format=2.2]
			S[table-format=2.2]
		} 
		\toprule
		Scenario
		& Device
		& \multicolumn{1}{c}{\makecell{Mean $[$\si{\meter}$]$}}
		& \multicolumn{1}{c}{\makecell{Standard\\deviation\\ $[$\si{\meter}$]$}} 
		& \multicolumn{1}{c}{\makecell{IQR $[$\si{\meter}$]$}}\\ 
		\midrule
		
		\multirow{2}{*}{LOS}									
		&3db Access  &    \LosThreedbMean & \LosThreedbStd & \LosThreedbIqr    \\ 
		&Decawave    &   \LosDwMean    &        \LosDwStd        &    \LosDwIqr    \\
		
		\midrule
		
		\multirow{2}{*}{\makecell[l]{NLOS with drywall}} 
		&3db Access  &   \DrywallThreedbMean    &        \DrywallThreedbStd        &    \DrywallThreedbIqr    \\ 
		&Decawave    &   \DrywallDwMean    &        \DrywallDwStd        &    \DrywallDwIqr    \\
		
		\midrule
		
		\multirow{2}{*}{\makecell[l]{NLOS with\\ concrete wall}} 
		&3db Access  &    \ConcreteThreedbMean    &        \ConcreteThreedbStd        &    \ConcreteThreedbIqr    \\ 
		&Decawave    &    \ConcreteDwMean    &        \ConcreteDwStd        &    \ConcreteDwIqr    \\
		
		\midrule
		
		\multirow{2}{*}{\makecell[l]{NLOS with\\ human body}} 
		&3db Access  &    \HumanThreedbMean    &        \HumanThreedbStd        &    \HumanThreedbIqr    \\ 
		&Decawave    &    \HumanDwMean    &        \HumanDwStd        &    \HumanDwIqr    \\ 
		
		\bottomrule
	\end{tabular}
	
	\centering
	\label{tab:distance-error-statistics}
\end{table}

Drywall is frequently used in modern buildings to delimit interior spaces. Surprisingly, it does not seem to cause a positive bias but a small negative one in both 3db and Decawave measurements, as can be seen in \figurename~\ref{fig:nlos_drywall_pdf_dist_error}. The errors caused by this type of obstruction are within several centimeters of LOS errors. This type of NLOS scenario is sometimes referred to in the literature as ``soft'' NLOS~\cite{venkatesh2007non}, since the LOS multipath component is still present in the CIR and the correct distance can be recovered. The fact that drywall does not introduce large errors is good news for proximity-detection and localization applications, because it means that ranging and localization errors will be small even if the devices are in different rooms, if they are separated by drywall.

Thicker obstacles such as a wall or the human body can affect the signal in multiple ways. First, through this type of obstacles, the signals usually travel at a lower speed than through the air, which causes a delay in the round-trip time and hence an error in the distance measurement. Second, these obstacles can attenuate the direct path component or block it altogether, case in which copies of the signal reflected on surrounding objects can cause errors in the TOA estimation algorithm. These scenarios are also known as ``hard'' NLOS~\cite{venkatesh2007non}. 

The aggregated distribution of ranging errors in hard NLOS scenarios (Figures~\ref{fig:nlos_wall_pdf_dist_error} and~\ref{fig:nlos_human_pdf_dist_error}) is often heavy-tailed and no longer Gaussian-shaped. However, in most cases, the error distribution \textit{at each test point} (i.e., at individual distances) is still approximately Gaussian, as shown in each boxplot from  \figurename~\ref{fig:nlos_wall_boxplots_error} and~\ref{fig:nlos_human_boxplots_error}. The biases depend on the particular environment and the multipath components that arrive at the receiver. Hence, at different distances, the bias can vary depending on how multipath components add up, which is why the \textit{aggregated} NLOS distributions can be multi-modal.

With concrete wall and human body shadowing, the ranging errors are between \SIrange{\fpeval{\ConcreteDwMean * 100}}{\fpeval{\HumanDwMean * 100}}{\centi\meter}. In both hard NLOS scenarios, Decawave devices had a standard deviation \SIrange{6}{7}{\centi\meter} lower than 3db devices. Only with human body shadowing the IQR of Decawave errors is \SI{\fpeval{(\HumanDwIqr - \HumanThreedbIqr) * 100}}{\centi\meter} higher than that of 3db because its error distribution, although shorter, has a fatter tail.

In conclusion, in all scenarios, both devices had mean errors within \SIrange{2}{5}{\centi\meter} of each other, with Decawave devices performing better in all scenarios except for the NLOS with human body shadowing one. In LOS and soft NLOS scenarios, the devices differed in the standard deviation and IQR by \SIrange{1}{2}{\centi\meter}, with Decawave devices obtaining a better performance in most cases. In the hard NLOS scenarios, Decawave devices had a lower spread than 3db devices by $1.23$--$2\times$, except for the IQR in the case with NLOS with human body, which was $\num[round-mode=places,round-precision=2]{\fpeval{\HumanDwIqr / \HumanThreedbIqr}}\times$ higher than the one of 3db devices. At individual test points, Decawave devices had $2\times$ lower spread than 3db devices. Compared to our previous work~\cite{flueratoru2020energy}, the performance of 3db devices has improved thanks to the refined calibration and to the correction of the firmware issues that previously caused large outliers in certain NLOS situations.

\subsection{Channel Diversity}
\label{ssec:ch-diversity}

The analysis so far was based only on distance measurements acquired on the \SI{6.5}{\giga\hertz} channel, since this was the only one available on the MDEK1001 devices. However, UWB devices can operate in more bands. The WiMedia Alliance defined 14 bands with \SI{500}{\mega\hertz} bandwidth in the range of \SIrange{3.1}{10.6}{\giga\hertz} for UWB communications\footnote{We will alternatively refer to the bands as (communication) channels.}. The use of the lower band between \SIrange{3.5}{4.5}{\giga\hertz} is often allowed only with interference mitigation techniques, while the \SIrange{6}{8.5}{\giga\hertz} band is less subject to regulations~\cite{uwb-regulations} and available in most countries. Since 3db Access devices can operate in the bands centered at $6.5$, $7$, and \SI{7.5}{\giga\hertz}, it is useful to compare the performance on these channels and investigate whether distance measurements could benefit from channel diversity.

We programmed the 3db Access devices to acquire measurements on the $6.5$, $7$, and \SI{7.5}{\giga\hertz} channels consecutively. The sampling period between measurements on successive channels is $T / 3$, where $T$ is the sampling period from \tablename~\ref{tab:ranging-setup}. 

\begin{figure}[t!]
	\centering
	\includegraphics[width=0.49\textwidth]{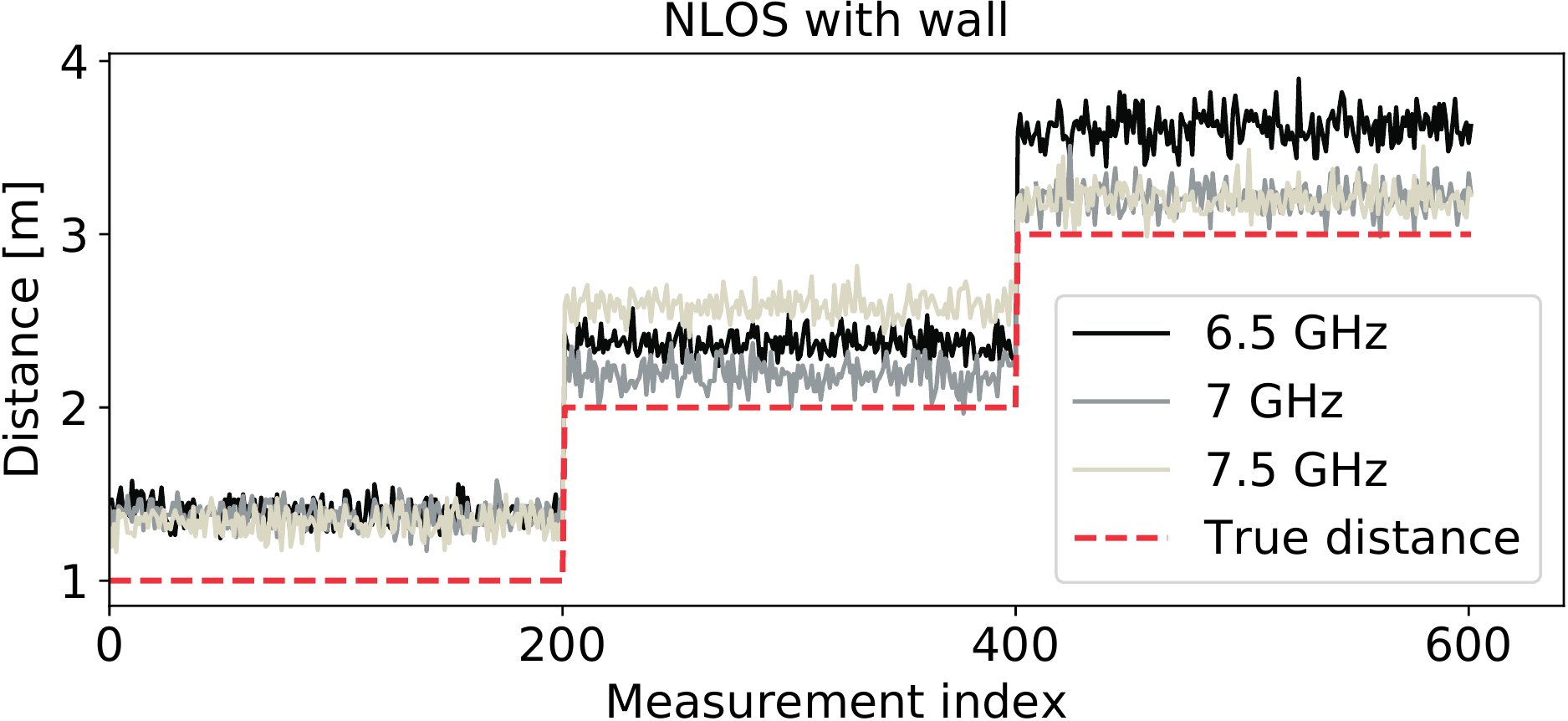}
	
	\caption{Impact of channel diversity on distance accuracy when the signal passes through a concrete wall: at different distances, some channels perform better than the others.}
	\label{fig:nlos_channel_diversity}
\end{figure}

We noticed that, while LOS errors have the same characteristics irrespective of the channel (we calibrated the devices to operate this way), in hard NLOS situations some channels can experience better conditions at different locations. \figurename~\ref{fig:nlos_channel_diversity} presents such an example for NLOS with concrete wall shadowing: at \SI{2}{\meter} distance, the \SI{7}{\giga\hertz} channel has lower errors than the others, while at \SI{3}{\meter} distance the $7$ and \SI{7.5}{\giga\hertz} channels had the highest accuracy. This can happen due to multipath interference, when copies of the signal traveling through multiple paths add up constructively or destructively at the receiver. The interference pattern depends on the frequency of the signal. Signals sent on different frequencies can have different propagation characteristics through obstacles. Since hard NLOS situations almost always cause positive biases, as we saw in the previous section, this prompts us to investigate whether taking the minimum or the mean of consecutive measurements (also called the min- and mean-select methods, respectively) acquired on different channels can improve the ranging accuracy.

\begin{figure}[t!]
	\centering
	\subfloat[]{%
		\label{fig:ch_div_los}%
		\includegraphics[width=0.48\linewidth]{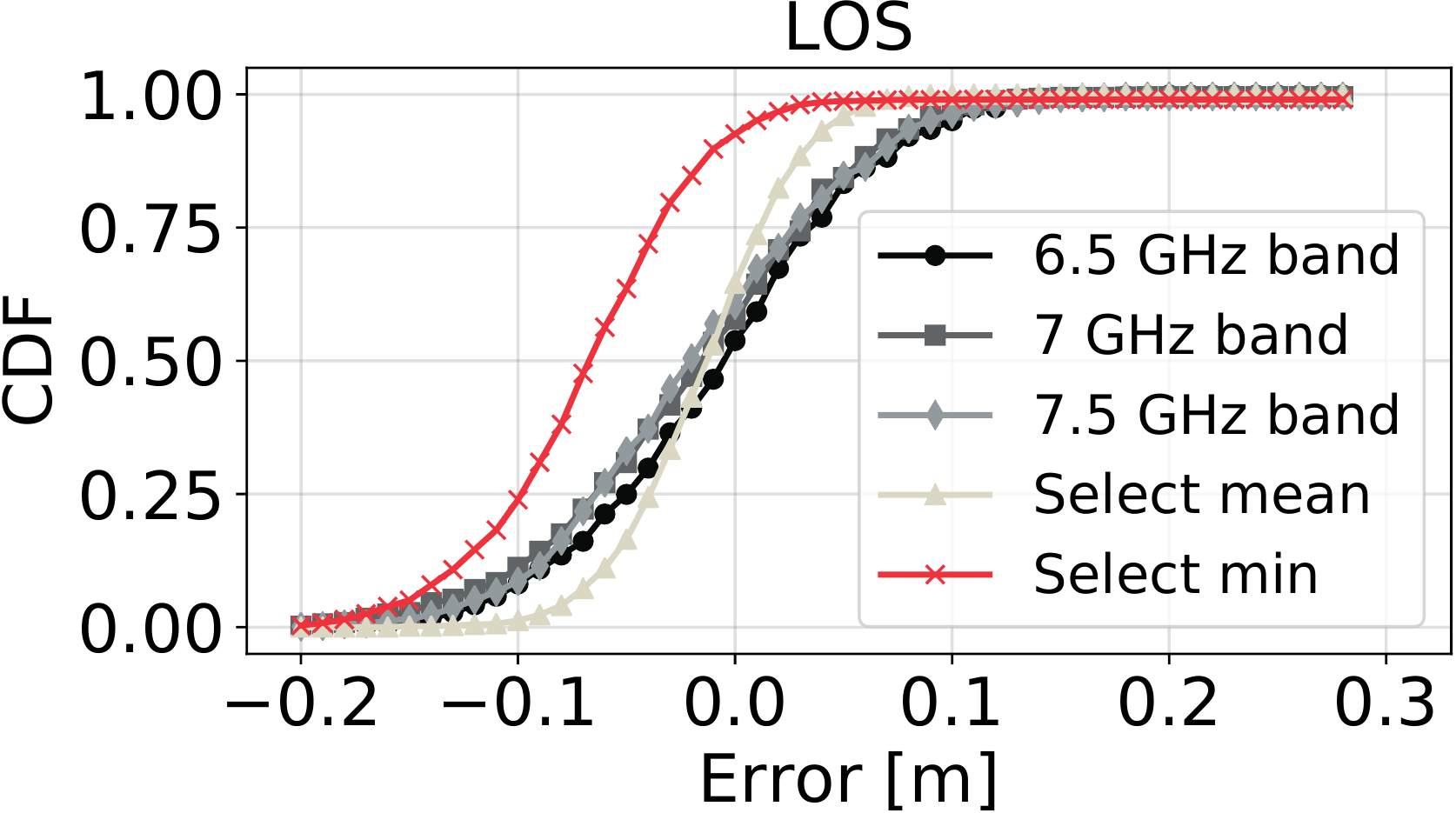}%
	}\quad%
	\subfloat[]{%
		\label{fig:ch_div_nlos_plexi}%
		\includegraphics[width=0.48\linewidth]{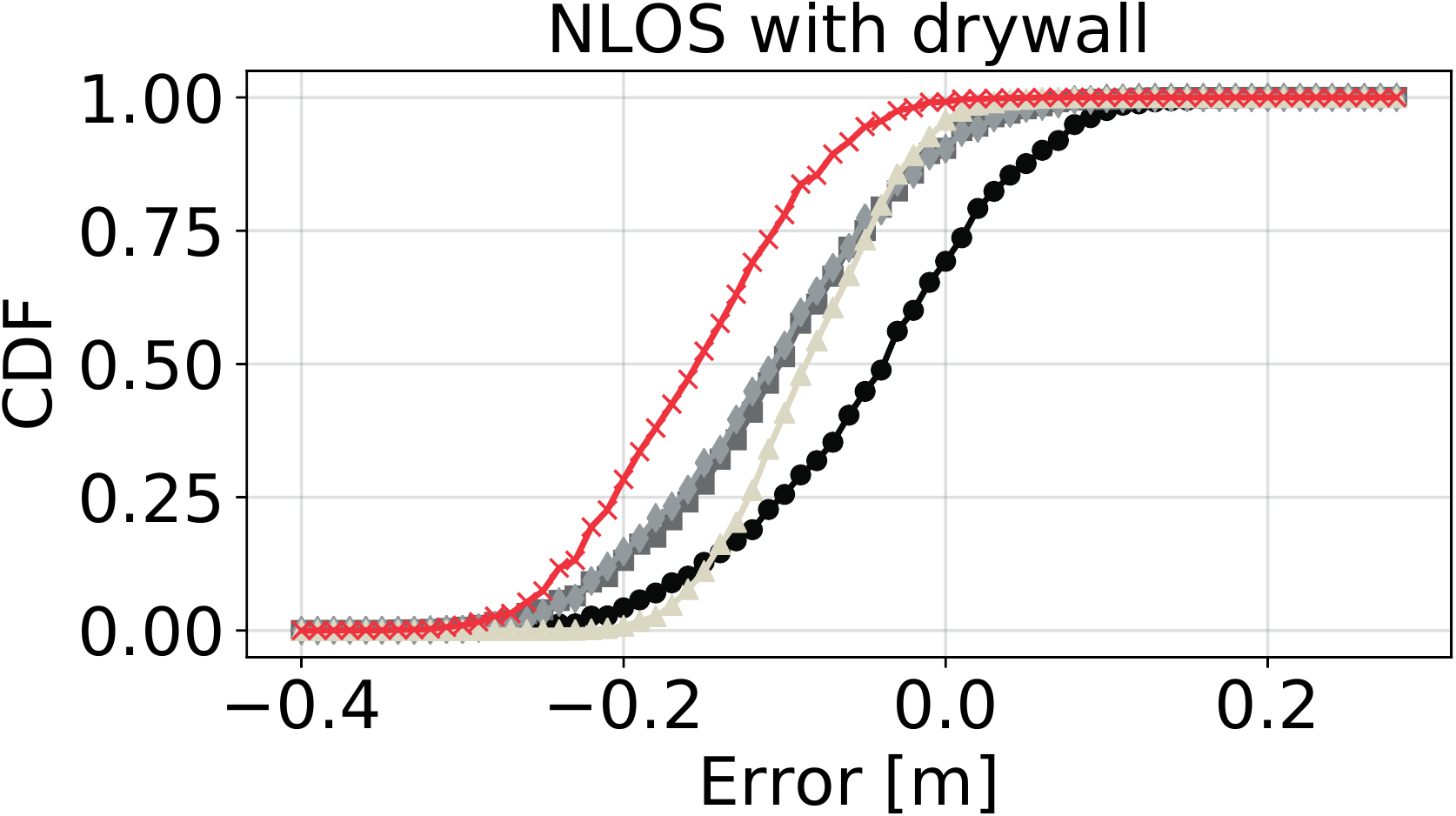}%
	} \\
	\subfloat[]{%
		\label{fig:ch_div_nlos_wall}%
		\includegraphics[width=0.48\linewidth]{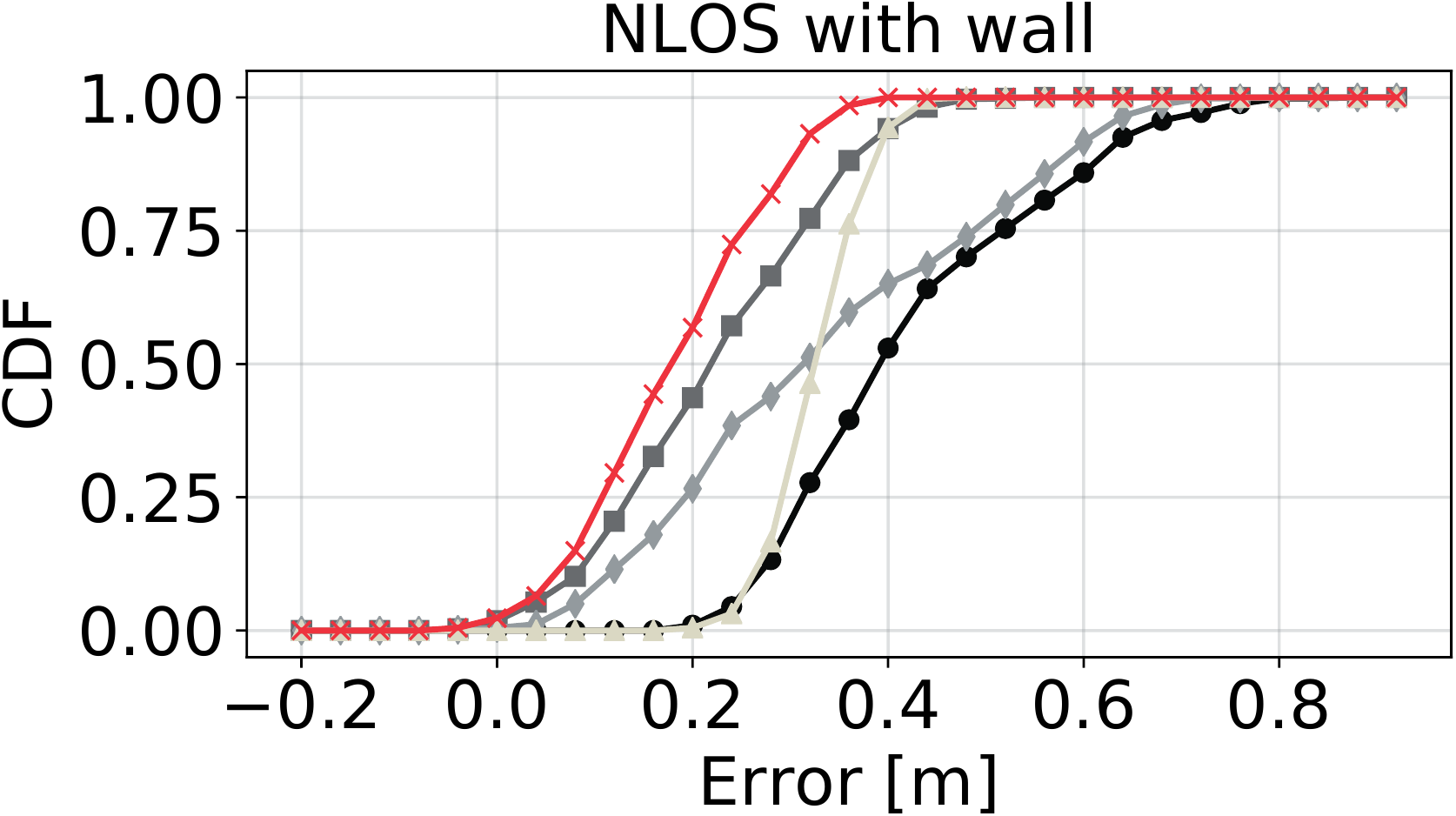}%
	}\quad%
	\subfloat[]{%
		\label{fig:ch_div_nlos_human}%
		\includegraphics[width=0.48\linewidth]{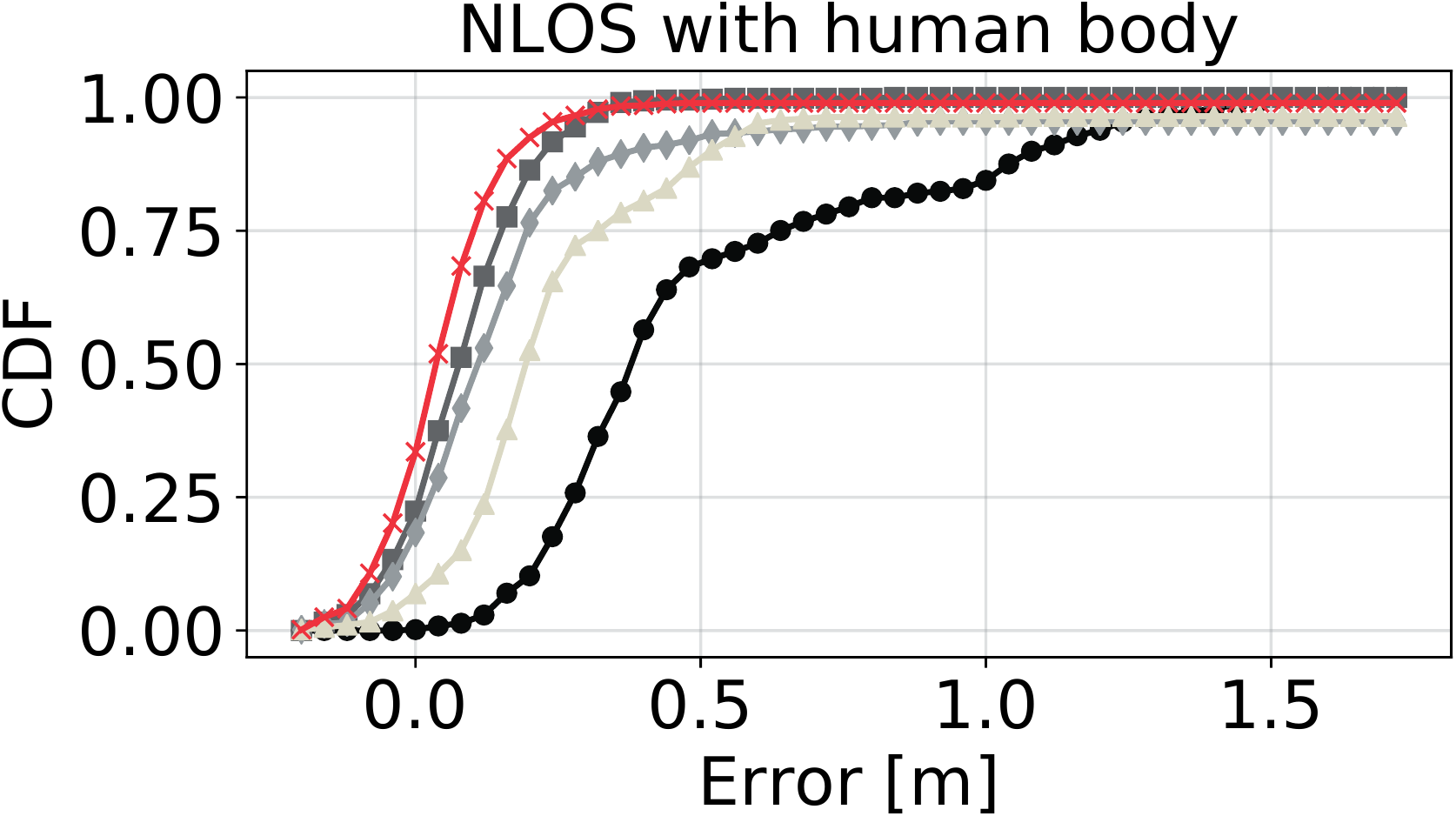}%
	}
	
	\caption{The cumulative distribution function (CDF) of 3db Access ranging errors when using only measurements from the $6.5$, $7$, and \SI{7.5}{\giga\hertz} band or the mean, median, or minimum of a set of consecutive measurements in all bands. The figure compares the CDF in (a) LOS, (b) NLOS with drywall, (c) NLOS with concrete wall, and (d) NLOS with human body. The legend in \figurename~\ref{fig:ch_div_los} is common to all subfigures. }
	\label{fig:dist_channel_diversity}
\end{figure}

\figurename~\ref{fig:dist_channel_diversity} shows the cumulative distribution function (CDF) of measurement errors in the individual bands, as well as of errors when we select either the mean or minimum of three consecutive measurements acquired on all channels. In a regular LOS scenario, all channels perform similarly. The mean-select method leaves the bias still centered around 0 and decreases the standard deviation with approximately \SI{2}{\centi\meter}. Instead, selecting the minimum measurement in LOS shifts the error distribution towards a negative mean, which decreases the accuracy. The same happens in the NLOS with drywall case.

In hard NLOS (\figurename~\ref{fig:ch_div_nlos_wall} and~\ref{fig:ch_div_nlos_human}), however, the min-select method achieves a median error of approximately \SI{18}{\centi\meter} with wall and \SI{5}{\centi\meter} with human obstructions. Therefore, this simple channel diversity technique reduces the bias of hard NLOS measurements by more than $2\times$ compared to using only the \SI{6.5}{\giga\hertz} channel. The mean-select method also reduces the error compared to individual channels in ``bad'' conditions but to a lesser degree than the min-select.

It is not always desirable to use channel diversity. First, acquiring measurements on all channels increases the number of messages and thus the energy consumption. Second, the min-select method increased the accuracy by $2\times$ in hard NLOS but decreased it in LOS. One method to take full advantage of channel diversity is to apply a NLOS detection technique~\cite{khodjaev_survey_2010} and acquire measurements on all channels only when the devices are in NLOS. In this way, additional measurements are triggered only when a higher ranging or localization accuracy is desired. Investigating the viability and efficiency of this method is left as future work.

\subsection{Localization}
\label{ssec:localization-hrp-lrp}

\begin{figure}[t!]
	\centering
	\includegraphics[width=0.49\textwidth]{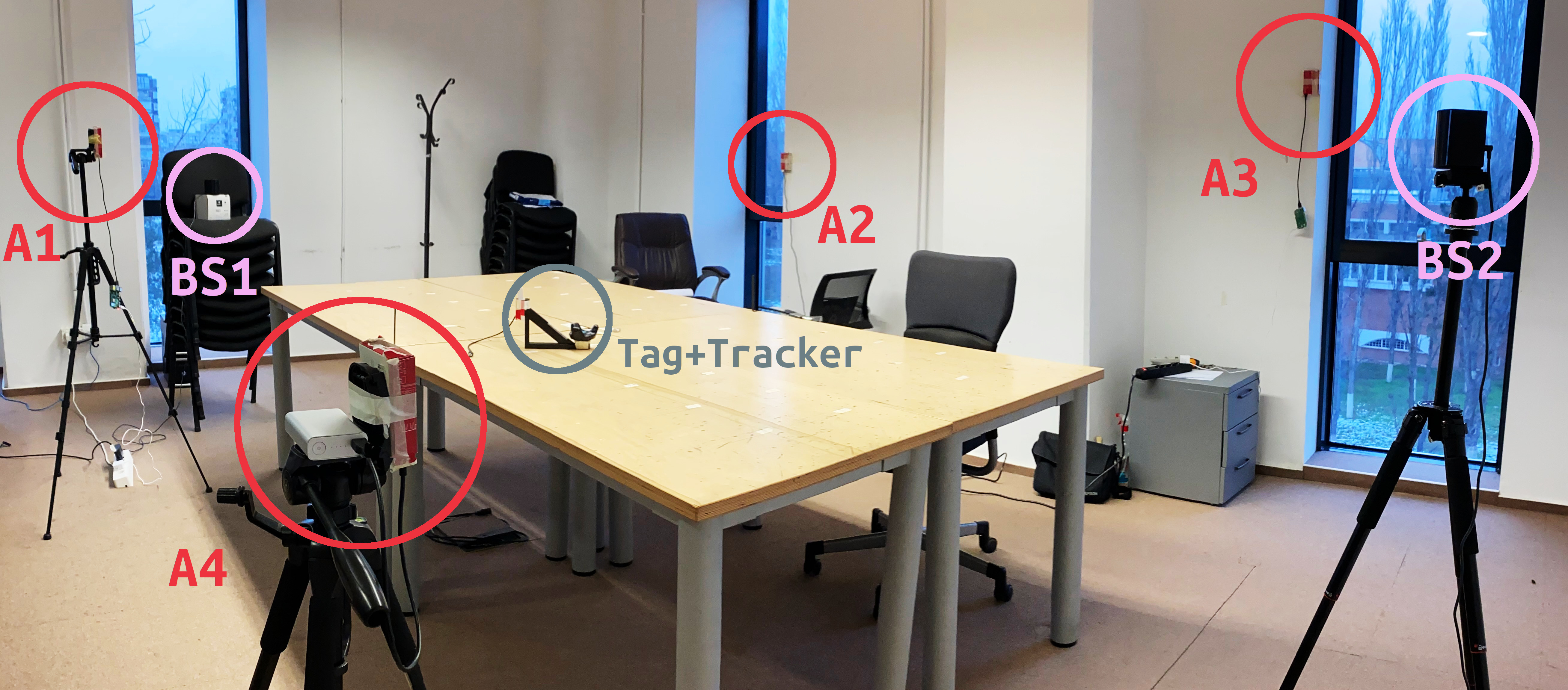}
	
	\caption{Localization setup: the four anchors (\texttt{A1} to \texttt{A4}) encompass an area of approximately $4.5 \times$\SI{3.6}{\meter} and the tracking area is on the table. The ground truth was acquired using two HTC Vive base stations (\texttt{BS1} and \texttt{BS2}) and a tracker that was colocated with the UWB tag, shown on the table.}
	\label{fig:localization_setup}
\end{figure}

\begin{figure*}[t!]
	\centering
	\subfloat{%
		\label{}%
		\includegraphics[width=0.4\linewidth]{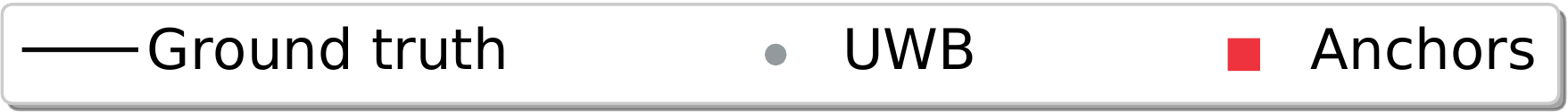}%
	}\\
	\setcounter{subfigure}{0}
	\subfloat[]{%
		\label{fig:los_3db}%
		\includegraphics[width=0.23\linewidth]{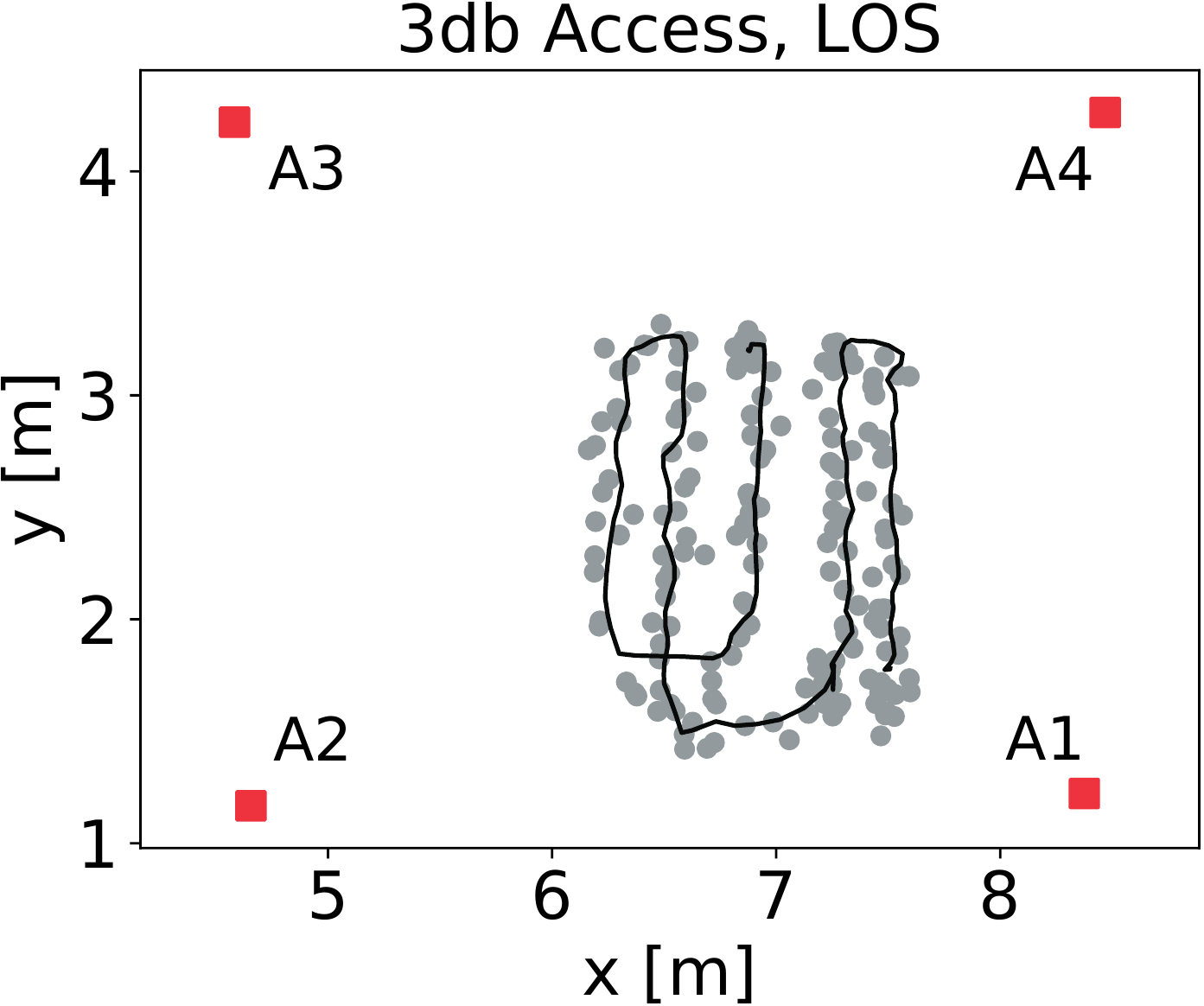}%
	}\quad%
	\subfloat[]{%
		\label{fig:los_dw}%
		\includegraphics[width=0.23\linewidth]{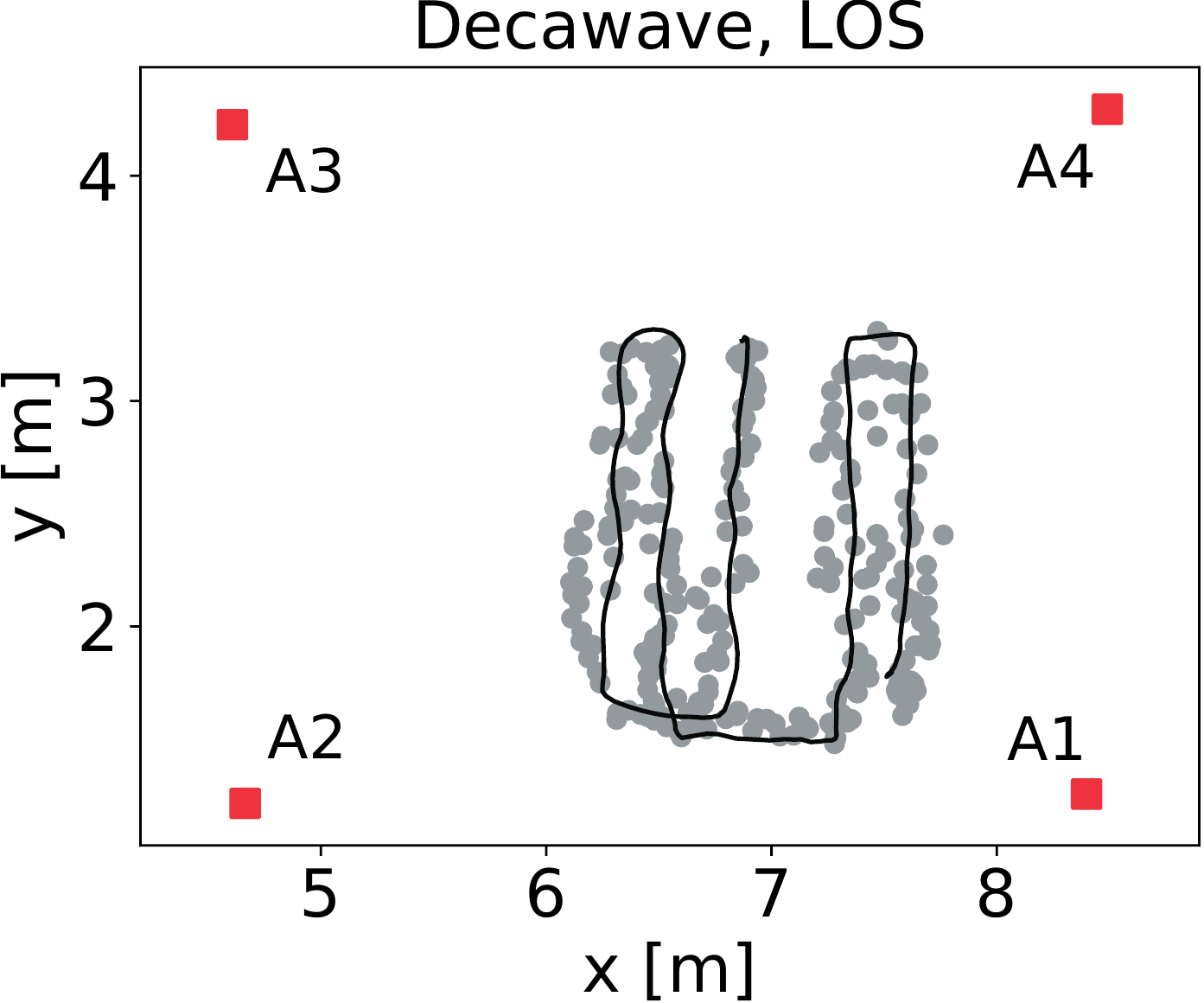}%
	}\quad%
	\subfloat[]{%
		\label{fig:nlos_3db}%
		\includegraphics[width=0.23\linewidth]{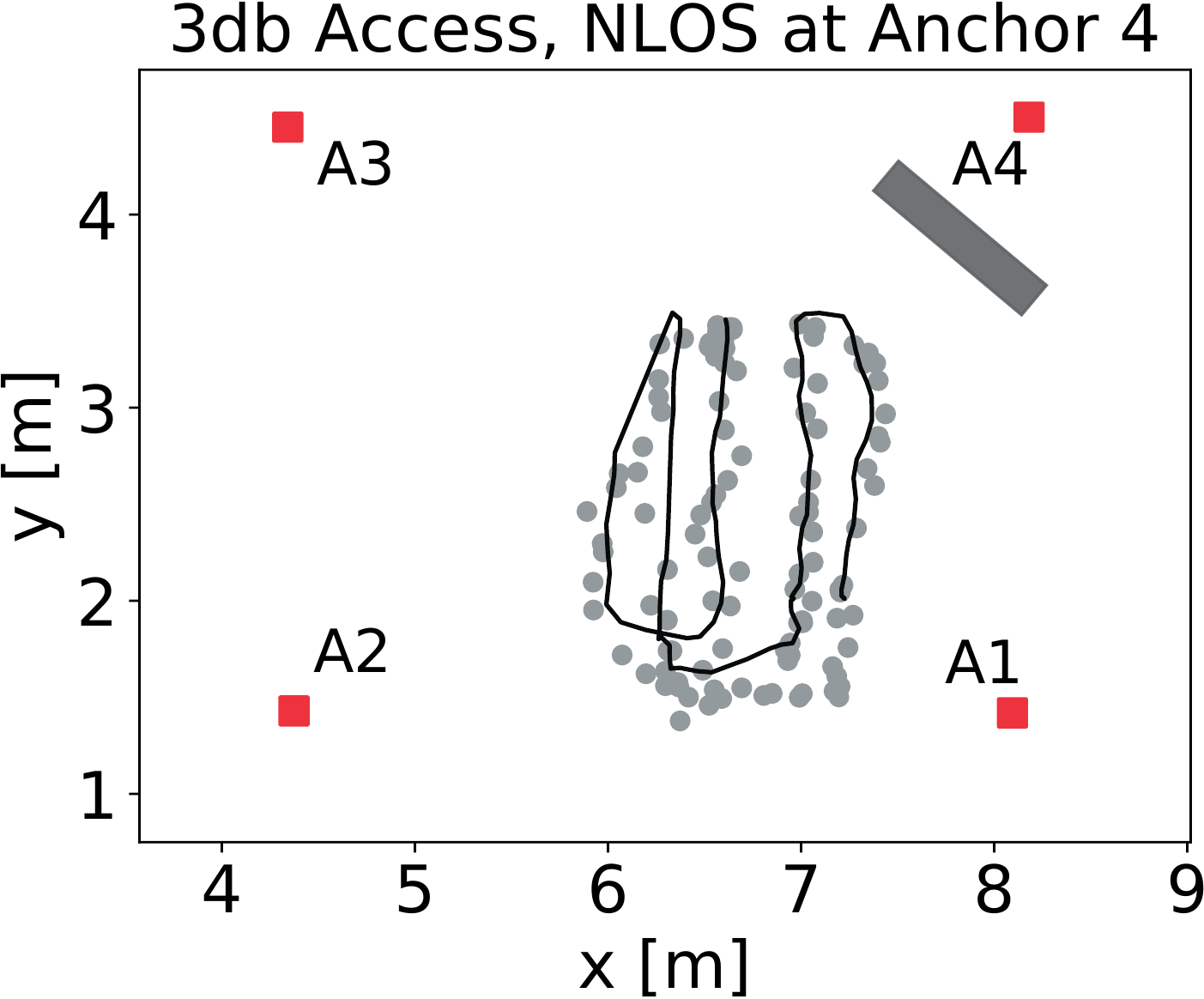}%
	}\quad%
	\subfloat[]{%
		\label{fig:nlos_dw}%
		\includegraphics[width=0.23\linewidth]{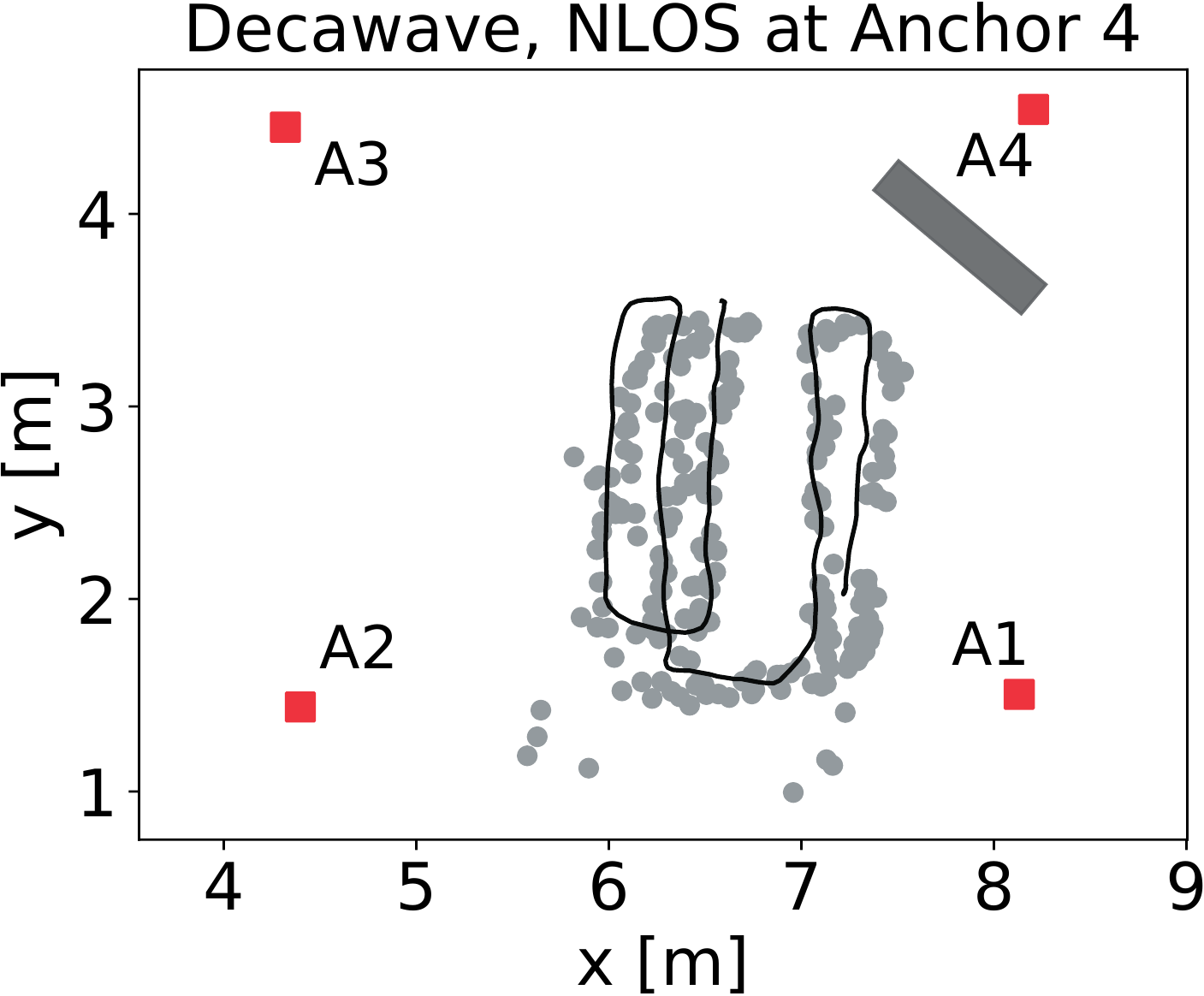}%
	}
	
	\caption{Localization errors of 3db Access and Decawave devices in LOS and NLOS, where anchor 2 (\texttt{A2}) was obstructed by a human body.}
	\label{fig:localization}
\end{figure*}

\begin{table*}[t!]
	\caption{Statistics of distance errors in localization recordings.}
	\centering
	\begin{tabular}{@{\extracolsep{4pt}}l l 
			S[table-format=2.1]
			S[table-format=2.1]
			S[table-format=2.1]
			S[table-format=2.1]
			S[table-format=2.1]
			S[table-format=2.1]
			S[table-format=2.1]
			S[table-format=2.1]@{}
		} 
		\toprule
		
		\multirow{2}{*}{Case}
		& \multirow{2}{*}{Device}
		& \multicolumn{4}{c}{Mean distance error [\si{\centi\meter}]} 
		& \multicolumn{4}{c}{\makecell[c]{Standard deviation of\\ distance error [\si{\centi\meter}]}} \\ [1ex] \cline{3-6} \cline{7-10}  \\ [-1.5ex]

		& & \multicolumn{1}{c}{\texttt{A1}} & \multicolumn{1}{c}{\texttt{A2}} & \multicolumn{1}{c}{\texttt{A3}} & \multicolumn{1}{c}{\texttt{A4}}
		& \multicolumn{1}{c}{\texttt{A1}} & \multicolumn{1}{c}{\texttt{A2}} & \multicolumn{1}{c}{\texttt{A3}} & \multicolumn{1}{c}{\texttt{A4}} \\
		
		\midrule
		
		\multirow{2}{*}{LOS}							
		& 3db Access 
		& 7.4        
		& 1.1        
		& 12.3       
		& 8.6        
		& 6.8        
		& 6.7        
		& 11.3       
		& 6.9        \\
		& Decawave   
		& 6.3        
		& 1.5        
		& 14.6       
		& 7.8        
		& 9.1        
		& 6.7        
		& 13.2       
		& 14.0		\\
		
		\midrule
		
		\multirow{2}{*}{\makecell[l]{NLOS at \texttt{A4}}}					
		& 3db Access 
		& 3.9        
		& 3.5        
		& 17.8       
		& 26.9       
		& 6.6        
		& 6.1        
		& 9.1        
		& 16.8 \\
		& Decawave    
		& 3.3        
		& 3.4        
		& 16.0       
		& 32.8       
		& 9.5        
		& 6.7        
		& 12.5       
		& 41.7 \\
		
		\bottomrule
	\end{tabular}
	
	\centering
	\label{tab:distance-error-statistics-loc}
\end{table*}

Because UWB devices have become popular for indoor localization, in this section we compare the localization performance of the Decawave and 3db Access devices experimentally. 
We placed four anchors over an area of approximately $4.5 \times$\SI{3.6}{\meter} shown in \figurename~\ref{fig:localization_setup} and at heights between \SIrange{1.2}{1.8}{\meter} (the anchors need to be at arm's length to acquire their ground truth location). We moved the tag by hand, at a height of approximately \SI{40}{\centi\meter} above the table, along predefined points marked on the table. Since the tag was moved by hand, the paths in the two recordings were not identical, but very similar nevertheless, as can be seen from \figurename~\ref{fig:localization}. This discrepancy should not significantly impact the comparison. During the first half of the trajectory, we oriented the tag towards anchors \texttt{A1} and \texttt{A4}, while in the second half we oriented it towards anchors \texttt{A2} and \texttt{A3}. We changed the orientation in order to vary the relative pose between the tag and the anchors, which can influence the ranging error~\cite{ledergerber_calibrating_2018}.

The tags initiated the SS-TWR to each anchor. The 3db tag performed distance measurements to anchors \texttt{A1} to \texttt{A4} in their index order. The order in which the Decawave tag interrogated anchors changed throughout the recording according to the proprietary localization algorithm of Decawave.

Ground truth locations were acquired by an HTC Vive motion capture system using the setup described in \cite{flueratoru2020htc}, which has an average accuracy of at least \SI{5}{\milli\meter}. The HTC Vive returns the location of a tracker which is colocated with the UWB tag. Then, a set of transformations is applied to recover the ground truth location of the tag. The anchor locations are also acquired using the HTC Vive system. 

We recorded measurements in two scenarios: one in which all anchors were in LOS with the tag and one in which the direct path to one of the anchors (\texttt{A4}) was blocked by a person, so the tag was at all times in NLOS with one anchor. Given the results from Section~\ref{ssec:distance-meas}, we expect a higher bias in the measurements coming from anchor \texttt{A4} but not necessarily the same bias from \tablename~\ref{tab:distance-error-statistics}, since the bias depends also on the particular room setup and environment. The NLOS distance error will introduce a \textit{localization} error, which can be partially compensated by the correct distances received from the other anchors. Even in LOS, distance measurements can be affected by orientation errors caused by the irregular antenna radiation pattern~\cite{ledergerber_calibrating_2018}.

During each recording, for each type of localization system (based on Decawave or 3db Access), we recorded the distances between each anchor and the tag, which were then given as input to a multilateration algorithm. As mentioned in Section~\ref{ssec:background-localization}, for both localization systems, we used the Gauss-Newton multilateration algorithm strengthened with a regularization term. We initialized the algorithm with $\delta=$\SI{1}{\milli\meter}, $k_{max} = 10$ iterations, $\bm{x}_r =$ the median of the anchors' locations, and $c = 10^{-1}$ (corresponding to a standard deviation of \SI{10}{\meter} around $\bm{x}_r$, suitable for our setup). Although the Decawave MDEK1001 kit has its own localization algorithm, we did not use it for the comparison since the algorithm is closed-source and therefore we could not apply it on the anchor-tag distances given by 3db Access devices.

\tablename~\ref{tab:distance-error-statistics-loc} presents the mean and standard deviation of distance errors between the tag and each anchor $A_i$:
\begin{equation}
e_d = \hat{d}_{ij} - d_{ij}.
\end{equation}
The true distance $d_{ij}$ is computed as the Euclidean distance between the location of anchor $A_i$ and each ground truth location of the tag, while $\hat{d}_{ij}$ is the measured distance between each anchor and the tag.

The average difference between the mean distance error of the two devices is \SI{1.15}{\centi\meter} and the average difference between the standard deviation of the distance error is \SI{2.82}{\centi\meter}, with Decawave devices having smaller errors in about \SI{50}{\percent} of the cases. As previously mentioned, in the \textit{localization} LOS scenario, the average bias is no longer null because of the changing orientation and the movement of the tag. All anchors have average errors under \SI{10}{\centi\meter} when they are in LOS with the tag, with the exception of \texttt{A3}. Anchor \texttt{A3} has higher distance errors than the others because of its close proximity to the concrete structure visible in \figurename~\ref{fig:localization_setup}. Although the structure does not completely obstruct the direct path between the tag and \texttt{A3}, it might cause the diffraction of the signal or other multipath effects, especially when the tag is close to the base station \texttt{BS1} (the distance errors are higher in that area). 

In NLOS, the effect of the human body shadowing is reflected in the distance error statistics of anchor \texttt{A4}. Similar to the ranging experiment, 3db Access devices have a mean ranging error lower by \SI{5.9}{\centi\meter} than Decawave devices. However, unlike in the ranging experiment, here the standard deviation of the Decawave distance errors in NLOS (i.e., between the tag and anchor \texttt{A4}) was $2.5\times$ higher than the one of 3db Access devices. It is worth noting, though, that even in the ranging experiment in NLOS with human body, 3db Access devices had $\num[round-mode=places,round-precision=2]{\fpeval{\HumanDwIqr / \HumanThreedbIqr}}\times$ lower IQR than Decawave devices.

\newcommand{\loctablecolwidth}{1.07}
\begin{table}[t!]
	\caption{Statistics of measurement-based localization errors.}
	
	\centering
	\begin{tabular}{@{\extracolsep{2pt}} l l 
			S[table-format=2.1, table-column-width=\loctablecolwidth cm]
			S[table-format=2.1, table-column-width=\loctablecolwidth cm]
			S[table-format=2.1, table-column-width=\loctablecolwidth cm]
			S[table-format=2.1, table-column-width=\loctablecolwidth cm]@{}
		} 
		\toprule
		
		\multirow{3}{*}{Case}
		& \multirow{3}{*}{Device}
		& \multicolumn{4}{c}{Localization error} \\
		
		&
		& \multicolumn{2}{c}{\makecell{Mean $[$\si{\centi\meter}$]$}}
		& \multicolumn{2}{c}{\makecell{Standard\\ deviation $[$\si{\centi\meter}$]$}} \\ [1ex] \cline{3-4} \cline{5-6} \\ [-1.5ex]
		
		& & \multicolumn{1}{c}{2D} & \multicolumn{1}{c}{3D} & \multicolumn{1}{c}{2D} & \multicolumn{1}{c}{3D} \\
		
		\midrule
		
		\multirow{2}{*}{LOS}							
		& 3db Access &    9.7     &    36.8    &    5.4     &    14.8    \\ 
		& Decawave   &    12.4    &    42.7    &    7.8     &    19.1    \\

		\midrule
		
		\multirow{2}{*}{\makecell[l]{NLOS}}					 
		& 3db Access &    18.9    &    72.2    &    10.5    &    34.1    \\ 
		& Decawave   &    22.9    &    89.1    &    23.5    &    43.7    \\
		
		\bottomrule
	\end{tabular}
	
	\centering
	\label{tab:localization-error-statistics}
\end{table}

\begin{figure}[t!]
	\centering
	\includegraphics[width=0.49\textwidth]{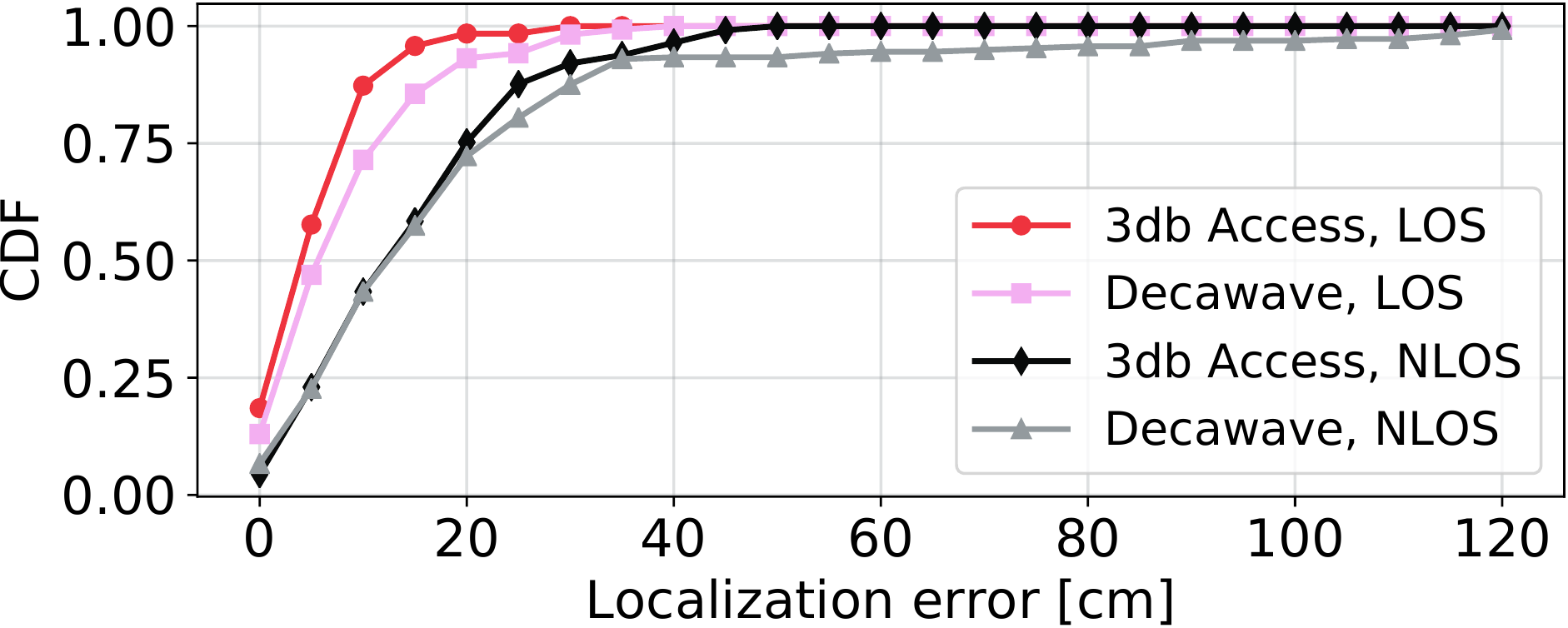}
	
	\caption{The CDF of 2D localization errors obtained from real measurements with 3db and Decawave devices.}
	\label{fig:cdf_localization_error}
\end{figure}

We compute the \textit{localization} error as the Euclidean distance between the true location and the estimated one, either in 2D or 3D. For the 3D case, the error is:
\begin{equation}
e = \sqrt{(x - \hat{x})^2 + (y - \hat{y})^2 + (z - \hat{z})^2},
\end{equation}
where $(x, y, z)$ are the Cartesian coordinates of the true location and $(\hat{x}, \hat{y}, \hat{z})$ are the Cartesian coordinates of the estimated location. For the 2D case, only the $x$ and $y$ coordinates corresponding to the plane parallel to the ground are taken into account. 

The localization algorithm always computes the location in 3D but we analyze the 2D and 3D errors separately because the selected multilateration algorithm is prone to large errors on the $z$ axis (height). This happens especially when measurements are noisy since the geometric dilution of precision (GDOP) on the $z$ axis is large. Therefore, it is important to distinguish errors in the 2D plan parallel to the ground, which usually need to be the smallest, from 3D errors. 

\tablename~\ref{tab:localization-error-statistics} presents the mean and standard deviation of the localization error of both devices in 2D and 3D. \figurename~\ref{fig:cdf_localization_error} shows the CDF of the localization error of the devices only in 2D. In LOS, the two devices have the mean and standard deviation of localization errors within at most \SI{5.9}{\centi\meter} of each other, leading to a similar performance. In the NLOS scenario, the mean and standard deviation of the localization errors of the Decawave-based localization system are higher by \SIrange{4}{16.9}{\centi\meter} than 3db's because of the higher distance errors between the tag and the NLOS anchor (\texttt{A4}). In 3D, the average localization errors are significantly higher than in 2D, with about \SI{30}{\centi\meter} in LOS and \SI{60}{\centi\meter} in NLOS, due to the measurement noise and high GDOP on the $z$ axis. In the 2D case, \SI{90}{\percent} of the LOS errors are under \SI{20}{\centi\meter}, while in NLOS \SI{75}{\percent} of the localization errors are under the same threshold.

\section{Error Modeling and Simulation}
\label{sec:simulation}

The localization results in Section~\ref{ssec:localization-hrp-lrp} are useful for comparing the two types of devices and for providing an estimate of the expected localization error in a small setup. We are now interested in evaluating the expected performance of a localization system that would be deployed on a larger scale (e.g., on the entire floor of an office building, in a shopping center, in a home). In such a setup, we expect a larger distance between anchors and possibly a lower anchor density than in our small-scale experiment. Moreover, while most of the existing literature assumes all anchors to be in the same room and preferably in LOS with the tag for the best accuracy, we argue that in real deployments such a requirement would be too costly in terms of price, setup effort, and maintenance. In such cases, administrators might prefer a localization system with lower accuracy but also lower setup costs.

In this section, we model the ranging errors obtained with 3db devices in the same LOS and NLOS scenarios as in Section~\ref{ssec:distance-meas}, but with augmented data sets. We argue that the proposed statistical models can be used to simulate realistic localization scenarios that would otherwise take days or weeks to implement and evaluate. Section~\ref{ssec:error-modeling} describes the proposed models and explains why a customized approach is needed to model errors obtained with different types of obstructions. In Section~\ref{ssec:building-deployment-simulation}, we evaluate the localization error of a simulated localization system based on 3db Access devices, where the anchors are in different rooms, which would be characteristic for a low-cost deployment. We also analyze the effect that different types of walls (made of concrete or gypsum) have on the localization error.

\subsection{Error Modeling}
\label{ssec:error-modeling}

\begin{table}[t!]
	\caption{Database of measurements with 3db Access Devices}
	\centering
	\begin{tabular}{l
			c
			S[table-format=1]
			S[table-format=5]
		} 
		\toprule
		Scenario
		& \makecell[c]{Distances\\ $[$\si{\meter}$]$}
		& \multicolumn{1}{c}{\makecell[c]{Nr.\\locations}}
		& \multicolumn{1}{c}{\makecell{Nr.\\ meas.}} \\ 
		\midrule
		
		LOS & 1, ..., 8 & 3 & 18344 \\
		
		\makecell[l]{NLOS with drywall} & 1, ..., 6 & 2 & 6600 \\
		
		\makecell[l]{NLOS with concrete wall} & 1, ..., 10.5 & 5 & 14868 \\
		
		\makecell[l]{NLOS with human body} & 1, ..., 10 & 3 & 8144  \\
		\bottomrule
	\end{tabular}
	\label{tab:database_info}
\end{table}

\begin{figure*}[t!]
	\centering
	
	\centering
	\subfloat[]{%
		\label{fig:fit_los}%
		\includegraphics[width=.235\linewidth]{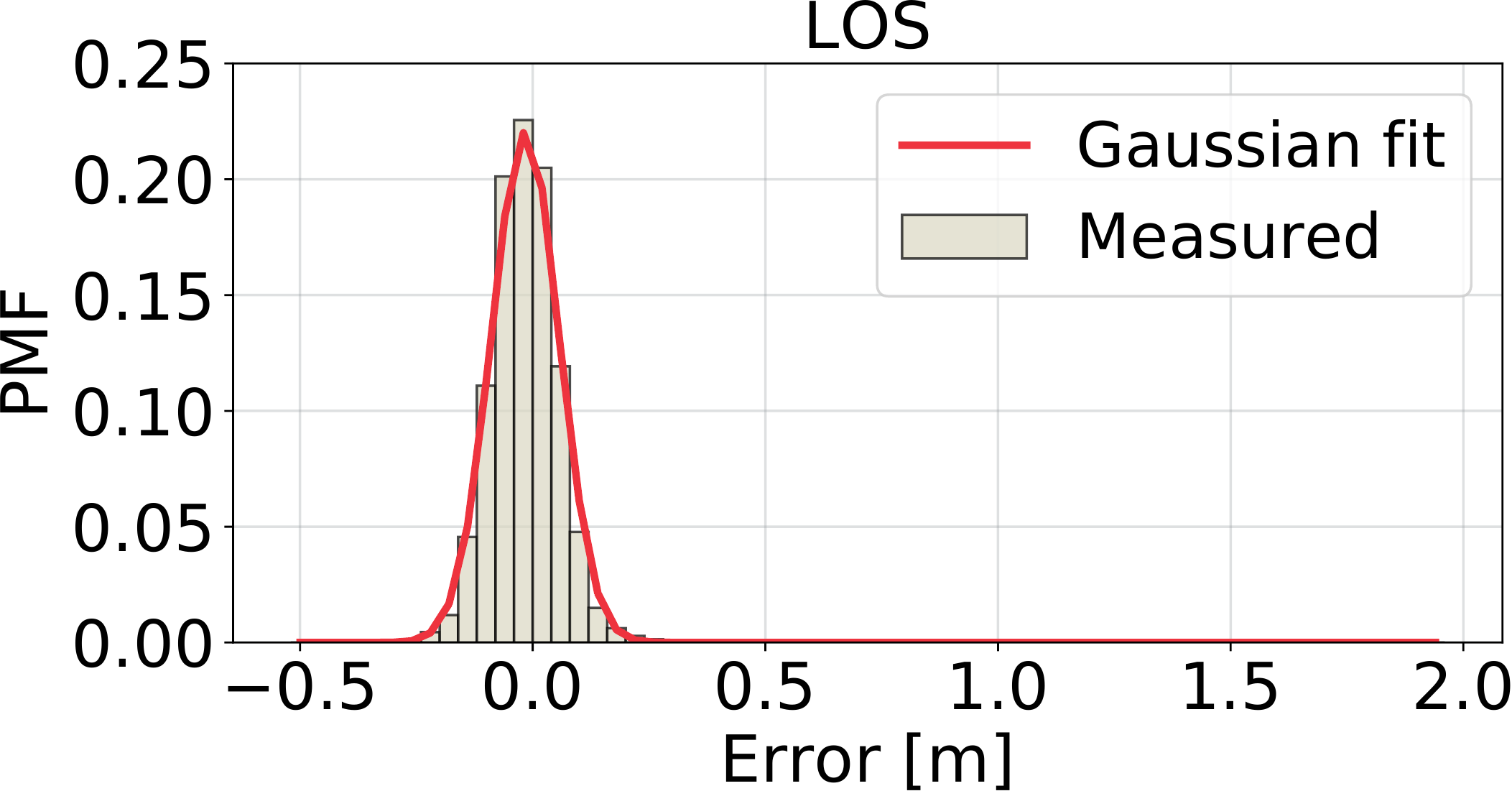}%
	}\quad%
	\subfloat[]{%
		\label{fig:fit_drywall}%
		\includegraphics[width=.235\linewidth]{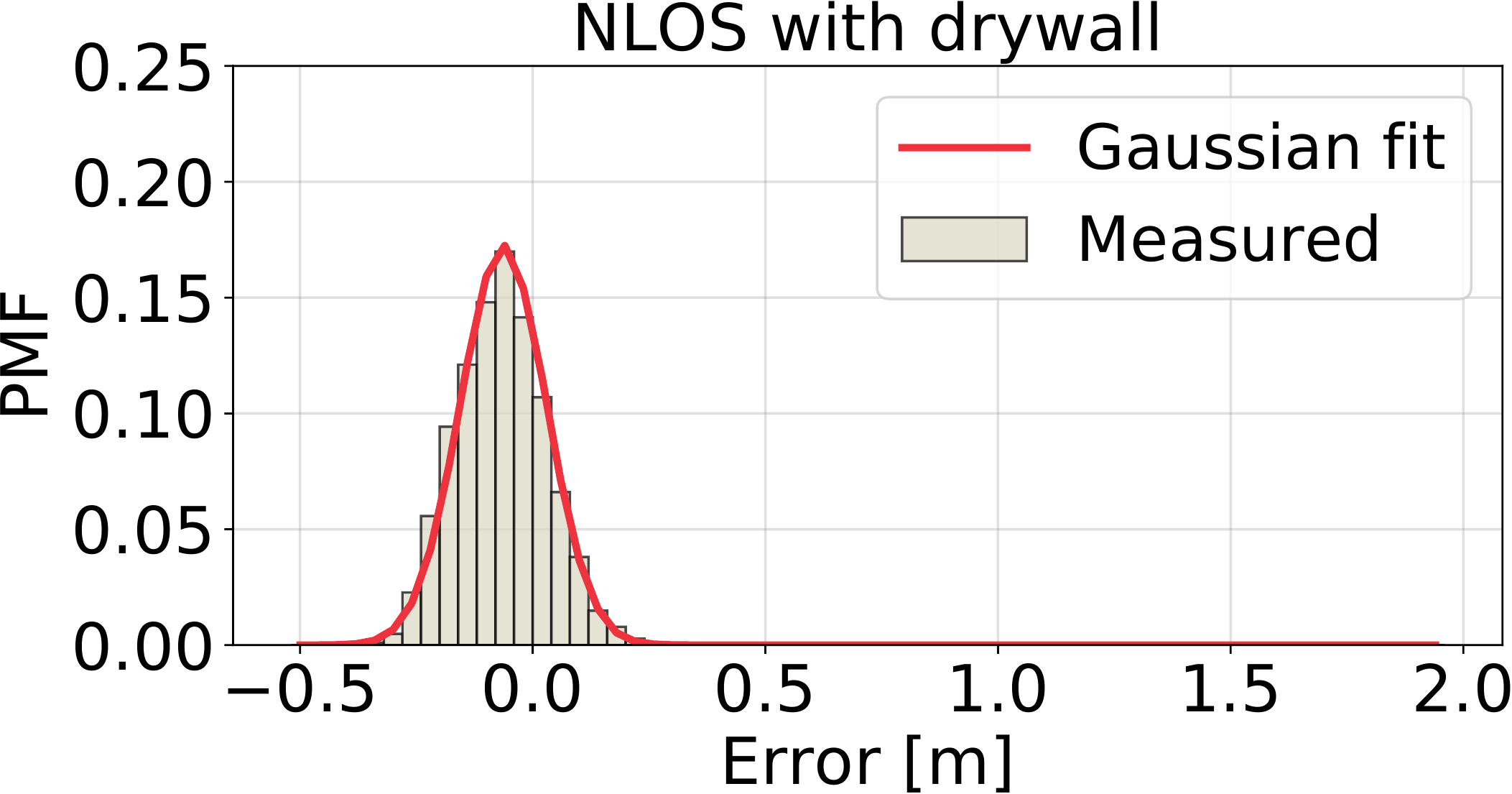}%
	}\quad%
	\subfloat[]{%
	\label{fig:fit_concrete_wall}%
		\includegraphics[width=.235\linewidth]{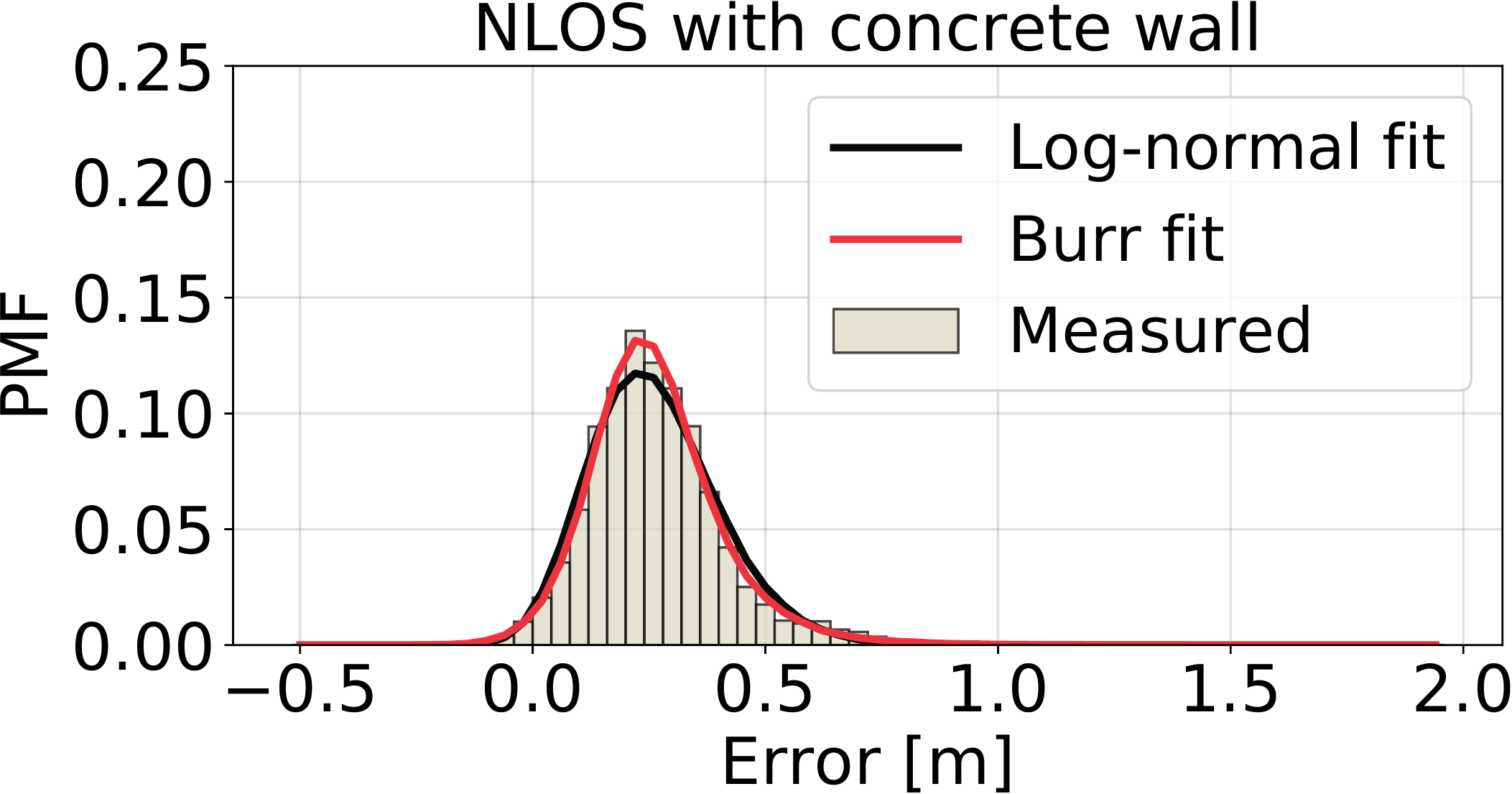}%
	}\quad%
	\subfloat[]{%
	\label{fig:fit_human}%
	\includegraphics[width=.235\linewidth]{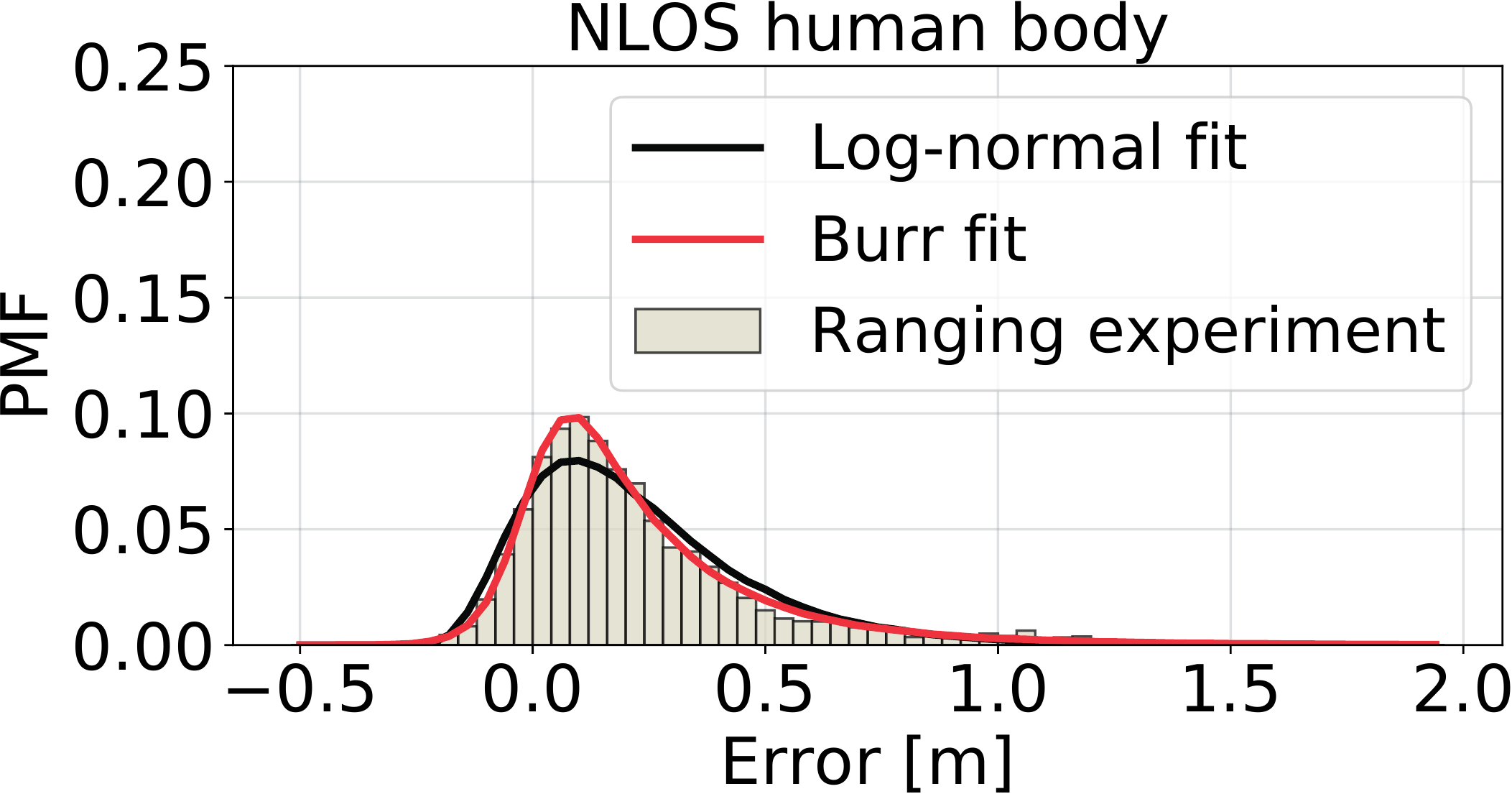}%
	}
	
	\caption{The histograms of distance errors based on data from the ranging experiment in LOS, NLOS with drywall, NLOS with concrete wall, and NLOS with human body, and the distributions that best model them.}
	\label{fig:simulation_fitting}
\end{figure*}

\newcommand{\LosMu}{0.004}
\newcommand{\LosSigma}{0.071}

\newcommand{\NlosDrywallMu}{-0.043}
\newcommand{\NlosDrywallSigma}{0.092}

\newcommand{\NlosConcreteBurrC}{9.64}
\newcommand{\NlosConcreteBurrD}{0.98}
\newcommand{\NlosConcreteBurrMu}{-0.46}
\newcommand{\NlosConcreteBurrSigma}{0.72}

\newcommand{\NlosConcreteLogS}{0.17}
\newcommand{\NlosConcreteLogMu}{-0.53}
\newcommand{\NlosConcreteLogSigma}{0.81}

\newcommand{\NlosHumanBurrC}{32.84}
\newcommand{\NlosHumanBurrD}{0.24}
\newcommand{\NlosHumanBurrMu}{-1.63}
\newcommand{\NlosHumanBurrSigma}{1.66}

\newcommand{\NlosHumanLogS}{0.44}
\newcommand{\NlosHumanLogMu}{-0.30}
\newcommand{\NlosHumanLogSigma}{0.50}

\newcommand{\eqtablewidth}{10.5cm}
\begin{table*}[t!]
	\caption{Parameters of the distributions used to model distance errors.}
	\centering
	\begin{tabular}[t]{@{} l l l
			S[table-format=1.2]
			S[table-format=1.3]
			S[table-format=1.3] @{}
		} 
		\toprule
		
		Scenario
		& Distribution
		& PDF
		& \multicolumn{1}{c}{Shape} & \multicolumn{1}{c}{$\mu$} & \multicolumn{1}{c}{$\sigma$} \\
		
		\midrule
		
		LOS & Gaussian & 
			\parbox{\eqtablewidth}{
				\begin{fleqn}
					\begin{equation}
					\label{eq:los-fit}
					f(x|\mu, \sigma) = \frac{1}{\sigma \sqrt{2\pi}} \exp^{-\frac{1}{2} \big(\frac{x - \mu}{\sigma}\big)^2}
					\end{equation}
				\end{fleqn}
			}
			& \multicolumn{1}{c}{--} & \LosMu & \LosSigma \\
			
		\midrule
		
		NLOS drywall & Gaussian & 
			See Eq.~\eqref{eq:los-fit}
			& \multicolumn{1}{c}{--} & \NlosDrywallMu & \NlosDrywallSigma \\

		\midrule
		
		\multirow{5}{*}{\makecell[l]{NLOS\\ concrete wall}} 
			& Burr XII & 
			\parbox{\eqtablewidth}{
			    \modi{
				\begin{fleqn}
					\begin{equation}
					f(x|c, d, \mu, \sigma) = cd \Bigg(\frac{x-\mu}{\sigma}\Bigg)^{c-1} \Big/ \Bigg(1 + \Big(\frac{x-\mu}{\sigma}\Big)^{c} \Bigg)^{d+1}, x>=0, c, d>0
					\label{eq:burr12}
					\end{equation}
				\end{fleqn}
				}
			}
			& \multicolumn{1}{c}{\makecell{$c = \NlosConcreteBurrC$ \\ $d = \NlosConcreteBurrD$ }} & \NlosConcreteBurrMu & \NlosConcreteBurrSigma \\
		
			& Log-normal & 
			\parbox{\eqtablewidth}{
				\begin{fleqn}
					\begin{equation}
					\label{eq:nlos-concretewall-fit_lognorm}
					f(x|s, \mu, \sigma) = \frac{1}{s (x - \mu) \sqrt{2\pi}} \exp \Bigg( -\frac{\ln^2 \Big(\frac{x - \mu}{\sigma}\Big) }{2s^2} \Bigg), x > 0, s>0
					\end{equation}
				\end{fleqn}
			} 
			& \multicolumn{1}{c}{$s = \NlosConcreteLogS$ } & \NlosConcreteLogMu & \NlosConcreteLogSigma \\
			
		\midrule
			
		\multirow{3}{*}{NLOS human} 
		& Burr XII &
		\modi{See Eq.~\eqref{eq:burr12}}
		& \multicolumn{1}{c}{\makecell{$c = \NlosHumanBurrC$ \\ $d = \NlosHumanBurrD$ }} & \NlosHumanBurrMu & \NlosHumanBurrSigma \\
		
		\\
		
		& Log-normal 
		&  See Eq.~\eqref{eq:nlos-concretewall-fit_lognorm}
		& \multicolumn{1}{c}{$s = \NlosHumanLogS$} & \NlosHumanLogMu & \NlosHumanLogSigma \\
		
		\bottomrule
	\end{tabular}
	
	\centering
	\label{tab:fitted_distr_params}
\end{table*}

In this section, we propose error models for distance measurements acquired in the same scenarios from Section~\ref{ssec:distance-meas}: LOS, NLOS with drywall, NLOS with a concrete wall, and NLOS with human body shadowing. For more statistically significant results, we augmented the former data sets with more measurements in at least two locations, so that the results are not biased by the multipath profile of a single room. 
\tablename~\ref{tab:database_info} lists the number of measurements from each scenario, the range of distances covered, and at how many locations we acquired measurements. The NLOS with concrete wall dataset includes measurements performed through walls with thickness between \SIrange{25.5}{67.5}{\centi\meter}. For the LOS and NLOS with human body scenarios, we also included the measurements of anchors in LOS and NLOS, respectively, with the tag. 

To model errors, we fitted some of the most well-known continuous distributions (the complete list is available at \cite{sicpy_cont_distr}) and computed the sum of squared errors (SSE) between the empirical PDF ($\tilde{y}_i$) and each fitted PDF at discrete points $i$:
\begin{equation}
SSE = \sum_{i}(y_i - \tilde{y}_i)^2.
\end{equation}

The best parameters $\hat{\bm{\theta}}$ for a specific probability distribution $p(\bm{x})$ are found by maximizing a likelihood function:
\begin{equation}
	\hat{\bm{\theta}} = \arg\max_{\bm{\theta}} p(\bm{x}|\bm{\theta}),
\end{equation}
over the entire parameter space. In other words, maximum likelihood estimation (MLE) selects the parameters under which the observed data is the most probable. The returned parameters are not guaranteed to be globally optimal. Where necessary, we provided good initial guesses for the optimization to improve the fit.

We present the distributions that minimized the SSE and their parameters obtained through MLE. If there were more distributions that achieved similar SSEs, we chose the most well-known and studied distributions. \tablename~\ref{tab:fitted_distr_params} shows the distributions that best fit experimental data obtained in the four scenarios, their PDF, as well as the parameters of the best fit. 

As illustrated in \figurename~\ref{fig:fit_los}, LOS errors can be modeled with a Gaussian distribution with a mean of \SI{\fpeval{\LosMu * 100}}{\centi\meter} and a standard deviation of \SI{\fpeval{\LosSigma * 100}}{\centi\meter}, whose PDF is given in Eq.~\eqref{eq:los-fit} from \tablename~\ref{tab:fitted_distr_params}. The calibration process presented in Appendix~\ref{ssec:distance-calib} removed biases caused by different channels, hardware, and distances such that in regular LOS conditions distance measurement errors are approximately centered around \SI{0}{\meter}.

Errors obtained in NLOS with a gypsum wall between the devices can also be modeled by a Gaussian distribution, as shown in \figurename~\ref{fig:fit_drywall}. Although errors obtained by 3db Access devices on the \SI{6.5}{\giga\hertz} channel through this obstacle had a left-sided tail (see \figurename~\ref{fig:los_pdf_dist_error}), when aggregating data from all channels and from an additional experiment, the errors converge to a Gaussian with a bias of \SI{\fpeval{\NlosDrywallMu * 100}}{\centi\meter} and a slightly larger standard deviation than in LOS, of \SI{\fpeval{\NlosDrywallSigma * 100}}{\centi\meter}. It is not clear why the gypsum wall causes negative biases. Its relative permittivity was found to be between $2.7$--$3.1$~\cite{zhekov2020dielectric}, higher than the relative permittivity of air, so the signal should travel at a lower speed through the obstacle, causing a delay. Since this delay is not reflected in the measurements, additional investigation is needed to determine whether other environmental factors are responsible for this bias. The main take-away is that gypsum walls introduce errors comparable to LOS propagation.

The scenarios where two devices are in NLOS with a concrete wall or with human body shadowing can be categorized as ``hard'' NLOS scenarios and introduce larger errors with heavier tails, as can be seen from \figurename~\ref{fig:fit_concrete_wall} and~\ref{fig:fit_human}, respectively.
We obtained the best fits for the Burr distributions type XII, also known as the Singh–Maddala distribution (Eq.~\eqref{eq:burr12}).
The Burr type XII is part of the family of log-logistic distributions used to model data that increases in an initial phase and then decreases, such as wealth distribution, survival analysis, or mortality rate~\cite{Kleiber2008}. Its shape is similar to the more well-known log-normal distribution but can better handle heavier tails~\cite{Kleiber2008}, as it is currently the case with our hard NLOS data. For the sake of completeness, we presented both the Burr and log-normal fits for the NLOS with concrete wall and human body shadowing scenarios, but also because the log-normal distribution has fewer parameters and is easier to interpret. The log-normal distribution has also been previously used in literature to model NLOS scenarios~\cite{molisch2003channel}. The PDF of the log-normal function is given in Eq.~\eqref{eq:nlos-concretewall-fit_lognorm} with a parametrization in which $s$, $\mu$, and $\sigma$ are also known as the shape, location, and scale parameters, respectively.
Given that Burr distributions are a better fit than the log-normal one for our experimental data, this suggests that such NLOS obstructions might introduce heavier tails than previously thought, especially in the case of human body shadowing where the Burr type III is a noticeably better fit than the log-normal (\figurename~\ref{fig:fit_human}). 

The chosen distributions can be used to simulate different localization scenarios. Our analysis shows that different types of obstructions can introduce very different errors and that a one-size-fit-all error model for NLOS propagation would likely lead to unrealistic results. Therefore, when evaluating the expected localization accuracy of a particular setup, different error models should be taken into account depending, for instance, on the crowdedness of the room or its wall structure. 

\subsection{Building Deployment}
\label{ssec:building-deployment-simulation}

\begin{figure}[t!]
	\centering		
	\includegraphics[width=0.37\textwidth]{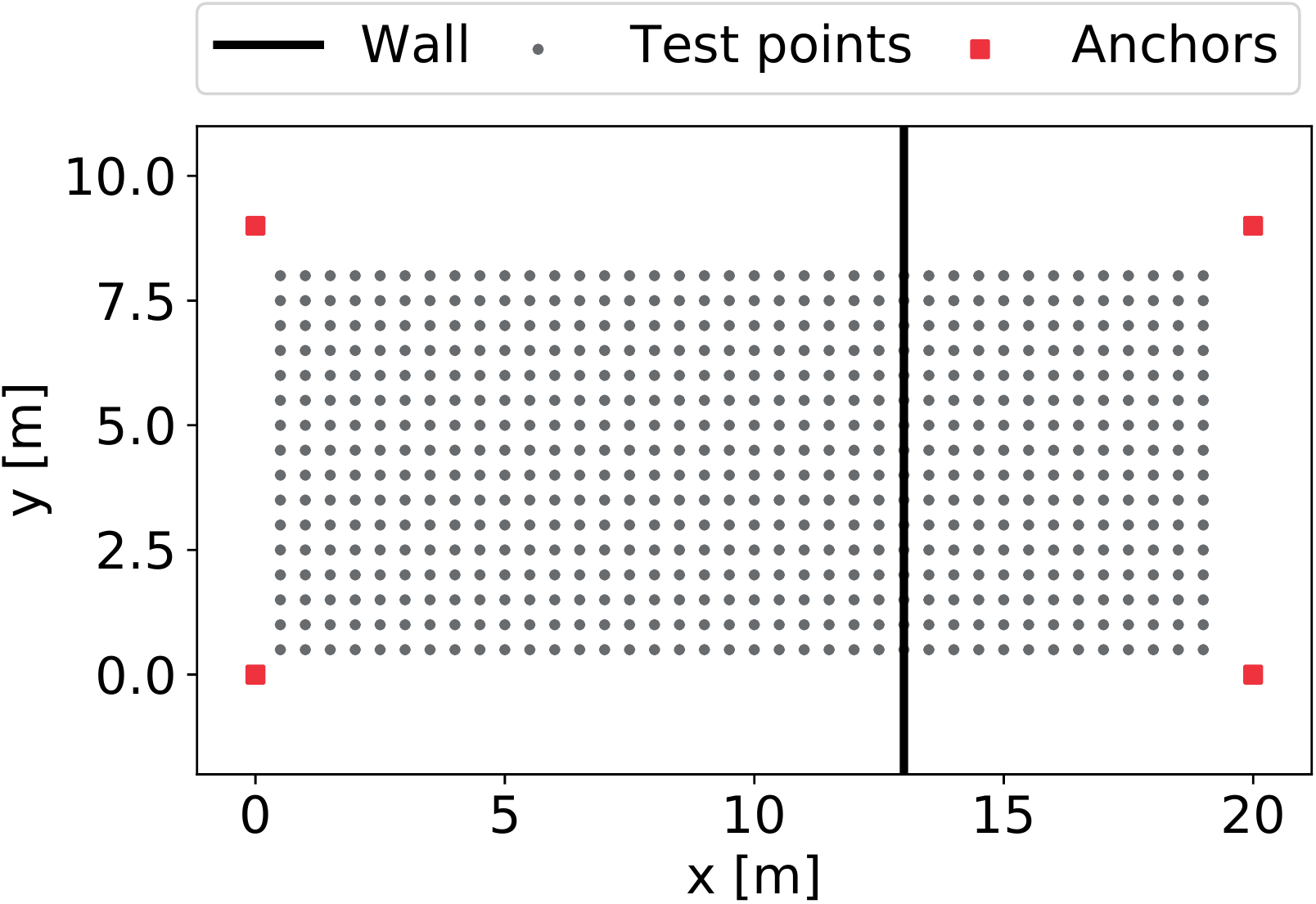}
	
	\caption{Simulation setup with four anchors. We consider a LOS case, in which there is no wall between the anchors, and two NLOS cases, in which the anchors are separated by a gypsum or concrete wall shown in the figure. We consider locations of the tag spread uniformly within the tracking area, in steps of \SI{0.25}{\meter} (the figure shows steps of \SI{0.5}{\meter} for better visibility).}
	\label{fig:simulation_setup}
\end{figure}

\begin{figure}[t!]
	\centering		
	\includegraphics[width=0.49\textwidth]{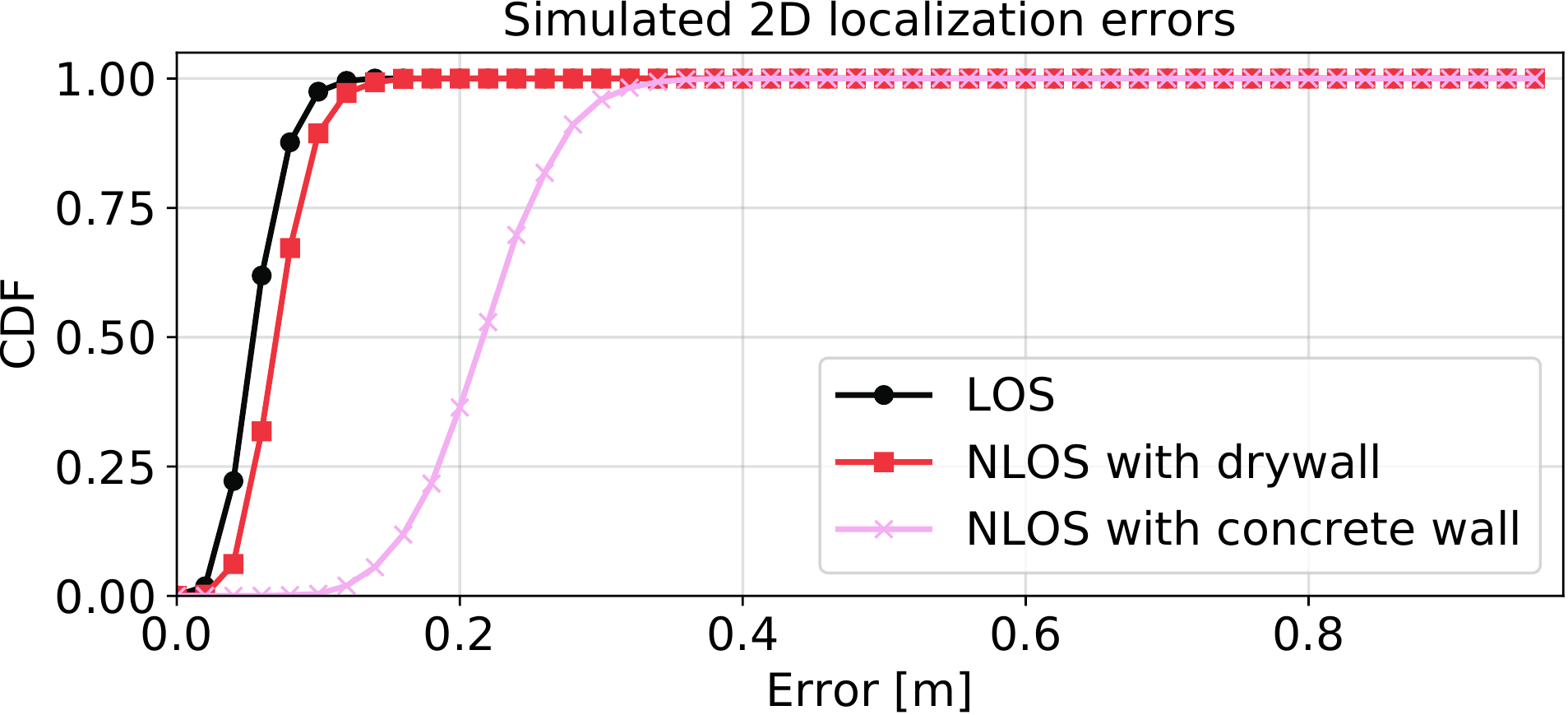}
	
	\caption{The CDF of simulated localization errors with four anchors when the anchors are at all times in LOS with the tag or when the anchors are in different rooms delimited by a drywall or a concrete wall.}
	\label{fig:simulation_localization_cdf}
\end{figure}

In this section, we illustrate an example in which the proposed statistical models can be used to evaluate the expected localization accuracy. We consider the setup from \figurename~\ref{fig:simulation_setup}, with four anchors placed in the corners of a space of $9\times$\SI{20}{\meter}, which is the area of our office space together with a meeting room. We consider a LOS scenario, in which there are no separating walls in the tracking area such that the anchors and the tag are at all times in LOS, and two NLOS scenarios when there is either a concrete or a drywall separating the area into two rooms of $9\times$\SI{13}{\meter} and $9\times$\SI{7}{\meter} (corresponding to the office and meeting room, respectively). The anchors are placed close to the ceiling: two (opposing) anchors at a height of \SI{3}{\meter} and the other two at \SI{2.7}{\meter}. We consider all possible locations of the tag over the tracking area in steps of \SI{0.25}{\meter}, similar to the points shown in \figurename~\ref{fig:simulation_setup}. At each test point, we simulate the measured distance between the tag and each anchor by adding an error term to the true distance. The error term is sampled from the distributions from \tablename~\ref{tab:fitted_distr_params}, based on the condition: LOS, NLOS with drywall, or NLOS with concrete wall.
We run the simulation for each scenario five times and compute the error statistics over the errors obtained in all runs.  

The CDFs of the 2D localization errors obtained in the three scenarios are presented in \figurename~\ref{fig:simulation_localization_cdf}. As expected based on the ranging experiments and error modeling, drywall does not introduce significant errors. The localization error when the space is separated by drywall is almost the same as if the wall were not there. The concrete wall, on the other hand, introduces a median 2D localization error of about \SI{25}{\centi\meter}. In some cases, this might be an acceptable localization error, given that half as many anchors are needed for the entire space than if four anchors (the minimum necessary for 3D localization) were deployed in each room. 

In a building with many rooms or delimitations (for instance, an office building), placing sets of four anchors in each room can be detrimental even if it improves the localization accuracy. First, devices in adjacent rooms (or even further away from each other) can be in communication range of each other so the anchors and users need to use different channels for communication and/or synchronize their transmissions in order to not interfere with each other. These issues are similar to challenges in the placement of base stations in cellular networks~\cite{andrews2011tractable}. Second, the tag would also need to select a subset of surrounding anchors with which to range based on the link quality between them (for instance, using the techniques described in~\cite{chen2011impact}) \textit{and} based on the geometry formed by the anchors, since the localization accuracy is the highest within the convex hull determined by the anchors~\cite{salman2011effects}. In addition, a higher number of deployed anchors increases the deployment costs. If the walls in a building introduce only small errors, these issues can be largely avoided by allowing walls within the tracking area encompassed by the minimum number of anchors. To establish whether this is the right approach for a given space, more work is needed, such as evaluating how two or more walls of different types influence the ranging error and developing a flexible, realistic simulator that outputs the best anchor configuration for a building plan while taking into account the cost--accuracy trade-off.	
	\section{Discussion}
\label{sec:discussion}

\subsection{Related work}
\label{ssec:previous-work}

\subsubsection{The Accuracy and Precision of UWB Ranging and Localization}

The reported NLOS bias and standard deviation of Decawave devices was $34 \pm$\SI{35}{\centi\meter} in office environments with different types of obstacles~\cite{jimenez2016comparing}, 
$15.6 \pm$\SI{7.4}{\centi\meter} with human body shadowing~\cite{tian_human_2019}, around $56\pm$\SI{20}{\centi\meter} when the path is blocked by concrete walls~\cite{schenck_information_2018}, and about $5 \pm$\SI{15}{\centi\meter} with shallow obstructions (such as plasterboard, wood, or steel)~\cite{schenck_information_2018}. The results for walls and shallow obstructions agree with our own observations but we obtained much larger errors with human body shadowing with both Decawave and 3db devices. In~\cite{delamare2019static}, an UWB localization system using Decawave devices achieved a localization accuracy of \SI{0.21}{\meter} in 2D and \SI{0.24}{\meter} in 3D in an industrial environment. 

In~\cite{ruiz2017comparing}, the authors compared the ranging and localization performance of three UWB devices which implement either the HRP PHY (Decawave DW1000), the LRP PHY (Ubisense), or a proprietary PHY (BeSpoon). Decawave achieved the best average localization error of \SI{0.5}{\meter}, followed by BeSpoon with \SI{0.71}{\meter}, and Ubisense with \SI{1.93}{\meter}. The article did not investigate the power and energy consumption of the devices. Ubisense devices have also been used in~\cite{shahi2012deterioration}, where they achieved sub-\SI{15}{\centi\meter} average error in LOS and about \SI{50}{\centi\meter} through a metallic enclosure. Another brand of LRP devices is PulsON (models P220 and the P400 series), but they implement a coherent interface. They have been used in~\cite{kumar2014experimental} for detecting a human target through the wall, and integrated in localization systems in~\cite{monica2014experimental, yan2015designing} that yielded localization errors under \SI{20}{\centi\meter}. PulsON and another brand of LRP devices from Multispectral Solutions Inc. (MSSI, now known as Zebra) were compared in~\cite{macgougan2009uwb}. PulsON and MSSI devices had average biases of \SI{10}{\centi\meter} and \SI{50}{\centi\meter}, respectively. 
The authors also noted the linear dependence on the errors with the distance and proposed a first-order linear model to correct them. 

Custom non-coherent UWB transceivers were proposed in~\cite{segura2010experimental} in an FPGA implementation and in~\cite{stoica2006low, fischer2008time} as integrated solutions, out of which only the former achieves errors lower than \SI{20}{\centi\meter}. To the best of our knowledge, the solutions are not compliant with the IEEE 802.15.4 standard nor are they available commercially.

A method for using channel diversity to improve the ranging accuracy (but only in LOS) was proposed in~\cite{kempke_surepoint_2016}. The authors leverage the constructive interference phenomenon to reduce the number of measurements needed for a single distance measurement, which can also decrease the energy consumption when applying diversity methods.

In our work, we focused on a basic localization algorithm, namely the regularized Gauss-Newton trilateration, since we focused on comparing the localization performance of two different devices. More complex algorithms which leverage constraints common in localization problems can increase the accuracy even further. For instance, in~\cite{beuchat2019enabling}, the authors proposed an optimization-based localization algorithm suitable for IoT devices and implemented it using Decawave UWB chipsets. The proposed method achieves $2$--$3\times$ higher localization accuracy than standard trilateration methods.

\subsubsection{The Energy Efficiency of UWB devices}

To the best of our knowledge, we are the first to compare the power and energy consumption of devices implementing the LRP and HRP PHYs as defined in the IEEE 802.15.4 standard~\cite{802-15-4}. Energy efficiency is starting to be a concern in UWB devices and, recently, concurrent ranging has been proposed as a solution to minimize the energy consumption in localization tasks at the application level~\cite{corbalan2020ultra}. Using this method, a tag can compute its location based on the time difference of arrival between multiple anchors' responses which are processed in a \textit{single} message (rather than $2\times N$ messages, where $N$ is the number of anchors, as in SS-TWR-based trilateration). In \cite{tiemann2019scalability}, the authors analyzed the energy efficiency of UWB localization systems depending on the association and synchronization demands. In the new DW3000 UWB chipset, Decawave claims to have improved the power consumption, but it is not yet clear by how much~\cite{dw3000_website}. There is also a newly-released, improved version of the 3db 3DB6830C chip, namely the ATA8352 chip~\cite{midas6-microchip}.

One option to decrease the energy consumption when the tag does not initiate the ranging is to use an ultra-low-power wake-up receiver to keep the tag active only before a message exchange and in the idle or deep sleep mode otherwise~\cite{niculescu2020energy}.

\subsubsection{NLOS Detection and Mitigation Techniques}

There is a large body of literature dedicated to NLOS detection and mitigation strategies~\cite{khodjaev_survey_2010}. Most works detect the NLOS condition using the statistical properties of the channel impulse response (CIR) of signals acquired in NLOS. These methods require knowledge of the CIR, which by default is not dumped by the device, and introduce additional processing times. Our open-source measurement database also includes the dumped CIRs of 3db Access devices, which could be used to evaluate the performance of these methods. It is noteworthy, however, that in our previous work~\cite{flueratoru2020energy}, the multipath components in the CIRs of 3db Access devices were found to be wider than those of Decawave because of their lower pulse bandwidth. Therefore, algorithms based on CIR characteristics might not generalize well to all types of devices. Other works proposed NLOS detection and mitigation methods that do not require prior knowldege on LOS/NLOS statistics, for instance based on sparse pseudo-input Gaussian processes (SPGP)~\cite{yang2018nlos} or on fuzzy theory~\cite{wen2017nlos}, which show promising results. Our measurement database contains also the CIRs of 3db devices, which can be used to test CIR-based NLOS detection and mitigation algorithms.

In~\cite{silva_ranging_2020}, the authors proposed a method to detect and correct NLOS biases when the signal propagates through a wall based on the wall's relative permittivity and thickness, which reduces the NLOS error by \SI{53}{\percent}. In the future, we will investigate whether this model fits the data from our experiments with both concrete and gypsum walls.

\subsubsection{Comparison with other Localization Technologies}
The power consumption of GNSS modules ranges from \SIrange{12}{72}{\milli\watt} for super-low-power modules~\cite{ublox-gps} to \SI{160}{\milli\watt} in a typical smartphone~\cite{carroll2010analysis}. The accuracy of GPS receivers is about \SI{2.5}{\meter} for high-end receivers~\cite{rychlicki2020analysis}, \SIrange{2}{5}{\meter} in smartphones~\cite{specht2019comparative}, and at least centimeter-level for real-time kinematic (RTK) GPS~\cite{feng2008gps}. Therefore, UWB devices can provide significantly higher localization accuracy indoors than GNSS receivers without enhancements; in terms of power consumption, LRP devices are comparable to super-low-power GNSS receivers, while the average power consumption of HRP devices exceeds the one of GNSS receivers in smartphones. However, given that the reception times might vary between the two technologies, their energy consumption per location might also be different and needs to be assessed in the future.

Bluetooth Low-Energy (BLE) has been also used in localization applications in recent years. Its power and energy consumption is similar to the one of 3db devices~\cite{siekkinen2012low} but its accuracy is at best \SIrange{0.7}{1}{\meter} with sensor fusion~\cite{robesaat2017improved}. Wi-Fi also achieves decimeter-level accuracy~\cite{kotaru2015spotfi} but has higher energy consumption~\cite{siekkinen2012low} than both BLE and UWB.

\subsection{Future Work}
\label{ssec:future-improvements}

As phone manufacturers started including UWB chips in smartphones, one vision would be the instrumentation of entire buildings with anchors to provide seamless positioning indoors. As briefly mentioned throughout the paper, there are still unsolved challenges to reach this goal. First, if anchors are placed in the same room in order to avoid NLOS measurements, buildings with many rooms or cubicles will need many localization cells (formed by the minimum number of anchors for 2D or 3D localization). If the cells are in range of each other, transmissions within multiple cells need to be synchronized or allocated to different bands, which increases the administration efforts. Moreover, since localization accuracy degrades at the edge of a cell, a hand-over protocol needs to be implemented at the tag to decide which anchors to select for localization at a given time. Allowing walls within one cell does not necessarily decrease the localization accuracy if the walls are shallow (such as gypsum walls) and can reduce the deployment effort and costs. However, more work is needed to model the ranging errors and maximum range through an arbitrary number of walls. Towards this end, a simulator for building deployments of UWB localization systems that recommends the optimal number of anchors and their placement for the desired accuracy--cost trade-off would be needed. 

In this work, we considered only range-based localization, where each distance measurement is obtained through at least two message exchanges between the tag and each anchor. This scheme does not scale well with many users and anchors because of the large number of messages involved and the need to schedule uplink transmissions. Moreover, this method is privacy-sensitive since anchors have access to at least part of the ranging information, which can be used to track the user with a certain precision. Instead, a GPS-like localization system based on the time-difference of arrival (TDOA) of signals, in which anchors act as satellites and passive tags use their broadcasts to locate themselves can, in theory, scale to an unlimited number of users and is more privacy-friendly. The drawback is that, in this case, anchors need to synchronize their transmissions. Wired synchronization introduces the lowest errors but is unlikely to be adopted because of the high deployment costs. Wireless synchronization, on the other hand, leads to localization errors in the range of decimeters even when anchors are in LOS with the tag~\cite{grobetawindhager2019snaploc} and requires anchors to be in LOS with each other. More work is needed, for instance, to determine whether a calibration protocol can allow wireless-synchronized TDOA systems where anchors are in different rooms to obtain a similar accuracy with the case in which anchors are in LOS with each other. 	
	\section{Conclusion}
\label{sec:conclusion}

This paper provided an outlook on the power and energy consumption, distance measurement statistics, and localization performance of 3db Access and Decawave devices, representative of the two types of UWB physical interfaces, LRP and HRP, respectively. 
\modi{Both devices have ranging and localization errors on the same order of magnitude. Decawave devices generally show better performance in LOS and through-the-wall NLOS conditions, while 3db devices have slightly better performance in NLOS with human body shadowing. For a similar maximum range, Decawave devices have $125\times$ higher energy consumption than 3db Access devices, while in the short-range mode (which decreases the range by at least $8\times$) they have $6.4\times$ higher energy consumption than 3db Access. Therefore, devices implementing the LRP PHY might be more suitable for ultra-low power applications, while the HRP PHY might be a better choice for the highest ranging accuracy.}

We evaluated the performance of 3db Access and Decawave devices in multiple LOS and NLOS (caused by a person, drywall, or concrete wall) scenarios and provided models for the error distribution of 3db Access distance measurements. These models can be further used to simulate realistic deployments of localization systems which would otherwise take days or even weeks to evaluate. 
We illustrated their applicability by simulating a localization scenario in which the anchors are placed in different rooms separated by drywall or a concrete wall. Results suggest that drywall causes negligible errors and anchors do not need to be in the same room to obtain high localization accuracy.  

More research is needed to evaluate the impact of multiple walls on the ranging accuracy and to create suitable models of the maximum range and distance measurement errors with an arbitrary number of walls. Using such models, a simulator of UWB localization systems could then recommend the minimum number of anchors for the desired localization accuracy and their placement. More work is also needed to select the best anchors at a particular moment based on their link to the tag and to synchronize UWB transmissions between clusters of anchors within range of each other.	
	\appendix[Distance Calibration]
\label{ssec:distance-calib}

Before comparing the ranging accuracy and precision of Decawave and 3db devices, a calibration step is also necessary since, as we will show, the raw distance error can depend on the hardware, on the channel used, and even on the distance. 

We collected distance measurements between different device pairs placed 2, 5, and \SI{10}{\meter} apart, for at least \SI{30}{\minute}, which was deemed a long enough time period to capture the long-term distribution of distance measurements. In a different experiment, we acquired measurements during \SI{24}{\hour} to obtain the long-term error distribution. The mean error obtained in windows longer than \SI{2}{\minute} was within $\pm$\SI{1}{\centi\meter} of the long-term mean error. We chose a longer window, of \SI{30}{\minute}, to obtain more stable distributions.

We recorded measurements on all the available channels (\SI{6.5}{\giga\hertz}, \SI{7}{\giga\hertz}, and \SI{7.5}{\giga\hertz}) using 3db devices and only on the \SI{6.5}{\giga\hertz} channel with Decawave, since it is the only available channel when using the default Decawave software for the MDEK1001 kit. Although the software compensates the antenna delay, in most cases distance measurements with Decawave devices still had a non-negative bias that was eliminated in the calibration process. 

\begin{figure}[t!]
	\centering
	\includegraphics[width=0.49\textwidth]{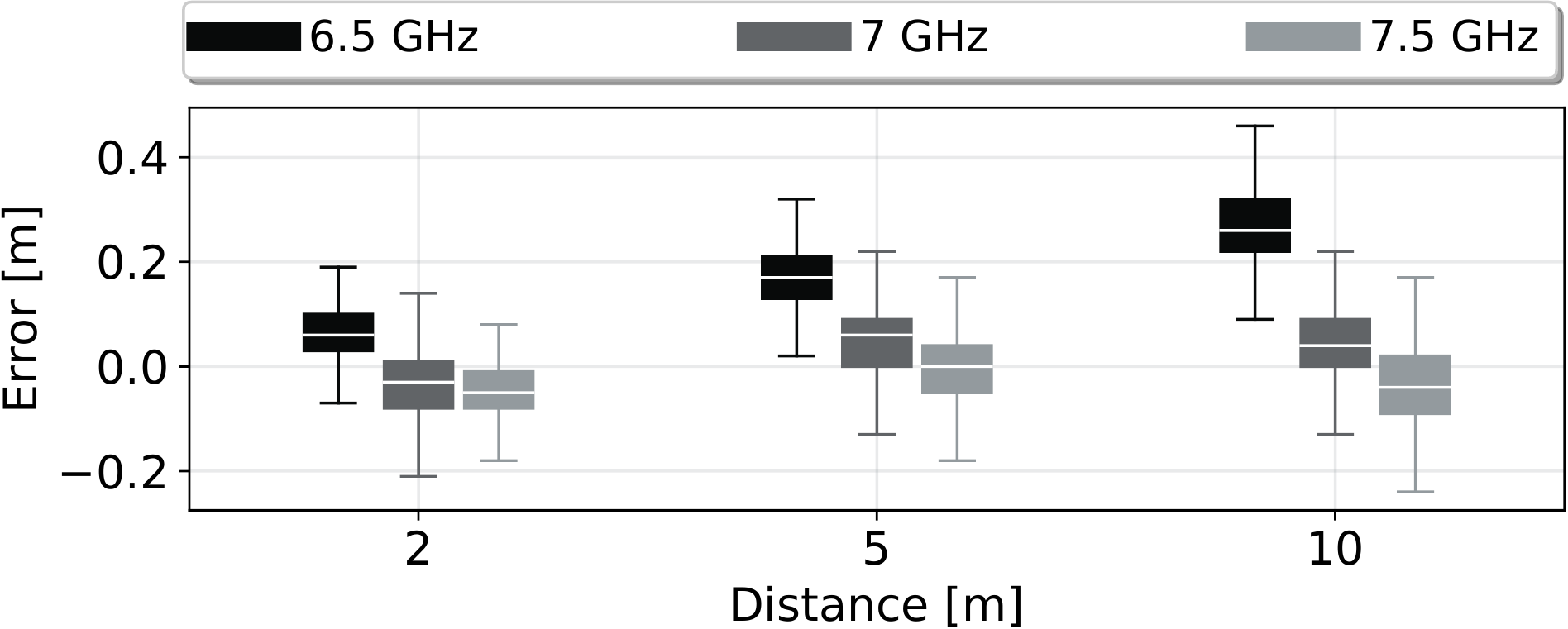}
	
	\caption{Comparison of the distribution of distance measurement errors of the same pair of 3db Access devices, on the $6.5$, $7$, and \SI{7.5}{\giga\hertz} channels, at distances of 2, 5, and \SI{10}{\meter} between the two devices. On the \SI{6.5}{\giga\hertz} channel, the mean error increases approximately linearly with the distance, while on the other two channels it is constant with the distance.}
	\label{fig:boxplots_020}
\end{figure}

Our findings suggest that the measurement bias of UWB devices varies with the distance, the channel, and the pair of devices used.
The dependency on the distance for the same channel can be seen in \figurename~\ref{fig:boxplots_020} for the \SI{6.5}{\giga\hertz} channel. 

The TOA estimation error increases with lower SNR~\cite{guvenc2005threshold}, so without proper calibration distance measurement errors increase with the measured distance. \figurename~\ref{fig:boxplots_020} also shows how the ranging error varies between channels because they have different amplitude saturation points. Fortunately, the channel is known at the time of the measurement and these errors can be compensated.

\begin{figure}[t!]
	\centering
	
	\centering
	\subfloat[]{%
		\label{fig:boxplots_alldevs-3db}%
		\includegraphics[width=.98\linewidth]{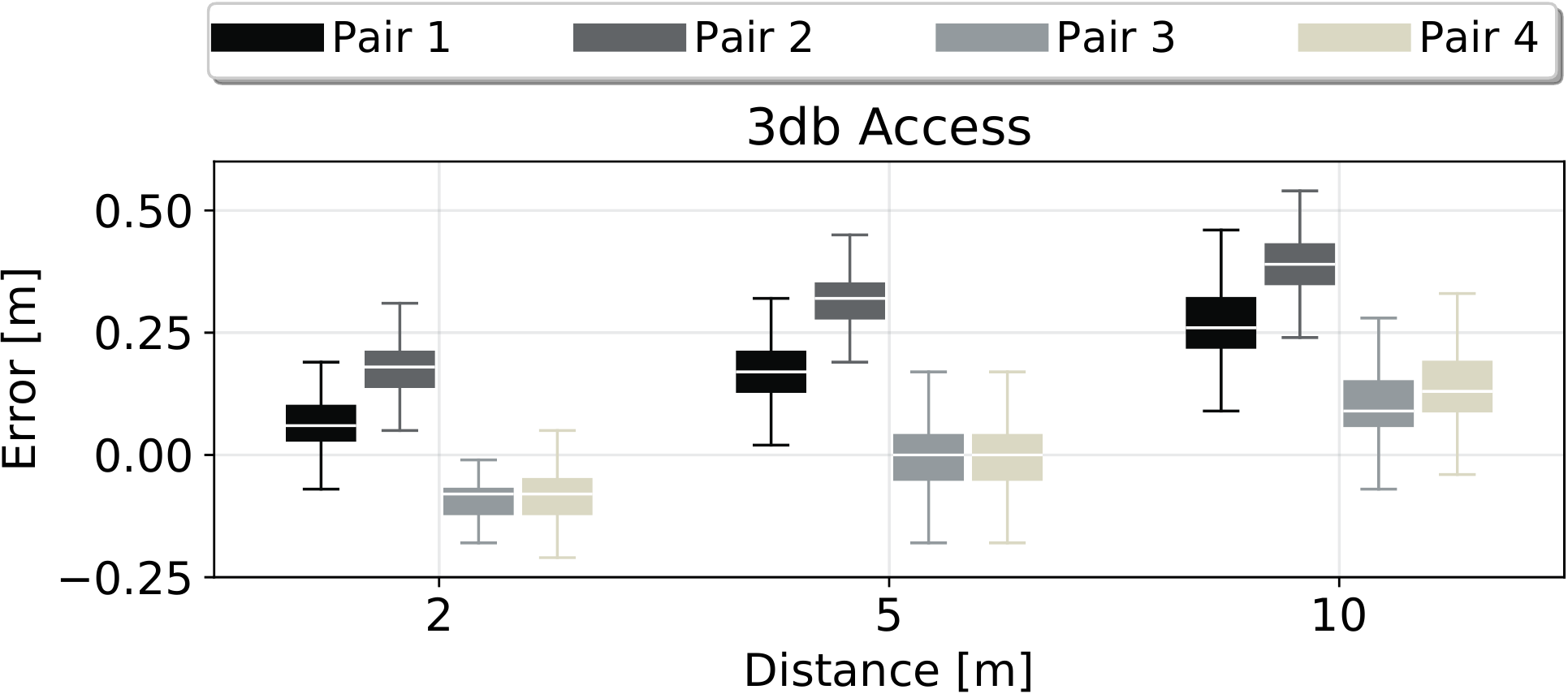}%
	} 
	
	\centering
	\subfloat[]{%
		\label{fig:boxplots_alldevs-dw}%
		\includegraphics[width=.99\linewidth]{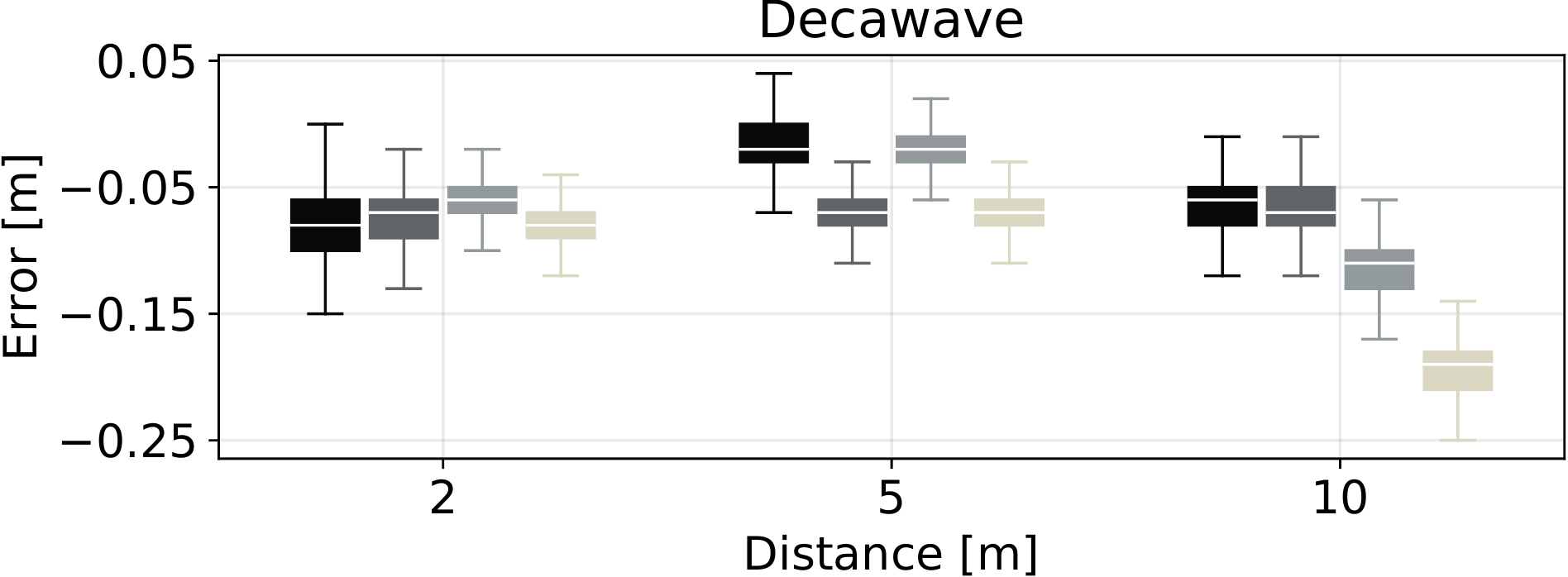}%
	}
	
	\caption{Comparison of the distribution of distance measurement errors of four pairs of (a) 3db Access and (b) Decawave devices on the \SI{6.5}{\giga\hertz} channel. The same transmitter (tag) was used in all pairs and only the receiver (anchor) was changed. The error varies with both the distance and the hardware.}
	\label{fig:boxplots_alldevs}
\end{figure}

The mean error also depends on the pair of devices used. This can occur due to the different clock offset or antenna delay of the hardware devices. Although these parameters can be estimated, the clock offset can still deviate in time or the antenna delay calibration might be imperfect. \figurename~\ref{fig:boxplots_alldevs} compares the error distributions of distance measurements acquired by different 3db and Decawave device pairs on the \SI{6.5}{\giga\hertz} channel. All pairs have a common transmitter (the tag used for localization in Section~\ref{ssec:localization-hrp-lrp}). For distances up to \SI{10}{\meter}, the mean error can vary with \SI{0.6}{\meter} for 3db devices and \SI{0.2}{\meter} for Decawave devices, but note that this range includes also the linear dependence of the distance error on the SNR.

An arbitrary non-negative measurement bias is not desirable in ranging or localization. Therefore, such errors are often eliminated during a calibration step. The desired measurement model in LOS is a zero-mean Gaussian, whose standard deviation is at least partly determined by the hardware capabilities. It is worth noting that while the mean distance error varies with up to \SI{0.6}{\meter} when we change the hardware, channel, or distance, the standard deviation of each error distribution is almost constant for each device (see Table \ref{tab:calibration-comp}). Since the raw error distribution is already Gaussian, we need to correct only its bias.

We consider each set of measurements $s_{ij}$ for a given pair of devices $p_i, i = 1,...,4$ and channel $c_j$, where $j = \{0, 1, 2\}$ corresponds, respectively, to channels at $6.5$, $7$, and \SI{7.5}{\giga\hertz}. Each set $s_{ij}$ contains an equal number of approximately 3000 measurements (equivalent to a recording time of \SI{30}{\minute}) taken at each distance $d_k \in \{2, 5, 10\}$ \si{\meter}. We assume a simple linear dependence of the measured distance on the true distance, which is found by minimizing the squared error:
\begin{equation}
E = \sum_{n=0}^{N}\lvert x_n \cdot p_0 + p_1 - y_n \lvert^2,
\end{equation}
where $N$ is the total number of measurements, $y_n$ is the measured distance at the true distance $x_n$ for all $n = 1,...,N$ and $p_0$ and $p_1$ are the polynomial coefficients.

Once the polynomial coefficients for a set $s_{ij}$ are computed, the measurements of that set can be corrected as follows:
\begin{equation}
x_c = \frac{x_m - p_1}{p_0},
\end{equation}
where $x_c$ and $x_m$ are, respectively, the calibrated and the raw measurements.

There are several caveats to this approach. First, we do not derive error coefficients for each factor that introduces errors (channel, pair, and true distance). Rather, the accumulated error is corrected for a particular set. 

Second, while it is feasible to obtain the calibration coefficients for a particular channel by performing measurements at different distances, the device-dependent calibration is harder to perform in a real application. Localization systems deployed inside buildings should be able to offer accurate locations even to unknown users, so one could not calibrate each anchor--tag pair. Even if all the users were known, for instance in a privately-deployed localization system, the tag population can easily reach hundreds or thousands devices, which again makes pair-wise device calibration impractical. Hardware-dependent distance errors are an interesting research topic but outside the scope of this paper, which is why we calibrated each pair of Decawave and 3db Access devices used (including the tag and each anchor in the localization experiment from Section~\ref{ssec:localization-hrp-lrp}).

Third, it is worth noting that the distance errors of Decawave devices do not linearly increase with the measured distance, as in the case of 3db Access. Instead, the dependency seems non-linear. Due to the lack of a more appropriate straight-forward model, we still apply a linear fitting with the mention that there might be better models for this distribution. Even so, the calibration reduces the measurement bias with \SI{8}{\centi\meter}, as shown in \tablename~\ref{tab:calibration-comp}, which presents the mean and the standard deviation of all errors before and after calibration. 

The calibration does not significantly reduce the \textit{overall} average error of 3db Access devices because the sets with a positive bias balance out those with a negative bias (predominantly on channels 1 and 2 which, for brevity, are not shown), but there is a benefit in the bias reduction of the individual sets. Moreover, the calibration reduces the standard deviation of measurement errors by more than half.

\begin{table}[t!]
	\caption{Distance error statistics before and after calibration.}
	\centering
	\begin{tabular}{l 
			l
			S[table-format=1.2]
			S[table-format=1.2]
		} 
		\toprule
		Device
		& Case
		& \multicolumn{1}{c}{\makecell{Mean [\si{\centi\meter}]}}
		& \multicolumn{1}{c}{\makecell{Standard\\deviation [\si{\centi\meter}]}} \\ 
		\midrule
		
		\multirow{2}{*}{3db Access}
		& Before calibration & -0.79 & 16.09 \\
		& After calibration & -0.05 & 6.54 \\
		
		\midrule
		
		\multirow{2}{*}{Decawave}
		& Before calibration & -7.56 & 4.79 \\
		& After calibration & 0.00 & 3.14\\
		\bottomrule
	\end{tabular}
	\centering
	\label{tab:calibration-comp}
\end{table}
	
	\section*{Acknowledgment}
	
	The authors gratefully acknowledge funding from European Union's Horizon 2020
	Research and Innovation programme under the Marie Sklodowska Curie grant agreement
	No.\ 813278 (A-WEAR: A network for dynamic wearable applications with privacy
	constraints, \url{http://www.a-wear.eu/}). This work does not represent the opinion of the
	European Union, and the European Union is not responsible for any use that might be
	made of its content.
	
	\ifCLASSOPTIONcaptionsoff
	\newpage
	\fi
	
	\bibliographystyle{ieeetr}
	\bibliography{main}

	\begin{IEEEbiography}[{\includegraphics[width=1in,height=1.25in,clip,keepaspectratio]{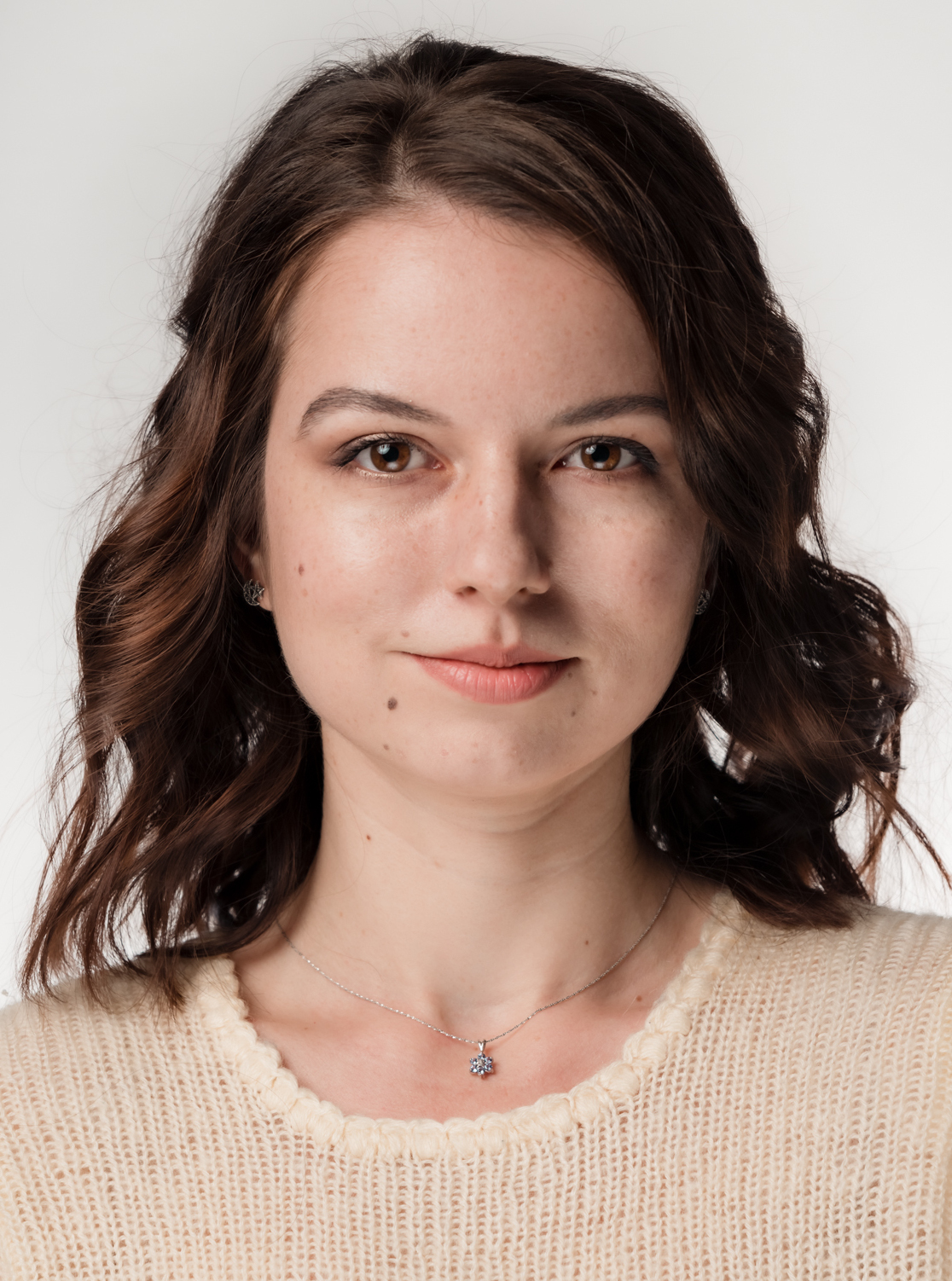}}]{Laura Flueratoru}
		obtained her Master’s degree in Electrical Engineering from ETH Zürich, Switzerland, in 2019, and her Bachelor's degree in Electronics and Telecommunications from University Politehnica of Bucharest, Romania, in 2017. She is currently pursuing a double Ph.D.\ degree at University Politehnica of Bucharest, Romania, and Tampere University, Finland, as a Marie Skłodowska-Curie Fellow in the European project A-WEAR.
		During her studies, she gained experience in both industry and research from internships at Freescale Semiconductor, École Polytechnique Fédérale de Lausanne (EPFL), Schindler Group. Her research interests include indoor localization, ultra-wideband communications, wireless and mobile communications, embedded systems, signal processing, and machine learning.
	\end{IEEEbiography}
	
	\begin{IEEEbiography}[{\includegraphics[width=1in,height=1.25in,clip,keepaspectratio]{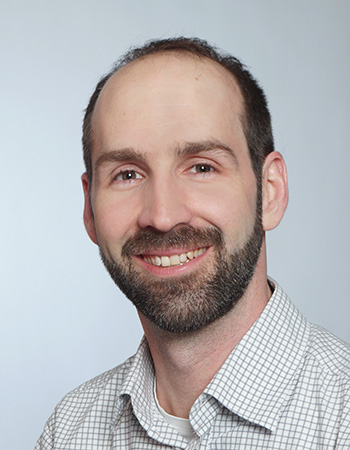}}]{Silvan Wehrli} received the  M.Sc.  degree  in  electrical  Engineering  and  Information  Technology  from ETH Z\"urich,  Switzerland, in 2005, and  the  Ph.D.  degree  from  the  Electronics  Laboratory,  ETH Z\"urich, Switzerland  for  his  thesis  “Integrated  Active  Pulsed  Reflector  for  an  Indoor Positioning System.” Currently, he is Vice President (VP) of Product Development at 3db Access. His expertise covers analog design, high-speed electronics and localization. Prior to 3db, Silvan used to work for Gigoptix/IDT in the area of high-speed optical communications. Silvan has 12+ years academic and industry experience. 
	\end{IEEEbiography}

	\begin{IEEEbiography}[{\includegraphics[width=1in,height=1.25in,clip,keepaspectratio]{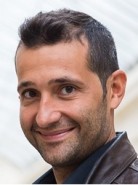}}]{Michele Magno}
		(Senior Member, IEEE) received the Master’s and Ph.D.\ degrees in electronic engineering from the University of Bologna, Italy, in 2004 and 2010, respectively. Currently, he is a Senior Researcher at ETH Z\"urich, Switzerland, where he is the Head of the Project-Based Learning Center. He has collaborated with several universities and research centers, such as Mid University Sweden, where he is a Guest Full Professor. He has published more than 150 articles in international journals and conferences, in which he got multiple best paper and best poster awards. The key topics of his research are wireless sensor networks, wearable devices, machine learning at the edge, energy harvesting, power management techniques, and extended lifetime of battery-operated devices.
	\end{IEEEbiography}
	
	\vfill
	
	\begin{IEEEbiography}[{\includegraphics[width=1in,height=1.25in,clip,keepaspectratio]{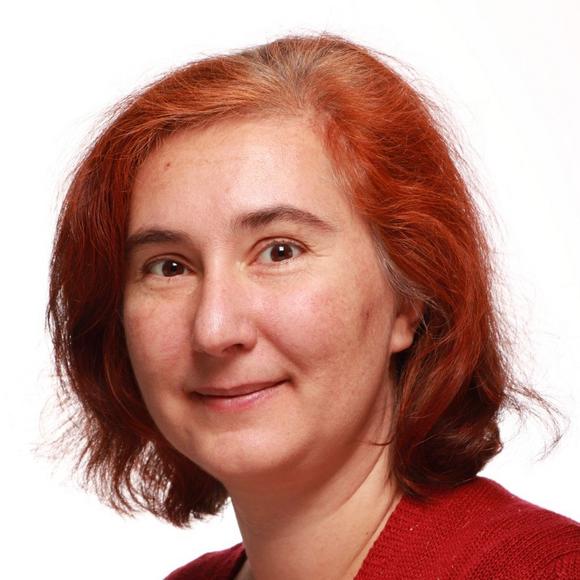}}]{Elena Simona Lohan}
		received an M.Sc.\ degree in Electrical Engineering from University Politehnica of Bucharest (1997), a D.E.A.\ degree in Econometrics at Ecole Polytechnique, Paris (1998), and a Ph.D.\ degree in Telecommunications from Tampere University of Technology (2003). She is now a full Professor at Electrical Engineering unit at Tampere University (TAU) and a Visiting Professor at Universitat Autonoma de Barcelona (UAB), Spain. She is leading a research group on Signal processing for wireless positioning. She is a co-editor of the first book on Galileo satellite system (Springer “Galileo Positioning technology”), co-editor of a Springer book on “Multi-technology positioning”, and author or co-author in more than 220 international peer-reviewed publications, 6 patents and inventions. She is also an associate Editor for RIN Journal of Navigation and for IET  Journal on Radar, Sonar, and Navigation. She is currently coordinating the A-WEAR MSCA European Joint Doctorate network in the field of wearable computing. Her expertise includes signal processing for wireless positioning and navigation, multipath and interference mitigation, and RF fingerprinting.
	\end{IEEEbiography}

	\begin{IEEEbiography}[{\includegraphics[width=1in,height=1.25in,clip,keepaspectratio]{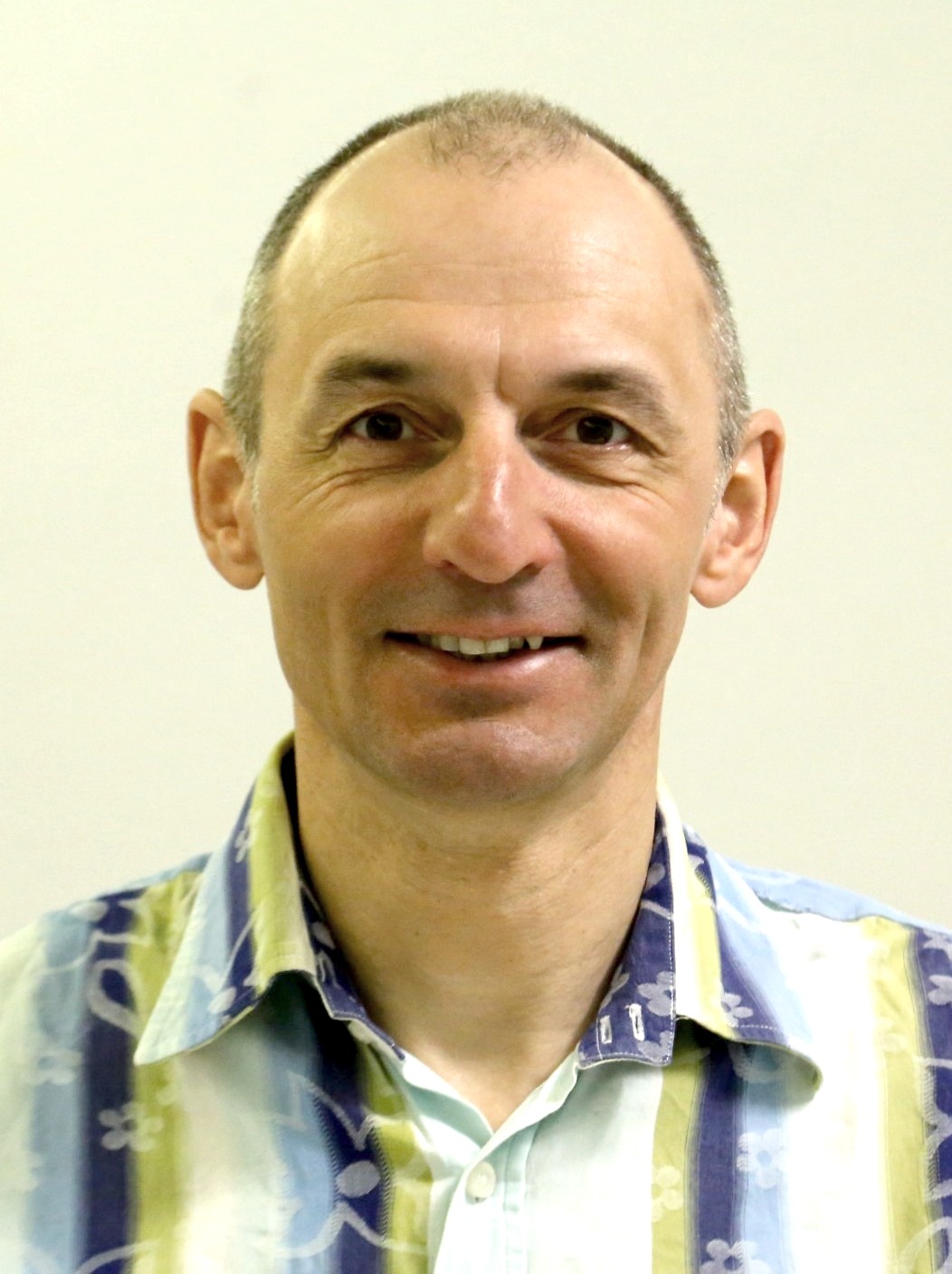}}]{Dragoș Niculescu} 
		obtained a Ph.D. in Computer Science from Rutgers University (New Jersey) in 2004, with a thesis on sensor networks routing and positioning. He spent five years as a researcher at NEC Laboratories America in Princeton, NJ, working on simulation and implementation of mesh networks, VoIP, and WiFi-related protocols. At University Politehnica of Bucharest he is currently teaching courses in Mobile Computing and Services for Mobile Networking; also researching mobile protocols, UWB, and 802.11 networking.
	\end{IEEEbiography}
	\vfill
\end{document}